\documentclass[12pt]{article}
\usepackage{titling}
\title{Doubly weighted M-estimation for nonrandom assignment and missing outcomes}
\author{Akanksha Negi$^\dagger$}
\date{November 23, 2020}
\usepackage{amsmath,amsfonts,bm,amsthm,amssymb,mathrsfs}
\usepackage{pdflscape,float, caption, threeparttable, booktabs}
\usepackage{graphicx, multirow, comment}
\usepackage{setspace, commath, float} 
\usepackage{mathtools}
\usepackage{siunitx, etoolbox}
\usepackage{caption,subcaption,sectsty}
\usepackage{times}
\usepackage[bottom]{footmisc}
\usepackage{multirow}
\usepackage[usenames,dvipsnames]{color}
\definecolor{crimson}{rgb}{0.76,0.13,0.28}
\definecolor{forestgreen}{rgb}{0.13, 0.55, 0.13}
\definecolor{dukeblue}{rgb}{0.0, 0.0, 0.61}
\definecolor{burgundy}{rgb}{0.5, 0.0, 0.13}
\definecolor{darkblue}{rgb}{0,0,.8}
\usepackage[round]{natbib}
\usepackage[hyperfootnotes=false]{hyperref}
\hypersetup{
colorlinks,
citecolor=black,
linkcolor=dukeblue,
urlcolor=black}
\usepackage{xcolor}
\usepackage{geometry}
\newcommand\independent{\protect\mathpalette{\protect\independenT}{\perp}}
\def\independenT#1#2{\mathrel{\rlap{$#1#2$}\mkern2mu{#1#2}}}

\newcommand\blfootnote[1]{%
\begingroup
\renewcommand\thefootnote{}\footnote{#1}%
\addtocounter{footnote}{-1}%
\endgroup
}

\newtheorem{theorem}{Theorem}

\newtheorem{corollary}{Corollary}
\newtheorem*{summary}{Summary}
\newtheorem{lemma}{Lemma}
\newtheorem{assumption}{Assumption}

\begin{document}
\maketitle	
\smallskip

\blfootnote{$^\ast$I am grateful to Jeffrey M. Wooldridge, Steven Haider, Ben Zou, and Kenneth Frank. Special thanks to Tim Vogelsang, Wendun Wang, Alyssa Carlson, Christian Cox, Tymon S{\l}oczy{\'n}ski, and seminar \& conference participants, for insightful comments and suggestions on earlier drafts of this paper.
}
\blfootnote{$^\dagger$Department of Econometrics and Business Statistics, Monash University. Email: \href{mailto:akanksha.negi@monash.edu}{akanksha.negi@monash.edu}, Webpage: \href{https://www.anegi.net/home}{www.anegi.net}}

\begin{spacing}{1.0}
\begin{abstract}
This paper proposes a new class of M-estimators that double weight for the twin problems of nonrandom treatment assignment and missing outcomes, both of which are common issues in the treatment effects literature. The proposed class is characterized by a `\textit{robustness}' property, which makes it resilient to parametric misspecification in either a conditional model of interest (for example, mean or quantile function) or the two weighting functions. As leading applications, the paper discusses estimation of two specific causal parameters; average and quantile treatment effects (ATE, QTEs), which can be expressed as functions of the doubly weighted estimator, under misspecification of the framework's parametric components. With respect to the ATE, this paper shows that the proposed estimator is \textit{doubly robust} even in the presence of missing outcomes. Finally, to demonstrate the estimator's viability in empirical settings, it is applied to \citet{calonico2017women}'s reconstructed sample from the National Supported Work training program.
\end{abstract}


\end{spacing}

\clearpage
\section{Introduction} \label{intro}
When interest lies in causal inference, the prevalence of missing data poses a major identification challenge.  A common issue is that the outcome of interest is missing for some proportion of the sample. In this case, the complete data method that drops observations with missing outcomes is widely used. While dropping is practically convenient, it not only leads to substantial loss of information but more importantly creates a nonrandom sample for estimation. In turn, dropping can generally lead to inconsistent treatment effect estimates. This paper proposes an estimator that double weights for the twin problems of nonrandom treatment assignment and missing outcomes by using information on covariates.

Weighting has been used extensively in both the missing data [\citet{horvitz1952generalization}, \citet{robins1994estimation}, \citet{robins1995semiparametric}, \citet{wooldridge2007inverse}] and treatment effect [\citet{rosenbaum1983central}, \citet{hahn1998role}, \citet{hirano2001estimation}, \citet{firpo2007efficient}, \citet{sloczynski2018general}] literatures. However, a weighting approach that corrects for general missingness in the outcome to estimate treatment effects using observational data is yet to be proposed. Previous studies have considered weighting to deal with specific missing data issues such as attrition and non-response in the presence of endogenous treatment selection [\citet{frolich2014treatment}, \citet{huber2014treatment}, \citet{fricke2020endogeneity}]. Typically, the identification argument in these papers is based on one or more instruments with discussion centered around estimation of average treatment effects.

This paper introduces inverse probability weighting alongside propensity score (PS) weighting in a general M-estimation framework to address two prevalent problems in the causal inference literature. Moreover, the objective function being solved is permitted to be non-smooth in the underlying parameters thereby covering both average and quantile treatment effects. A key feature of the proposed estimator is its \textit{robustness} to parametric misspecification in either a conditional model of interest (such as mean or quantile) or the two weighting functions. In addition, the ATE estimator which uses the proposed strategy is shown to be `\textit{doubly robust}' [\citet{sloczynski2018general}] even in the presence of missing outcomes. 

The key identifying assumptions for consistency of the doubly weighted estimator of a population level parameter are unconfoundedness\footnote{This is a widely used assumption in the treatment effects literature and is known by a variety of names such as exogeneity, ignorability, selection on observables, and conditional independence assumption (CIA).} and missing at random. Put differently, the two restrictions imply that the treatment assignment and missing outcomes mechanisms are as good as randomly assigned after conditioning on covariates. With respect to missingness, the mechanism also allows sample observability to depend on the treatment status.  As such it allows for differential non-response, attrition, and even non-compliance to the extent that conditioning variables predict it.

For many observational studies, unconfoundedness may be a reasonable assumption. Previous literature has found several situations where such an assumption is tenable, especially when pre-treatment values of the outcome variable are available. For example, \citet{lalonde1986evaluating} and \citet{hotz2006evaluating} have shown that controlling for pre-training earnings alone reduces significant bias between non-experimental and experimental estimates. The literature assessing teacher impact on student achievement has reported similar findings with pre-test scores [\citet{chetty2014measuring}, \citet{kane2008estimating}, and \citet{shadish2008can}], indicating the plausibility of unconfoundedness in such settings.

Estimation then follows in two steps. The first step estimates the treatment and missing outcome probabilities using binary response maximum likelihood\footnote{As a practical matter, researchers typically follow the convention of estimating these probabilities as flexible logit functions.} and second step plugs in the estimated probabilities as weights to solve a general objective function. Given the parametric nature of the first and second steps, this paper highlights a \textit{robustness} property which allows the estimator to remain consistent for a parameter of interest under misspecification of either a conditional model or the two probability weights. Consequently, the asymptotic theory in this paper distinguishes between these two halves. The first half focuses on misspecification of either a conditional expectation function (CEF) or a conditional quantile function (CQF), whereas the second half considers misspecification in the weighting functions. 

As illustrative examples, the paper discusses robust estimation of two specific causal parameters, namely, the ATE and QTEs, expressed as functions of the doubly weighted estimator. Consistent estimation of the ATE is achievable under both misspecification scenarios. Of particular interest is the case when the conditional mean function is misspecified. %In this case, consistent estimation of ATE relies on double weighting and results from the generalized linear model literature. 
For estimation of quantile treatment effects, the paper considers three different parameters, namely, conditional quantile treatment effect (CQTE), a linear approximation to CQTE, and unconditional quantile treatment effect (UQTE), each of which may be of interest to the researcher depending on whether features of the conditional or unconditional outcomes distribution are of interest. %In the event that the underlying CQF is assumed to be correct, the doubly weighted estimator is shown to be consistent for the true CQTE, otherwise, delivers a consistent weighted linear approximation to the true CQTE (using results from Angrist et al. (2006)). 
%In addition, this paper underscores the importance of double weighting for a parameter like UQTE where covariates, which serve to remove biases due to nonrandom assignment and missing outcomes, enter the estimating equation only through the two probability models. 
Simulations show that the doubly weighted ATE and QTE estimates have the lowest finite sample bias compared to alternatives which ignore one or both problems.\footnote{Such as the unweighted estimator which drops missing outcomes and does not weight or the ps-weighted estimator which drops the missing data and weights by the propensity score to correct for nonrandom assignment.}

Finally, the proposed method is applied to estimate average and distributional impacts of the National Supported Work (NSW) training program on earnings for the Aid to Families with Dependent Children (AFDC) target group. The sample is obtained from \citet{calonico2017women} who recreate Lalonde's within-study analysis for the AFDC women. The idea behind choosing this empirical application is to utilize the presence of experimental and non-experimental comparison groups for evaluating whether the strategy of double weighting brings us close the experimental benchmark relative to other alternatives. The paper finds that the empirical bias for the doubly weighted estimate is much smaller than that for the unweighted estimate. %It seems that in this particular application, the missing outcomes problem is less consequential than nonrandom assignment for obtaining estimates close to the experimental benchmark. 

The rest of this paper is structured as follows. Section \ref{model} describes the basic potential outcomes framework and provides a short description of the population models with an introduction to the naive unweighted estimator. Section \ref{id} discusses the treatment assignment and missing outcome mechanisms which leads us directly to the identification lemma. Section \ref{theory} develops the first half of the asymptotic theory for the doubly weighted estimator with a focus on misspecification of a conditional feature of interest. This half also requires the weights to be correct for delivering parameter identification. In contrast, section \ref{strongid} considers the other half where a conditional model of interest is correctly specified but the weights may be misspecified. Identification here relies on the parameter solving a conditional problem. 
Section \ref{dr} studies the specifics of robustness for estimating ATE and QTEs in rigorous detail. Section \ref{sims} provides supporting Monte Carlo evidence under three interesting cases of misspecification; correct conditional model with misspecified weights, misspecified conditional model with correct weights, and misspecified model and weights. Section \ref{app} applies the proposed method to job training data from \citet{calonico2017women} and section \ref{end} concludes with directions for future research. 

\section{Potential outcomes and the population models} \label{model}
Consider the standard Neyman-Rubin causal model. Let $Y(1)$ and $Y(0)$ denote potential outcomes corresponding to the treatment and control states and let $W$ be an indicator for whether an individual received the treatment. Then observed outcome is
\begin{align} \label{eq:1}
Y=\text{ }Y(0)\cdot (1-W)+Y(1)\cdot W
\end{align}
Also, let $\mathbf{X}$ be a vector of pre-treatment characteristics which includes an intercept.\footnote{As mentioned in \citet{negiwooldridge2020}, $\mathbf{X}$ may include functions of covariates such as levels, squares, and interactions which will be chosen by the researcher. The dimension of the covariate vector is assumed fixed and does not grow with the sample size.}
Some feature of the distribution of $(Y(g),\mathbf{X})\subset \mathfrak{R}^{\text{M}}$ is assumed to depend on a finite $P_g\times 1$ vector $\bm{\theta_g}$, contained in a parameter space $\bm{\Theta_g}\subset \mathfrak{R}^{P_g}$.\footnote{For generality, the dimension of $\bm{\theta_g}$ is allowed to be different for the treatment and control group problems and is also different than the dimension of $\mathbf{X}$, where $\mathbf{X}\in\mathfrak{X} \subset \mathfrak{R}^{\text{dim}(\mathfrak{X})}$} Let $q(Y(g), \mathbf{X}, \bm{\theta_g})$ be an objective function that depends on outcomes, covariates, and the parameter vector, $\bm{\theta_g}$. Then, the parameter of interest is defined to be a solution to the following M-estimation problem.

\begin{assumption}{(Identification of $\bm{\theta_g^0}$)}\label{as:id} 
The parameter vector $\bm{\theta_g^0}\in \bm{\Theta_g}$ is a unique solution to the population minimization problem 
\begin{equation} \label{eq:id}
	\underset{\bm{\theta_g}\in\bm{\Theta_g}}{\mathrm{min}}\mathbb{E}\left[q(Y(g),\mathbf{X},\bm{\theta_g})\right]
\end{equation} for each $g=0,1$. \qed
\end{assumption} 
Examples include the smooth ordinary least squares (OLS) function, $q(Y(g),\mathbf{X},\bm{\theta_g})= (Y(g)-\mathbf{X}\bm{\theta_g})^2$ or the non-smooth conditional quantile regression (CQR) of \citet{KoenkerBassett1978}, $q(Y(g),\mathbf{X},\bm{\theta_g})= c_{\tau}(Y(g)-\mathbf{X}\bm{\theta_g})$\footnote{For a random variable $u$, $c_{\tau}(u) = (\tau-\mathbf{1}\{u<0\})u$ is the asymmetric loss function for estimating quantiles and $\mathbf{1}\{\cdot\}$ is an indicator function.}. Other examples of $q(\cdot)$ can be log-likelihood and quasi-log-likelihood (QLL) functions.

An implicit point in assumption \ref{as:id} is that $\bm{\theta_g^0}$ is not assumed to be correctly specified for a conditional feature like a conditional mean, variance, or even the full conditional distribution. It simply requires $\bm{\theta_g^0}$ to \textit{uniquely} minimize the population problem in (\ref{eq:id}). If $\bm{\theta_g^0}$ is correctly specified for any of the above mentioned quantities, then the parameter is of direct interest to researchers. However, if $\bm{\theta_g^0}$ is misspecified for any of these distributional features, assumption \ref{as:id} guarantees a unique pseudo true solution, $\bm{\theta_g^\ast}$ [\citet{white1982maximum}]. In the case of misspecification, determining whether $\bm{\theta_g^\ast}$ is meaningful will depend on the conditional feature being studied and the estimation method used. For example, in the case of OLS, $\bm{\theta_g^0}$ will index a linear projection if one is agnostic about linearity of the CEF.  \citet{Angristetal2006} establish analogous approximation properties for quantiles, where a misspecified CQF can still provide the best weighted mean square approximation to the true CQF.  

Let $`S$' be a binary indicator such that $S=1$ if the outcome is observed and $S=0$ otherwise. The objective of this paper is to consistently estimate $\bm{\theta_g^0}$. In the presence of missing outcomes, a common empirical strategy is to solve the following M-estimation problems for the treatment and control groups, respectively. \vspace{-1em}
\begin{equation}\label{eq:fullselect} 
\begin{split}
&\underset{\bm{\theta_1}\in\bm{\Theta_1}}{\text{min}}\sum_{i=1}^{N}S_i\cdot W_{i}\cdot q(Y_i(1),\mathbf{X}_i,\bm{\theta_1}) \\
&\underset{\bm{\theta_0}\in\bm{\Theta_0}}{\text{min}}\sum_{i=1}^{N}S_i\cdot (1-W_{i})\cdot q(Y_i(0),\mathbf{X}_i,\bm{\theta_0}) 
\end{split}
\end{equation}
Let us refer to the estimator that solves (\ref{eq:fullselect}) as the unweighted M-estimator and denote it as $\bm{\hat{\theta}_{g}^u}$. This estimator uses the available sample after dropping the missing data to estimate $\bm{\theta_g^0}$. Using the reverse analogy principle, $\bm{\hat{\theta}_{g}^u}$ will be consistent for $\bm{\theta_g^0}$ if it solves the population analogue of (\ref{eq:fullselect}), which may not be true. As an example, consider
\begin{equation*}
	\begin{split}
		&Y(g) = \mathbf{X}\bm{\theta_g}+U(g), \text{ } g=0,1 \\
		&\mathbb{E}[\mathbf{X}^\prime U(g)] = \mathbf{0}
	\end{split}
\end{equation*}
In this case, even if the treatment is randomly assigned, missingness may still be correlated with the treatment, observable factors, or both. Hence, the population first order condition for the selected sample, $\mathbb{E}[S\cdot W \cdot \mathbf{X}^\prime U(g)]$, is not zero even though $\mathbb{E}[\mathbf{X}^\prime U(g)] = \mathbf{0}$. So identification of $\bm{\theta_g^0}$ is now confounded on two grounds; nonrandom assignment which renders the treatment and control groups incomparable
and missing outcomes which violates the ‘random sampling’ assumption. The next
section discusses the identification approach taken in this paper.

\section{Identification of parameter of interest} \label{id}
Without imposing any structure on the assignment and missingness mechanisms in the population, estimating $\bm{\theta_g^0}$ remains difficult. To proceed with identification, I assume that the treatment is unconfounded on covariates.\footnote{Like most other assumptions, unconfoundedness is non-refutable. For methods that indirectly test for its validity, see \citet{huber2015test}, \citet{de2014testing}, and \citet{heckman1989choosing}.} Formally, 

\begin{assumption}{(Strong ignorability)} \label{as:uc}
Assume,
\begin{equation}\label{eq:uc} 
\{Y(0), Y(1)\independent W\} \bf{|} \ \mathbf{X}
\end{equation}
\begin{enumerate}
\item[i)] The vector of pre-treatment covariates, $\mathbf{X}$, is always observed for the entire sample.
\item[ii)] For all $\mathbf{x}\in \mathfrak{X} \subset \mathfrak{R}^{\mathrm{dim}(\mathfrak{X})}$, define $p(\mathbf{x}) = \mathbb{P}(W=1|\mathbf{X}=\mathbf{x})$ such that $p(\mathbf{x})>\kappa$ for a constant $\kappa>0$. \qed
\end{enumerate}
\end{assumption} 

Equation (\ref{eq:uc}) indicates that conditioning on covariates is enough to parse out any systematic differences that may exist between the treatment and control groups. One advantage of unconfoundedness is that, intuitively, it has a better chance of holding once we control for a rich set of variables in $\mathbf{X}$.\footnote{For example, \citet{hirano2001estimation} control for a rich set of prognostic factors to justify unconfoundedness while estimating the effects of right heart catheterization (RHC) on survival rates of patients.} Note that unconfoundedness not only includes cases where the treatment is a deterministic function of the covariates, for example stratified (or block) experiments, but also cases where the treatment is a stochastic function of covariates. Part i) requires that we observe these covariates for all individuals. Part ii) is an overlap condition which ensures that for all values of $\mathbf{x}$ in $\mathfrak{X}$, we observe units in both the treatment and control groups.\footnote{Methods for checking overlap involve calculating normalized sample average differences for each covariate and checking the empirical distribution of propensity scores.} 

With respect to the missing outcomes mechanism, I assume selection on observables 

\begin{assumption}{(Missing at Random (MAR))} \label{as:mar} Assume,
\begin{equation} \label{eq:ignorability}
\{Y(0), Y(1)\independent S\}{\bf{|}}\ \mathbf{X}, W
\end{equation}
\begin{enumerate}
\item[i)] In addition to $\mathbf{X}$, $W$ is always observed for the entire sample. 
\item[ii)] For each $(\mathbf{x},w)\in (\mathbf{X},W) \subset \mathfrak{R}^{\mathrm{dim}(\mathfrak{X})+1}$, define, $r(\mathbf{x},w) \equiv \mathbb{P}(S=1|\mathbf{X}=\mathbf{x},W=w)$ such that $\eta<r(\mathbf{x},w)<1$ for a constant $\eta>0$ and $w=0,1$. \qed
\end{enumerate}
\end{assumption}	

Equation (\ref{eq:ignorability}) states that conditional on covariates and the treatment status, the individuals whose outcomes are missing do not differ systematically from those who are observed. This implies that adjusting for $\mathbf{X}$ and $W$ renders the outcomes as good as randomly missing. In the statistics literature, this assumption is known as MAR and represents a mechanism wherein missingness only depends on observables and not on the missing values of the variable itself [\citet{little2019statistical}]. Special cases covered under this mechanism are patterns such as missing completely at random (MCAR) and exogenous missingness considered in \citet{wooldridge2007inverse}. Allowing the missingness probability to be a function of the treatment indicator is particularly useful in cases of differential nonresponse. For instance, in NSW, people assigned to the treatment group were less likely to drop out of the program compared to the control group. In such cases, covariates alone may not be sufficient for predicting missingness. To the extent that being observed in the sample is predicted by $\mathbf{X}$ and $W$, assumption \ref{as:mar} can accommodate non-observability due to sampling design, item non-response, and attrition in a two period panel.\footnote{For the case of attrition, one must assume that second period missingness is ignorable conditional on initial period covariates and the treatment status.} 

Part i) of the above assumption ensures that $\mathbf{X}$ and $W$ are fully observed and part ii) again imposes an overlap condition. It states that there is a positive probability of observing people in the sample for a given $\mathbf{X}$ and $W$.

Then solving the doubly weighted population problem given below is the same as solving the original M-estimation problem in (\ref{eq:id}). The following lemma establishes this equality 
 
\begin{lemma}{(Identification)}\label{lemma:id}
Given assumptions \ref{as:id}, \ref{as:uc}, \ref{as:mar}, assume i) $q(Y(g),\mathbf{X},\bm{\theta_g})$ is a real valued function for all $(Y(g), \mathbf{X})\subset \mathfrak{R}^M$ ii) $\mathbb{E}\left[\left. |q(Y(g),\mathbf{X},\bm{\theta_g})| \right. \right]< \infty$ for all $\bm{\theta_g} \in \bm{\Theta_g}$, $g=0,1$, then 
\begin{equation}\label{c}
\mathbb{E}\big[\omega_g\cdot q\left(Y(g), \mathbf{X}, \bm{\theta_g}\right)\big]= \mathbb{E}\big[q\left(Y(g), \mathbf{X}, \bm{\theta_g}\right)\big] 
\end{equation}
where $\displaystyle \omega_1 = \frac{S\cdot W}{r(\mathbf{X},W)\cdot p(\mathbf{X})}$,  $\displaystyle \omega_0 = \frac{S\cdot (1-W)}{r(\mathbf{X},W)\cdot (1-p(\mathbf{X}))}$. \qed 
\end{lemma}

The proof uses two applications of the law of iterated expectations (LIEs) with unconfoundedness and MAR to arrive at the above result. It implies that one can now address the identification issue due to nonrandom assignment and missing outcomes by solving the doubly weighted population problem.\footnote{Define {\small $q\left(Y, \mathbf{X}, \bm{\theta}\right)=q(Y(1), \mathbf{X}, \bm{\theta_1}) \text{ for } W=1$ and $q(Y(0), \mathbf{X}, \bm{\theta_0}) \text{ for } W=0 $}, then $\mathbb{E}[\omega_g\cdot q\left(Y, \mathbf{X}, \bm{\theta}\right)]\equiv \mathbb{E}[\omega_g\cdot q\left(Y(g), \mathbf{X}, \bm{\theta_g}\right)]$ which makes it  function of the observed random vector $\{(Y_i,\mathbf{X}_i,W_i,S_i): i=1,2,\ldots,N\}$.} 

\section{Asymptotic theory under weak identification} \label{theory}
Lemma \ref{lemma:id} is important for us as it helps to illustrate the role of double weighting in dealing with the two issues at hand. However, to operationalize this argument, we first need to estimate $r(\mathbf{X},W)$ and $p(\mathbf{X})$ before introducing the estimator and studying its asymptotic properties. 

The following assumptions posit that we have a correctly specified model for the two probabilities and that we estimate them using binary response maximum likelihood. Since both $W$ and $S$ are binary responses, estimation of $\bm{\gamma_0}$ and $\bm{\delta_0}$ using MLE will be asymptotically efficient under correct specification of these functions. Consistency and asymptotic normality for $\bm{\gamma_0}$ and $\bm{\delta_0}$ follow from theorems 2.5 and 3.3 of \citet{newey1994large}.   

\begin{assumption}{(Correct parametric specification of propensity score)} \label{as:correctps}
Assume that i) There exists a known parametric function $G(\mathbf{X},\bm{\gamma})$ for $p(\mathbf{X})$ where $\bm{\gamma}\in \bm{\Gamma}\subset \mathfrak{R}^I$ and $0<G(\mathbf{X},\bm{\gamma})<1$ for all $\mathbf{X} \in \mathcal{X}$, $\bm{\gamma}\in \bm{\Gamma}$; ii) There exists $\bm{\gamma_0}\in \bm{\Gamma}$ s.t. $p(\mathbf{X})=G(\mathbf{X},\bm{\gamma_0})$; iii) $\bm{\hat{\gamma}}$ is the binary response maximum likelihood estimator that solves 
\begin{equation}\label{psll}
	\underset{\bm{\gamma}\in \bm{\Gamma}}{\textup{max}}\sum_{i=1}^{N}\{W_i\textup{log}G(\mathbf{X}_i,\bm{\gamma})+(1-W_i)\textup{log}(1-G(\mathbf{X}_i,\bm{\gamma}))\}
\end{equation}
\end{assumption} \qed

\begin{assumption}{(Correct parametric specification of missing outcomes probability)} \label{as:correctsp}
Assume that i) There exists a known parametric function $R(\mathbf{X},W,\bm{\delta})$ for $r(\mathbf{X},W)$ where $\bm{\delta}\in \bm{\Delta}\subset \mathfrak{R}^{K}$ and $R(\mathbf{X},W,\bm{\delta})>0$ for all $\mathbf{X} \in \mathcal{X}$, $\bm{\delta}\in \bm{\Delta}$; ii) There exists $\bm{\delta_0}\in \bm{\Delta}$ s.t. $r(\mathbf{X},W) = R(\mathbf{X},W,\bm{\delta_0})$; iii) $\bm{\hat{\delta}}$ is the binary response maximum likelihood estimator that solves 
\begin{equation}\label{missll}
	\underset{\bm{\delta}\in \bm{\Delta}}{\textup{max}}\sum_{i=1}^{N}\{S_i\textup{log}R(\mathbf{X}_i,W_i,\bm{\delta})+(1-S_i)\textup{log}(1-R(\mathbf{X}_i,W_i,\bm{\delta}))\}
\end{equation} 
\end{assumption} \qed

The influence function representations for $\bm{\hat{\gamma}}$ and $\bm{\hat{\delta}}$ can then be written as \vspace{-0.5em}
\begin{equation}\label{if}
	\begin{split}
	\sqrt{N}\left(\bm{\hat{\gamma}}-\bm{\gamma_0}\right) = \mathbb{E}\left(\mathbf{d}_i\mathbf{d}_i^\prime\right)^{-1}N^{-1/2}\sum_{i=1}^{N}\mathbf{d}_i+o_p(1) \\
	\sqrt{N}\left(\bm{\hat{\delta}}-\bm{\delta_0}\right) = \mathbb{E}\left(\mathbf{b}_i\mathbf{b}_i^\prime\right)^{-1}N^{-1/2}\sum_{i=1}^{N}\mathbf{b}_i+o_p(1) 
	\end{split}	
\end{equation} 
where $\mathbf{d}_i$ and $\mathbf{b}_i$ are scores of the binary response log-likelihood problems in (\ref{psll}) and (\ref{missll}) evaluated at the probability limits $\bm{\gamma_0}$ and $\bm{\delta_0}$, respectively. The \textit{doubly weighted} estimator is then defined as:
\begin{equation}\label{estimator}
\bm{\hat{\theta}_g} = \underset{\bm{\theta_g}\in\bm{\Theta_g}}{\text{argmin}}\sum_{i=1}^{N} \widehat{\omega}_{ig} \cdot q(Y_i(g),\mathbf{X}_i,\bm{\theta_g})
\end{equation}
where $ \widehat{\omega}_{i1}=\frac{S_i\cdot W_i}{R(\mathbf{X}_i,W_i,\bm{\hat{\delta}})\cdot G(\mathbf{X}_i,\bm{\hat{\gamma}})}$ and $ \widehat{\omega}_{i0} = \frac{S_i\cdot (1-W_i)}{R(\mathbf{X}_i,W_i,\bm{\hat{\delta}})\cdot (1-G(\mathbf{X}_i,\bm{\hat{\gamma}}))}$ are the estimated weights for solving the treatment and control group problems, respectively.\footnote{When necessary, the estimated weights will also be denoted as $\omega_g(\bm{\hat{\delta}}, \bm{\hat{\gamma}})\equiv\widehat{\omega}_g$.} 

Given the two-step nature of the estimation problem; first step uses binary response MLE for estimating the probability weights and second step solves an objective function using the first-step weights, the asymptotic theory utilizes results for two-step estimators with a non-smooth objective function to establish the large sample properties of $\bm{\hat{\theta}_g}$.
The following theorem fills in the primitive regularity conditions for applying the uniform law of large numbers.

\begin{theorem}{(Consistency)} \label{theorem:consistency} 	Suppose assumption \ref{as:id} holds and that i) $\{(Y_i, \mathbf{X}_i, W_{i}, S_i);\  i=1,2,\ldots, N\}$ are $i.i.d$ draws satisfying assumptions \ref{as:uc} and \ref{as:mar}; ii) $\bm{\Theta_g}$ is compact for $g=0,1$; iii) $G(\mathbf{X},\bm{\gamma})$ satisfies assumption \ref{as:correctps} and is continuous for each $\bm{\gamma}$ on the support of $\mathbf{X}$. Similarly, $R(\mathbf{X},W,\bm{\delta})$ satisfies assumption \ref{as:correctsp} and is continuous for each $\bm{\delta}$ on the support of $(\mathbf{X},W)$; iv) $q(Y(g),\mathbf{X},\bm{\theta_g})$ is continuous at each $\bm{\theta_g}\in\bm{\Theta_g}$ with probability one; v) $ \mathbb{E}\bigg[\underset{\bm{\theta_g}\in\bm{\Theta_g}}{\textup{sup}}|q(Y(g), \mathbf{X},\bm{\theta_g})|\bigg]  <\infty $.
Then, $\bm{\hat{\theta}_g}\overset{p}{\rightarrow} \bm{\theta_g^0} $. \qed
\end{theorem}

The proof follows from verifying the conditions in Lemma 2.4 of \citet{newey1994large}. Under the dominance condition given in v), uniform convergence of sample averages holds quite generally. 

For establishing asymptotic normality, I provide primitive conditions for the general case of non-smooth objective functions. Let the score of $q(Y(g),\mathbf{X},\bm{\theta_g})$ at the true parameter, $\bm{\theta_g^0}$, be denoted as $\mathbf{h}(Y(g),\mathbf{X},\bm{\theta_g^0}) \equiv \mathbf{h}_g$ and suppose it exists with probability one. Let the population problem be denoted as 
\begin{equation*}
Q_0(\bm{\theta_g})\equiv \mathbb{E}\left[\omega_{g}\cdot q(Y(g),\mathbf{X},\bm{\theta_g})\right] 
\end{equation*}
and the sample analogue be given as
\begin{equation*}
Q_N(\bm{\theta_g})\equiv \frac{1}{N\hat{\rho}_g}\sum_{i=1}^{N}\widehat{\omega}_{ig}\cdot q(Y_i(g),\mathbf{X}_i,\bm{\theta_g})
\end{equation*}
where $\hat{\rho}_g= N_g/N$ and $N\hat{\rho}_g \rightarrow \infty$ as $\hat{\rho}_g \rightarrow \rho_g$.\footnote{The sampling fractions $N_0 = \sum_{i=1}^{N}S_i\cdot W_i$ and $N_1 = \sum_{i=1}^{N}S_i\cdot (1-W_i)$ are random which implies that $N=N_0+N_1$ is also random as opposed to being fixed ahead of time.} For the sake of asymptotics, we may ignore the division by $\hat{\rho}_g$. The main condition needed for establishing asymptotic normality is stochastic equicontinuity of the empirical process
\begin{equation}\label{ep}
	\bm{v}_N(\bm{\theta_g}) \equiv \frac{1}{\sqrt{N}}\sum_{i=1}^{N}\bigg\{\widehat{\omega}_{ig}\mathbf{h}_{ig}(\bm{\theta_g})-\mathbb{E}\big[\widehat{\omega}_{ig}\mathbf{h}_{ig}(\bm{\theta_g})\big]\bigg\} 
\end{equation} which will be sufficient to guarantee uniform convergence of the objective function to its population counterpart. 

\begin{theorem}{(Asymptotic Normality)} \label{theorem:normality} 
	In addition to the conditions mentioned in Theorem \ref{theorem:consistency}, assume
	i) $\bm{\theta_g^0}\in$ $int(\bm{\Theta_g})$; ii) $q(Y(g),\mathbf{X},\bm{\theta_g})$ is continuously differentiable on $int(\bm{\Theta_g})$ with probability one; iii) $\frac{1}{N}\sum_{i=1}^{N}\widehat{\omega}_{ig}\cdot \mathbf{h}(Y_i(g),\mathbf{X}_i,\bm{\hat{\theta}_g})= o_p(N^{-1/2})$;  iv) {\small $\mathbb{E}\bigg[\underset{\bm{\theta_g}\in\bm{\Theta_g}}{\textup{sup}}\lVert \mathbf{h}(Y(g), \mathbf{X},\bm{\theta_g})\rVert^2\bigg]  <\infty $}; v) $G(\cdot, 
	\bm{\gamma})$ and $R(\cdot,\bm{\delta})$ are both twice continuously differentiable on $int(\Gamma)$ and $int(\Delta)$, respectively; vi) {\small $\mathbb{E}\bigg[\underset{\bm{\delta}\in\bm{\Delta}}{\textup{sup}}\lVert \mathbf{b}(\mathbf{X}, W, S,\bm{\delta})\rVert^2\bigg]<\infty$}, {\small $\mathbb{E}\bigg[\underset{\bm{\gamma}\in\bm{\Gamma}}{\textup{sup}}\lVert \mathbf{d}(\mathbf{X}, W,\bm{\gamma})\rVert^2\bigg]<\infty$}; vii) $\mathbb{E}\big[\omega_{g}\cdot\mathbf{h}(Y(g),\mathbf{X},\bm{\theta_g})\big]$ is continuously differentiable on $int(\bm{\Theta_g})$; viii) {\small $\mathbf{H_g} \equiv \nabla_{\bm{\theta_g}} \mathbb{E}\left[\omega_{g}\cdot\mathbf{h}(Y(g),\mathbf{X},\bm{\theta_g^0})\right] %= \nabla_{\bm{\theta_g}} \mathbb{E}\left[\mathbf{h}(Y(g),\mathbf{X},\bm{\theta_g^0})\right]
	$} 
	is nonsingular; ix) $\{\bm{v}_N(\bm{\theta_g}): N\geq 1\}$ is stochastically equicontinuous.
	 Then, 
	 \begin{equation*}
	 \sqrt{N}(\bm{\hat{\theta}_g}-\bm{\theta_g^0})\overset{d}{\rightarrow} N\left(\bm{0}, \mathbf{H}^{-1}_{\mathbf{g}}\mathbf{\Omega_g}\mathbf{H}^{-1}_{\mathbf{g}}\right)\end{equation*}
	where $\mathbf{\Omega_g} = \mathbb{E}\big(\mathbf{l}_{ig}\mathbf{l}_{ig}^{\prime}\big) - \mathbb{E}\left(\mathbf{l}_{ig}\mathbf{b}_i^{\prime}\right)\mathbb{E}\left(\mathbf{b}_i\mathbf{b}_i^{\prime}\right)^{-1}\mathbb{E}\big(\mathbf{b}_i\mathbf{l}_{ig}^{\prime}\big)-\mathbb{E}\left(\mathbf{l}_{ig}\mathbf{d}_i^{\prime}\right)\mathbb{E}\left(\mathbf{d}_i\mathbf{d}_i^{\prime}\right)^{-1}\mathbb{E}\big(\mathbf{d}_i\mathbf{l}_{ig}^{\prime}\big)$ for each $g=0,1$ and $\mathbf{l}_{ig}\equiv \omega_{ig}\mathbf{h}_{ig}$ is score of the weighted objective function evaluated at $\bm{\theta_g^0}$.\qed
\end{theorem} 
Sufficient primitive conditions for stochastic equicontinuity may be found in \citet{andrews1994empirical}. The asymptotic variance expression derived  above offers some interesting insights. First, the middle term, $\bm{\Omega_g}$, represents the variance of the residual from the population regression of the weighted score, $\mathbf{l}_{ig}$, on the two binary response scores, $\mathbf{b}_i$ and $\mathbf{d}_i$. Note that even though $\bm{\Omega_g}$ would involve covariance between the two MLE scores, that term is zero on account of the two scores being conditionally independent.

Second, the expression for $\bm{\Omega_g}$ has an efficiency implication for the second step estimate, $\bm{\hat{\theta}_g}$. When a researcher is only willing to assume identification of $\bm{\theta_g^0}$ in the unconditional sense, it is potentially more efficient to estimate the two weights even when they are known. To show this formally, let us assume that $p(\mathbf{X})$ and $r(\mathbf{X}, W)$ are known and $\bm{\tilde{\theta}_g}$ is the doubly weighted estimator that uses known weights, $\omega_g$.  Then,
\begin{corollary}{(Efficiency gain with estimated weights)}\label{theorem:gain}  Under the assumptions of theorem \ref{theorem:normality}, 
	\begin{equation*}
		\begin{split}
		\mathrm{Avar}\big[\sqrt{N}\big(\bm{\tilde{\theta}_g}-\bm{\theta_g^0}\big)\big]- \mathrm{Avar}\big[\sqrt{N}\big(\bm{\hat{\theta}_g}-\bm{\theta_g^0}\big)\big]&= \mathbf{H}^{-1} _{\mathbf{g}}\mathbf{\Sigma_g}\mathbf{H}_{\mathbf{g}}^{-1}- \mathbf{H}_{\mathbf{g}}^{-1}\mathbf{\Omega_g}\mathbf{H}^{-1} _{\mathbf{g}}\\
		&=\mathbf{H}^{-1} _{\mathbf{g}}\mathbf{\left(\Sigma_g-\Omega_g\right)}\mathbf{H}^{-1} _{\mathbf{g}} 
		\end{split}
		\end{equation*}
	is positive semi-definite and where $\mathbf{\Sigma_g} = \mathbb{E}(\mathbf{l}_{ig}\mathbf{l}_{ig}^\prime)$. \qed
\end{corollary}
In other words, we do no worse, asymptotically, by estimating the weights even when we actually know them. This result can be seen an extension of \citet{wooldridge2007inverse} to the case when one has two sets of probability weights being estimated in the first stage.\footnote{In the missing data literature, this result has also been called the ``efficiency puzzle''. \citet{prokhorov2009gmm} study this puzzle in a GMM framework using an augmented set of moment conditions, where the first set of moments correspond to the weighted objective function and the second set belongs to the missing outcomes (or selection in their case) problem.} %A similar kind of result has been observed in ... 

\section{A conditional feature of interest is correctly specified} \label{strongid}
The asymptotic results in the previous section were derived under the assumption that some feature of the conditional distribution of outcomes may be misspecified. This was implicit in defining $\bm{\theta_g^0}$ as a solution to the unconditional M-estimation problem. Examples include estimating a misspecified linear conditional mean or quantile function. In contrast, this section highlights the other half of the asymptotic theory which is formalized using a strong version of the identification assumption and allowing the weights to be misspecified. 

\begin{assumption}{(Strong identification of $\bm{\theta_g^0}$)}\label{as:strongid} 
The parameter vector $\bm{\theta_g^0}\in \bm{\Theta_g}$ is the unique solution to the population minimization problem 
\begin{equation} \label{eq:exogeneity}
\underset{\bm{\theta_g}\in\bm{\Theta_g}}{\mathrm{min}}\mathbb{E}\left[q(Y(g),\mathbf{X},\bm{\theta_g})|\mathbf{X}\right]; \text{ } g=0,1
\end{equation} 
under unconfoundedness (defined in \ref{as:uc}) and MAR (defined in \ref{as:mar}) for each $\mathbf{X} \in \mathfrak{X} \subset \mathfrak{R}^{\text{dim}(\mathfrak{X})}$. \qed
\end{assumption}

The above can be seen as a strengthening of the identification assumption in section \ref{theory} since LIE implies that $\bm{\theta_g^0}$ is also a solution to the unconditional M-estimation problem. By requiring $\bm{\theta_g^0}$ to solve (\ref{eq:exogeneity}), assumption \ref{strongid} is intended for situations where a conditional feature of interest is correctly specified. An implication of this strengthened identification is that $\bm{\theta_g^0}$ now solves the conditional score of the objective function i.e. $\mathbb{E}\big[\mathbf{h}(Y(g),\mathbf{X},\bm{\theta_g^0})|\mathbf{X}\big] = \mathbf{0}$. 

For instance, the conditional score will be zero in the case of estimating a correctly specified CEF with either OLS or quasi maximum likelihood estimation (QMLE) in the linear exponential family (LEF). This would also hold for a correctly specified CQF estimated either using quantile regression or QMLE in the tick exponential family [\citet{Komunjer2005}].

Delineating these two identification scenarios is important for determining which causal parameter can be estimated consistently under each setting. As we will see in the next section, it is possible to estimate the ATE under both cases of misspecification. However the same cannot be said for QTE parameters. In addition to assumption \ref{as:strongid}, the asymptotic results in this half do not rely on correct specification of weights. In other words, assuming $R(\cdot, \cdot, \bm{\delta})$ and $G(\cdot, \bm{\gamma})$ to be correctly specified is rather restrictive and not required for the doubly weighted estimator to be consistent for $\bm{\theta_g^0}$.

\begin{assumption}{(Parametric specification of propensity score)} \label{as:missps} Assume that conditions i) and iii) of assumption \ref{as:correctps} hold where condition ii) is defined for some $\bm{\gamma^\ast} \in \bm{\Gamma}$ such that plim$(\bm{\hat{\gamma}})=\bm{\gamma^\ast}$. \qed
\end{assumption} 

\begin{assumption}{(Parametric specification of missingness probability)} \label{as:missp} Assume that conditions i) and iii) of assumption \ref{as:correctsp} hold where condition ii) is defined for some $\bm{\delta^\ast} \in \bm{\Delta}$ such that plim$(\bm{\hat{\delta}})=\bm{\delta^\ast}$. \qed
\end{assumption}
Note that assumptions \ref{as:missps} and \ref{as:missp} do not require the parametric models for the two probabilities to be correctly specified. Nevertheless, we continue to assume that $\bm{\hat{\gamma}}$ and $\bm{\hat{\delta}}$ solve the same binary response problem as in Assumptions \ref{as:correctps} and \ref{as:correctsp} with probability limits given by pseudo true values $\bm{\gamma^\ast}$ and $\bm{\delta^\ast}$, respectively [\citet{white1982maximum}].
To show that $\bm{\theta_g^0}$ is still a solution to the doubly weighted population problem with misspecified weights, a sketch of the argument is given below. Consider,
\begin{equation}\label{eq:missweightedpop}
\mathbb{E}\left[\omega_g^\ast\cdot q(Y(g),\mathbf{X},\bm{\theta_g})\right]
\end{equation}
where $\omega_g^\ast$ are asymptotic weights which use $G(\mathbf{X}, \bm{\gamma^\ast})$ and $R(\mathbf{X},W,\bm{\delta^\ast})$. Using LIE along with unconfoundedness and MAR, I can rewrite the above expectation as 
\begin{equation*}
\mathbb{E}\left[\xi_g(\mathbf{X})\cdot \mathbb{E}\{q(Y(g),\mathbf{X},\bm{\theta_g})|\mathbf{X}\}\right]
\end{equation*}
where $\xi_g(\mathbf{X})$ is a function of weights for $g=0,1$. The strong identification assumption implies 
\begin{equation*}
	\begin{split}
	& \mathbb{E}\big[q(Y(g),\mathbf{X},\bm{\theta_g^0})|\mathbf{X}\big]
	\leq \mathbb{E}\left[q(Y(g),\mathbf{X},\bm{\theta_g})|\mathbf{X}\right], \  \forall \ \bm{\theta_g}  \in \bm{\Theta_g}
	\end{split}
\end{equation*}
Further, since $\xi_g(\mathbf{X}) > 0$,
\begin{equation*}
	\mathbb{E}\left[\omega_g^\ast\cdot q(Y(g),\mathbf{X},\bm{\theta_g^0})\right] \leq \mathbb{E}\left[\omega_g^\ast\cdot q(Y(g),\mathbf{X},\bm{\theta_g})\right], \text{ } \bm{\theta_g} \in \bm{\Theta_g}
\end{equation*}
where the inequality is strict when $\bm{\theta_g}\neq \bm{\theta_g^0}$. Therefore, solving the doubly weighted problem identifies the parameter even if the weights are wrong.  In general, the parameter that solves (\ref{eq:missweightedpop}) will be different from the one that solves the same problem with correct weights.\footnote{When $R(\mathbf{X}, W, \bm{\delta^\ast}) = r(\mathbf{X},W)$ and $G(\mathbf{X}, \bm{\gamma^\ast})=p(\mathbf{X})$, then solving (\ref{eq:missweightedpop}) will be the same as solving the problems in section \ref{id}.}  But as long as $\bm{\theta_g^0}$ is a unique solution, solving (\ref{eq:missweightedpop}) will identify it.

The following two theorems establish consistency and asymptotic normality of the doubly weighted estimator.

\begin{theorem}{(Consistency under strong identification)} \label{theorem:consistency_2}
Under assumptions \ref{as:uc}, \ref{as:mar}, \ref{as:strongid}, \ref{as:missps}, and \ref{as:missp} with regularity conditions (1), (2) and (3) of Theorem \ref{theorem:consistency}, $
\bm{\hat{\theta}_g}\overset{p}{\rightarrow} \bm{\theta_g^0} \text{ as } N\rightarrow \infty$
where $\bm{\hat{\theta}_g}$ is the doubly-weighted estimator that solves (\ref{eq:missweightedpop}). \qed
\end{theorem}

\begin{theorem}{(Asymptotic Normality under strong identification)} \label{theorem:normality_2}
Under the assumptions of theorem \ref{theorem:consistency_2} and the regularity conditions of theorem \ref{theorem:normality} where MLE estimators $\bm{\hat{\gamma}}$ and $\bm{\hat{\delta}}$ have probability limits given by $\bm{\gamma^\ast}$ and $\bm{\delta^\ast}$, then
$\sqrt{N}(\bm{\hat{\theta}_g}-\bm{\theta_g^0})\overset{d}{\rightarrow} N\big(\bm{0}, \mathbf{H}_{\bm{g}}^{-1}\bm{\Omega_g}\mathbf{H}_{\bm{g}}^{-1}\big) $
where $\bm{\Omega_g}=\mathbb{E}\big(\mathbf{l}_{ig}\mathbf{l}_{ig}^{\prime}\big)$ with $\mathbf{H}_{\bm{g}}$ and $\mathbf{l}_{ig}$ defined in Theorem \ref{theorem:normality} except with asymptotic weights given by $\omega^\ast_{ig}$. \qed
\end{theorem}
Substantively, there is no real difference in the proof of the above theorem compared to those derived in section \ref{theory} except that now $\bm{\hat{\gamma}}$ and $\bm{\hat{\delta}}$ are converging to probability limits that could be potentially different from those indexing the true treatment and missing outcome probabilities. A consequence of the objective function solving the conditional problem is reflected in the asymptotic variance expression. Compared to the previous section, $\bm{\Omega_g}$ now is simply the variance of weighted score of the objective function without first stage adjustment of the estimated probabilities.  This is because under assumption \ref{as:strongid}, $\mathbb{E}\big(\mathbf{l}_{ig}\mathbf{b}_i^\prime\big) =\mathbb{E}\big(\mathbf{l}_{ig}\mathbf{d}_i^\prime\big)=\mathbf{0}$. A sketch of the proof for $\mathbb{E}\big(\mathbf{l}_{ig}\mathbf{b}_i^\prime\big) =\mathbf{0}$ is provided below. The argument for $\mathbb{E}\big(\mathbf{l}_{ig}\mathbf{d}_i^\prime\big)$ follows analogously. 
\begin{equation*}
	 \mathbb{E}\big(\mathbf{l}_{ig}\mathbf{b}_i^\prime\big)\equiv \mathbb{E}\big(\omega^\ast_{ig}\mathbf{h}_{ig}\mathbf{b}_i^\prime\big)
	= \mathbb{E}\big[\zeta_g(\mathbf{X}_i)\cdot \mathbb{E}\big(\mathbf{h}(Y_i(g), \mathbf{X}_i,\bm{\theta_g^0})|\mathbf{X}_i\big)\big] = \mathbf{0}
\end{equation*}
where $\zeta_g(\mathbf{X})$ is a function of weights. The first equality uses the definition of $\mathbf{l}_{ig}$ with misspecified weights and second equality applies LIEs with unconfoundedness and MAR. In other words, the reason for obtaining a simpler expression for $\bm{\Omega_g}$ is because the correlation between weighted score of the objective function and the two binary response scores is zero when $\bm{\theta_g^0}$ is correctly specified for a conditional feature of interest and we use an appropriate method to estimate it. 

A simpler expression for $\bm{\Omega_g}$ also means that we can no longer exploit these correlations between scores to obtain asymptotic efficiency for estimating $\bm{\theta_g^0}$. Again, let $\bm{\tilde{\theta}_g}$ be the doubly weighted estimator that uses true weights, $\omega_g$, then 
\begin{corollary}{(No gain with estimated weights under strong identification)}\label{theorem:gain_2} Under the assumptions of theorem \ref{theorem:normality_2} 
	\begin{equation*}
	\mathrm{Avar}\big[\sqrt{N}\big(\bm{\tilde{\theta}_g}-\bm{\theta_g^0}\big)\big] = \mathrm{Avar}\big[\sqrt{N}\big(\bm{\hat{\theta}_g}-\bm{\theta_g^0}\big)\big] = \mathbf{H}_{\bm{g}}^{-1}\bm{\Omega_g}\mathbf{H}_{\bm{g}}^{-1}
	\end{equation*} 
\end{corollary} \qed

Hence knowledge of the weights does little when for instance we have a correctly specified CEF or CQF and we use either OLS or QR to estimate the parameters indexing these conditional models of interest.

A special case of weights misspecification is when $\omega_g^\ast$ is a constant. This is plausible since $R(\mathbf{X}, W, \bm{\delta^\ast})$ and $G(\mathbf{X}, \bm{\gamma^\ast})$ are allowed to be any bounded positive functions of $\mathbf{X}$ and $W$. In other words, the unweighted estimator, $\bm{\hat{\theta}_{g}^u}$, which does not weight to correct for either problem is also consistent for $\bm{\theta_g^0}$ under the results of theorem \ref{theorem:consistency_2}. In fact, assumptions \ref{as:missps} and \ref{as:missp} suggest that any weighted estimator will suffice for estimating $\bm{\theta_g^0}$. In this case, one may turn to asymptotic efficiency to guide our choice between weighting or not weighting at all. The following result says that if the objective function satisfies the generalized conditional information matrix equality (GCIME), the unweighted estimator is asymptotically more efficient than any of its weighted counterpart (correctly specified weights or not). 

\begin{corollary}{(Efficiency gain with unweighted estimator under GCIME)} \label{theorem:gcime}
Under assumptions of theorem \ref{theorem:normality_2} if we additionally suppose that the objective function satisfies GCIME in the population which is defined as:
{\small \begin{equation}\label{eq:gcime}
\mathbb{E}\big[\mathbf{h}(Y(g), \mathbf{X}, \bm{\theta_g^0})\mathbf{h}(Y(g), \mathbf{X}, \bm{\theta_g^0})^\prime|\mathbf{X}\big] = \sigma_{0g}^2\cdot \bm{\nabla}_{\bm{\theta_g}}\mathbb{E}\big[\mathbf{h}(Y(g), \mathbf{X}, \bm{\theta_g^0})|\mathbf{X}\big] = \sigma_{0g}^2 \cdot \mathbf{A}(\mathbf{X}, \bm{\theta_g^0})
\end{equation}}
Then, {\small $\mathrm{Avar}\big[\sqrt{N}\big(\bm{\hat{\theta}_g}-\bm{\theta_g^0}\big) \big]= \mathbf{H}_{\bm{g}}^{-1}\bm{\Omega_g}\mathbf{H}_{\bm{g}}^{-1} \ \  \text{ and } \ \ 
	\mathrm{Avar}\big[\sqrt{N}\big(\bm{\hat{\theta}_{g}^u}-\bm{\theta_{g}^0}\big) \big]= (\mathbf{H}_{\bm{g}}^{\mathbf{u}})^{-1}\bm{\Omega}_{\bm{g}}^{\mathbf{u}} (\mathbf{H}_{\bm{g}}^{\mathbf{u}})^{-1} $}
and, \hspace{10em} {\small \[\mathrm{Avar}\big[\sqrt{N}\big(\bm{\hat{\theta}_g}-\bm{\theta_g^0}\big)\big] -\mathrm{Avar}\big[\sqrt{N}\big(\bm{\hat{\theta}_{g}^u}-\bm{\theta_{g}^0}\big) \big]\]} is positive semi-definite. \qed 
\end{corollary} 

The proof of this theorem follows from noting that we can express the difference in the two asymptotic variances as the expected outer product of population residuals from the regression of $\mathbf{B}_i $ on $\mathbf{D}_i$, which are weighted versions of square root of matrix $\mathbf{A}_i$ (See appendix \ref{prf} for details). Hence the difference is positive semi-definite.

We know GCIME is known in a variety of estimation contexts. In the case of full maximum likelihood, GCIME holds for $q(Y(g),\mathbf{X},\bm{\theta_g}) = -\ln{f_g(Y|\mathbf{X},\bm{\theta_g})}$ where $f_g(\cdot|\cdot)$ is the true conditional density with $\sigma_{0g}^2=1$. For estimating conditional mean parameters using QMLE in the linear exponential family (LEF), GCIME holds if $\mathrm{Var}(Y(g)|\mathbf{X})=\sigma_{0g}^2\cdot v[m(\mathbf{X},\bm{\theta_g^0})]$. In other words, GCIME will be satisfied if $\mathrm{Var}(Y(g)|\mathbf{X})$ satisfies the generalized linear model assumption, irrespective of whether the higher order moments of the conditional distribution correspond to the chosen QLL or not. For estimation using nonlinear least squares, GCIME will hold for $q(Y(g),\mathbf{X},\bm{\theta_g}) = [Y(g)-m(\mathbf{X}, \bm{\theta_g})]^2$ with the homoskedasticity assumption. Hence in all these cases the unweighted estimator will be more efficient than its weighted counterpart. But when GCIME is not satisfied, the two may not be easy to rank. 

\section{Estimation of treatment effects} \label{dr}
The asymptotic theory can now be used to discuss estimation of specific causal estimands like ATE and QTEs which can be expressed as functions of the doubly weighted estimator, $\bm{\theta_g^0}$.  

%Given the importance of ATE in applied work, the following section discusses how the doubly weighted framework along with results from the generalized linear model literature help us in proposing a doubly robust (DR) estimator for the ATE. 

\subsection{Average treatment effect} 
As discussed in \citet{sloczynski2018general}, DR estimators remain consistent for the population ATE despite misspecification in either the conditional mean function or the propensity score, but not both. The current doubly weighted framework along with results developed in sections \ref{theory} and \ref{strongid}  allow us to extend this result to the case with missing outcomes. 

Let $m(\mathbf{X},\bm{\theta_g})$ be a parametric model for the conditional mean which is said to be correctly specified for the CEF if for some $\bm{\theta_g^0}\in \bm{\Theta_g}$
\begin{equation*}
\mathbb{E}[Y(g)|\mathbf{X}] = m(\mathbf{X},\bm{\theta_g^0})
\end{equation*}
or equivalently, $Y(g) = m(\mathbf{X},\bm{\theta_g^0})+U(g)$ such that $\mathbb{E}[U(g)|\mathbf{X}] = 0$. Then, let us consider the following two scenarios in turn. 

\subsection{Double robustness}
\paragraph{First half: Correct conditional mean}
When the conditional mean function is correct, there is more than one estimation method that can be used to consistently estimate $\bm{\theta_g^0}$, namely, nonlinear least squares (NLS) and QMLE with LEF. For both these examples, results from section \ref{strongid} dictate that weighting is not needed for consistency. The fact that one could weight by the misspecified weights and still consistently estimate $\bm{\theta_g^0}$ is what forms the `\textit{first part}' of the DR result with double weighting.

Once $\bm{\theta_g^0}$ has been estimated by solving the sample version of the NLS or QMLE problem, ATE can be estimated as follows,
\[\hat{\Delta}_{\text{ate}} = \frac{1}{N}\sum_{i=1}^{N}m(\mathbf{X}_i, \bm{\hat{\theta}_1})- \frac{1}{N}\sum_{i=1}^{N}m(\mathbf{X}_i, \bm{\hat{\theta}_0})\]
If in addition to having a correct conditional mean, I also assume the error variance of the outcomes to be homoskedastic $\big(\mathbb{E}[U^2(g)|\mathbf{X}]=\text{Var}[U(g)|\mathbf{X}] = \sigma^2_{0g}\big)$, then the NLS estimator that does not weight at all is the preferred alternative from an efficiency perspective. This is due to GCIME being satisfied with NLS under homoskedasticity.

\paragraph{Second half: Correct weights}
If one acknowledges misspecification in the conditional mean model, there is no general way of consistently estimating the ATE. However, a useful mean fitting property of QMLEs in LEF along with double weighting can be used here to obtain consistent estimates of the unconditional means, $\mathbb{E}[Y(g)]$, despite misspecification in the conditional means, $\mathbb{E}[Y(g)|\mathbf{X}]$.\footnote{The property of QMLEs that we are most familiar with is the one where parameters in a correctly specified conditional mean can be consistently estimated if we choose $m(\mathbf{X},\bm{\theta_g})$ so that it's range corresponds to the chosen LEF density (or QLL function), irrespective of the range and nature of the outcomes. This property is used in the first half of DR.}

In the generalized linear model (GLM) literature, the link function, $h^{-1}(\cdot)$, relates the mean of the distribution to a linear index as follows 
\begin{equation}\label{eq:glm}
h^{-1}(\mathbb{E}[Y(g)|\mathbf{X}]) = \mathbf{X}\bm{\theta_g}
\end{equation}The estimation strategy then is to choose $m(\mathbf{X},\bm{\theta_g})$ to be the function, $h(\cdot)$, with the QLL corresponding to a choice of LEF density. 
Then the population first order conditions from solving this QMLE problem give us
\begin{equation}\label{glmfoc}
	\mathbb{E}\left[\frac{\bm{\nabla_{\theta_g}}h(\mathbf{X}\bm{\theta_g^\ast})^{\prime}\cdot (Y(g)-h(\mathbf{X}\bm{\theta_g^\ast}))}{v[h(\mathbf{X}\bm{\theta_g^\ast})]}\right]=\mathbf{0}
\end{equation}
where $v[h(\cdot)]$ is variance of the mean function and $\bm{\theta_g}^{\ast}$ denotes the pseudo true parameter indexing the misspecified conditional mean model [\citet{white1982maximum}]. In particular, by choosing $h^{-1}(\cdot)$ to be the canonical link for the QLL associated with the density, the gradient in numerator of (\ref{glmfoc}) cancels with the variance term in the denominator. Note that this occurs only when one uses the canonical link function. 

Such cancellation of terms ensures that if one includes an intercept in $\mathbf{X}$, the misspecified mean model fits the overall mean of the distribution  (see \citet{wooldridge2010econometric} chapter 13 for more detail) so that, 
\begin{equation*}
	\mathbb{E}[Y(g)]=\mathbb{E}[h(\mathbf{X}\bm{\theta_g^\ast})]
\end{equation*}
With nonrandom assignment and missing outcomes, solving the sample GLM FOC in (\ref{glmfoc}) would still not be sufficient for consistently estimating $\bm{\theta_g^\ast}$. Therefore, one would instead solve the doubly weighted FOC given below. 
\begin{equation}\label{glmsamplefoc}
	\begin{split}
	\sum_{i=1}^{N}\widehat{\omega}_{i1}\cdot \mathbf{X}_i^\prime \cdot\big[Y_i-h(\mathbf{X}_i\bm{\hat{\theta}_1})\big] &= \mathbf{0} \\
	\sum_{i=1}^{N}\widehat{\omega}_{i0}\cdot \mathbf{X}_i^\prime \cdot \big[Y_i-h(\mathbf{X}_i\bm{\hat{\theta}_0})\big] &= \mathbf{0}
	\end{split}
\end{equation}
The role played by weighting is crucial here for $\bm{\hat{\theta}_g}$ to be consistent for the pseudo true parameter $\bm{\theta_g^\ast}$. This forms the `\textit{second half}' of the DR result with double weighting.\footnote{Section \ref{prf} in the online appendix provides a detailed proof of how population GLM FOCs identify the unconditional means (and hence the ATE).}

If $h(\cdot)$ is the identity function, the first order conditions above can be recognized as those belonging to OLS with the line of best fit passing through the mean of $Y$. This is because OLS is a QMLE with normal QLL and identity link function, typically used for outcomes with unrestricted support. Other combinations of QLLs and canonical link functions can be found in Table 2 of \citet{negiwooldridge2020} and have to be chosen depending on the range and nature of $Y$.

\begin{summary} DR estimation of ATE with double weighting \\
	
	\textbf{Case 1: Correct mean, misspecified weights}
	\begin{enumerate}
		\item[1.] Consistent estimates for the conditional mean parameters, $\bm{\theta_g^0}$, can be obtained by either using NLS or QMLE in LEF.
		\item[2.] A consistent estimator of ATE is obtained as
		\begin{equation*}
		\hat{\Delta}_{\textup{ate}} = \frac{1}{N}\sum_{i=1}^{N}m(\mathbf{X}_i, \bm{\hat{\theta}_1})-\frac{1}{N}\sum_{i=1}^{N}m(\mathbf{X}_i, \bm{\hat{\theta}_0})
		\end{equation*}
	\end{enumerate}
	
	\textbf{Case 2: Misspecified mean, correct weights}
	\begin{enumerate}
		\item[1.] Depending upon the range and nature of the outcome, $Y$, choose an appropriate QLL associated with an LEF density. Choose the mean function, $m(\mathbf{X},\bm{\theta_g}) = h(\mathbf{X}\bm{\theta_g})$, where $h(\cdot)$ is the inverse canonical link function associated with the chosen  density. Using this combination of mean function and QLL, use the moment conditions in (\ref{glmsamplefoc}) to obtain consistent estimates, $\bm{\hat{\theta}_g}$.
		\item[2.] Consistent estimates of ATE can then be obtained as follows
		 \begin{equation*}
			\hat{\Delta}_{\textup{ate}} = \frac{1}{N}\sum_{i=1}^{N}h(\mathbf{X}_i\bm{\hat{\theta}_1})-\frac{1}{N}\sum_{i=1}^{N}h(\mathbf{X}_i\bm{\hat{\theta}_0})
			\end{equation*}
	\end{enumerate}	
\end{summary}
where $\mathbf{X}$ includes an intercept and $\bm{\hat{\theta}_g}$ solves the GLM first order conditions.

\subsection{Quantile treatment effects}
Unlike the case of ATE, it is generally not possible to obtain UQTE by averaging CQTE over the distribution of $\mathbf{X}$. In this section, I use double weighting to illustrate estimation of three different quantile estimands, namely, UQTE, CQTE, and a weighted linear approximation (LP) to the true CQTE, each of which may be of interest to the researcher depending on whether features of the conditional or unconditional outcomes distribution are of interest. Whether $\bm{\theta_g^0}$ indexes the true CQF or an approximation depends on what is being assumed about the conditional quantile model and the estimation method used. 

Let's assume that the two potential outcomes are continuous in $\mathfrak{R}$. It is typical to define the $\tau^{th}$ quantile of $Y(g)$ as
\begin{equation*}
	\mathcal{Q}_{\tau,g} = \text{inf}\{y: F_g(y)\geq \tau\}, \text{ } 0<\tau<1 
\end{equation*}
Then the UQTE for the $\tau^{th}$ quantile is defined as the difference in the marginal quantiles of the outcomes distributions,
\begin{equation*}
	\text{UQTE}_{\tau} = \mathcal{Q}_{\tau,1} -\mathcal{Q}_{\tau,0} 
\end{equation*}
Similarly, one may define the $\tau^{th}$ conditional quantile of $Y(g)$ for $\mathbf{X}=\mathbf{x}$ as,
\begin{equation*}
	\mathcal{Q}_{\tau,g}(\mathbf{x}) = \text{inf}\{y: F_g(y|\mathbf{x})\geq \tau\}, \text{ } 0<\tau<1 
\end{equation*}
where $F_g(\cdot|\mathbf{x})$ denotes the conditional distribution function of $Y(g)$ given $\mathbf{X}=\mathbf{x}$. Then, CQTE for the $\tau^{th}$ quantile for some subgroup defined by $\mathbf{X}$ is 
\begin{equation*}
	\text{CQTE}_{\tau}(\mathbf{X}) = \mathcal{Q}_{\tau,1}(\mathbf{X})- \mathcal{Q}_{\tau,0}(\mathbf{X})
\end{equation*}
Let $\mathfrak{q}_{\tau}(\mathbf{X},\bm{\theta_g}(\tau))$ be a parametric model for the $\tau^{th}$ conditional quantile of $Y(g)$ which is said to be correctly specified if for some $\bm{\theta_g^0}(\tau)\in\bm{\Theta_g}$
\begin{equation}\label{def:correctq}
	\mathcal{Q}_{\tau,g}(\mathbf{X}) = \mathfrak{q}_{\tau}(\mathbf{X},\bm{\theta_g^0}(\tau)) 
\end{equation}	

\paragraph{Estimation of CQTE$_\tau$:}
Incidentally, much like conditional mean, if CQF$_\tau$ is correctly specified, there are two methods that will ensure consistent estimation of the CQF parameters, $\bm{\theta_g^0}(\tau)$. The first is CQR of \citet{KoenkerBassett1978} and the second is a class of QML estimators that use a special `\textit{tick-exponential}' family of distributions to suggest consistent estimators of conditional quantile parameters. This QMLE class has been proposed by \citet{Komunjer2005}. The method is analogous to estimating a correctly specified conditional mean function using QMLE in the linear exponential family. 

For estimation that uses CQR, $\bm{\theta_g}(\tau)$ will actually solve the stronger conditional problem,
\begin{equation} \label{checkfunc}
\bm{\theta_g^0}(\tau) = \underset{\bm{\theta_g}\in\bm{\Theta_g}}{\mathrm{argmin}} \ \mathbb{E}\left[c_{\tau}(Y(g)-\mathfrak{q}_{\tau}(\mathbf{X},\bm{\theta_g}(\tau)))|\mathbf{X}\right]
\end{equation}  
For estimation via QMLE, as long as the CQF is correct and we choose an appropriate QLL then,
\begin{equation}\label{qml_komunjer}
\bm{\theta_g^0}(\tau) = \underset{\bm{\theta_g}\in\bm{\Theta_g}}{\mathrm{argmin}} \ \mathbb{E}\left[-\ln{\{\phi^{\tau}(Y(g),\mathfrak{q}_{\tau}(\mathbf{X},\bm{\theta_g}(\tau)))\}}|\mathbf{X}\right]
\end{equation} 
where $\phi^{\tau}(\cdot, \cdot)$ is the density that belongs to the tick-exponential family.\footnote{$\phi^{\tau}(y,\eta)=\phi^{\tau}(y,\eta) = exp\left[-(1-\tau)[a(\eta)-b(y)]\mathbf{1}\{y\leq \eta\}+\tau[a(\eta)-c(y)]\mathbf{1}\{y>\eta\}\right]$ is a probability density and $\eta$ is the $\tau$-quantile of $\phi^{\tau}$ such that $\int_{-\infty}^{\eta} \phi^{\tau}(y,\eta) dy = \tau$. \citet{Komunjer2005} shows that CQR of \citet{KoenkerBassett1978} is a special case of this QMLE class.} As dictated by results in section \ref{strongid}, weighting the QR or QML objective functions, irrespective of whether the weights are correctly specified or not will also deliver a consistent estimator of $\bm{\theta_g}(\tau)$. 

Once we have obtained $\bm{\hat{\theta}_g}$ either by solving the QR or QML problem, the $\tau^{th}$ conditional quantile treatment effect for  subgroup $\mathbf{X}$ can be estimated as $\widehat{\text{CQTE}}_\tau(\mathbf{X}) = \mathfrak{q}_{\tau}(\mathbf{X},\bm{\hat{\theta}_1}(\tau)) - \mathfrak{q}_{\tau}(\mathbf{X}, \bm{\hat{\theta}_0}(\tau))$.

\paragraph{Estimation of LP to CQTE$_\tau$:}
The traditional literature on conditional quantile estimation has focused on correct specification. However, \citet{Angristetal2006} establish an approximation property of CQR that is analogous to the approximation property of linear regression. The main implication of such a result is that solving CQR with $\mathfrak{q}_{\tau}(\mathbf{X}, \bm{\theta_g}(\tau)) = \mathbf{X}\bm{\theta_g}(\tau)$ would still identify a weighted linear approximation to CQF$_\tau$. Therefore, the difference in LPs of $\tau$-quantile CQFs is interpretable as identifying an LP to the CQTE$_\tau$. 

As before, weighting becomes crucial in the presence of nonrandom assignment and missing outcomes for identifying the LP parameters. 
\begin{equation}\label{eq:ACV}
	\begin{split}
	\bm{\hat{\theta}_g}(\tau) = \underset{\bm{\theta_g}\in \bm{\Theta_g}}{\text{argmin}}\sum_{i=1}^{N} \widehat{\omega}_{ig}\cdot c_{\tau}(Y_i-\mathbf{X}_i\bm{\theta_g}(\tau))
	\end{split}	
\end{equation}
In other words, one would need to weight the CQR problem with correct weighting functions for $\bm{\hat{\theta}_g}(\tau)\overset{p}{\rightarrow}\bm{\theta_g^\ast}(\tau)$, which indexes the true LP to CQF$_\tau$ for group $g$. Then,  
\begin{equation}
	\widehat{\text{LP}}[\text{CQTE}_\tau(\mathbf{X})]= \mathbf{X}[\bm{\hat{\theta}_1}(\tau) - \bm{\hat{\theta}_0}(\tau)]
\end{equation}

\paragraph{Direct estimation of UQTE$_\tau$:}
As mentioned in the beginning of this section, estimating UQTE$_\tau$ from CQTE$_\tau$ is generally not possible even if we assume a correct model for the conditional quantiles of $Y(g)$. In other words, one cannot obtain unconditional quantiles from averaging conditional quantiles over the distribution of $\mathbf{X}$. In this case, we can directly estimate $\mathcal{Q}_{\tau,g}$ by running a quantile regression of the outcome on an intercept (similar to \citet{firpo2007efficient}).\footnote{\citet{firpo2007efficient} uses propensity score weighting to directly estimate unconditional quantiles in the presence of nonrandom assignment.} In the present case, the solution to the doubly weighted objective function gives us, 
\begin{equation*}
	\begin{split}
	\hat{\theta}_g(\tau) &= \underset{\theta_g\in \Theta_g}{\text{argmin}} \sum_{i=1}^{N} \widehat{\omega}_{ig}\cdot c_{\tau}(Y_i-\theta_g(\tau))
	\end{split}
\end{equation*}
such that $\hat{\theta}_g(\tau) \overset{p}{\rightarrow}\mathcal{Q}_{\tau,g}$. Weighting by $G(\cdot)$ and $R(\cdot)$ is crucial here since these functions serve to remove biases arising due to nonrandom assignment and missing outcomes. One can then obtain the unconditional quantile treatment effect as, \[\widehat{\text{UQTE}}_{\tau} = \hat{\theta}_1(\tau)-\hat{\theta}_0(\tau)\]
An alternative method of estimating UQTE$_\tau$ is to use recentered influence functions suggested by \citet{firpo2009unconditional} (see appendix \ref{uqterif}).

The next section discusses results from a Monte Carlo study which evaluates the finite sample behavior of doubly weighted ATE and QTE estimators under three different misspecification scenarios.

\section{Simulations}\label{sims}
This section compares the empirical distributions of ATE and QTEs using unweighted, ps-weighted, and d-weighted estimators.\footnote{Details of the simulation design are given in section \ref{sims_appndx} of the online appendix.} The discussion is centered around three common misspecification scenarios that are interesting from an empirical standpoint. These cases are enumerated in tables \ref{tab:ate_cases} and \ref{tab:quant_cases} for estimating ATE and QTEs, respectively. Two of them describe situations implicit in the first and second half of the asymptotic theory, whereas the third case considers all three parametric components of the framework to be misspecified.  Even though the theory developed in this paper is silent for the third case, simulation results appear to be promising. 

\subsection{Average treatment effect: Results}
Case (1) in Table \ref{tab:ate_cases} considers a misspecified mean function but correct probability weights. This is the principal case covered in section \ref{theory} wherein weighting is crucial. As one can see, the empirical distribution of the doubly weighted estimator is centered on the true ATE whereas that for the unweighted estimator is shifted to the right (see figure \ref{fig:ate}, Case 1). 

Case (2) looks at what happens when everything, conditional mean and the two weights, is misspecified. The theory in this paper does not address this particular case. However, this characterizes an interesting possibility given that misspecification of all components is a valid concern. The simulation results do offer some insight here. The doubly weighted estimator seems to be the only choice that delivers the true ATE on average whereas the others distributions are shifted away from the truth (see figure \ref{fig:ate}, Case 2).

Finally, case (3) depicts the possibility of a correctly specified conditional mean function but misspecified weights. Here weighting does not have any bite in resolving the identification issue, beyond what is already achieved from having a correct mean function. In figure \ref{fig:ate}, case 3, the empirical distributions of the estimated ATE for the unweighted, ps-weighted, and d-weighted estimators all coincide and are centered on the true ATE. 

[\textit{Figure \ref{fig:ate} here}]

\subsection{Quantile treatment effects: Results}
As discussed earlier, there are really three parameters worth discussing when one talks about QTEs; CQTE, LP to CQTE, and UQTE. Misspecification in the CQF shifts attention to consistently estimating a linear projection to the true CQTE. First case in Table \ref{tab:quant_cases} considers exactly such a scenario. Using the results in \citet{Angristetal2006}, I interpret the solution to the doubly weighted problem given in (\ref{eq:ACV}) as providing a consistent weighted linear projection to the true CQF which is then used to estimate an LP to the true CQTE. Case 1 of Figure \ref{fig:qte} plots the bias in estimated LP relative to the true LP as a function of $X_1$ for the three estimators. Note that weighting here is crucial for consistently estimating the LP.  The relative bias of the doubly weighted estimator is the lowest amongst all and coincides with the line of no bias. Case 2 considers the situation when along with a misspecified CQF, the weights are also wrong. We still find the proposed estimator performing the best in terms of bias.

Finally figure \ref{fig:cqte} considers a correctly specified CQF in which case we can estimate the CQTE.\footnote{See section \ref{sims_appndx} of the online appendix for details regrading plotting the CQTE curve.}  One can observe in the figure that the estimated function using double weighting coincides with the true CQTE irrespective of how we weight. All three estimators; unweighted, ps-weighted, and doubly weighted will be consistent for the true CQTE. Misspecification in the weights will not affect this result. 

I also consider direct estimation of UQTE which does not require parametric specification of the CQF since it is simply a difference of the marginal quantiles. So the two weights are the only relevant components of the framework which will affect consistency of UQTE. In figure \ref{fig:uqte}, case 1, when both weights are correct, not weighting and double weighting both bring us close to the true parameter. For the second case where both probability models are misspecified, double weighting does a little worse than not weighting at all. However, the results at other quantiles reflect more favorably upon double weighting (see section \ref{figs} of the online appendix for results at 50th and 75th quantiles). Propensity score weighting performs the worst in both cases suggesting instances where weighting for nonrandom assignment after dropping data that is missing may not be the preferred alternative. 

[\textit{Figure \ref{fig:qte} here}]
[\textit{Figure \ref{fig:cqte} here}]
[\textit{Figure \ref{fig:uqte} here}]

\section{Returns to job training} \label{app}
In this section, I apply the proposed estimator to the Aid to Families with Dependent Children (AFDC) sample of women from the National Supported Work program compiled by \citet{calonico2017women} (CS, thereafter). NSW was a transitional and subsidized work experience program which was implemented as a randomized experiment in the United States between 1975-1979. CS replicate \citet{lalonde1986evaluating}'s within-study analysis for the AFDC women in the program, where the purpose of such an analysis is to evaluate how training estimates obtained from using non-experimental identification strategies (for example, CIA) compare to experimental estimates. To compute the non-experimental estimates, CS combine the NSW experimental sample with two non-experimental comparison groups drawn from PSID, called PSID-1 and PSID-2.\footnote{The PSID-1 sample constructed by CS involves keeping all female household heads continuously from 1975-1979 who were between 20 and 55 years of age in 1975 and were not retired in 1975. The sample labeled PSID-2 further restricts PSID-1 to include only those women who received AFDC welfare in 1975.} In this paper, I utilize the within-study feature of this empirical application to estimate how close the doubly weighted estimates get to the experimental estimate compared with ps-weighting and unweighted estimates.

To construct these empirical bias measures, I first augment the CS sample to allow for women who had missing earnings information in 1979. This renders 26\% of the experimental and 11\% of the PSID samples missing. I then combine the experimental treatment group of NSW with three distinct comparison groups present in the CS dataset, namely, the experimental control group, and the two PSID samples, to compute the unweighted, ps-weighted, and d-weighted training estimates.\footnote{For details regarding sample construction and estimation of weights, see section \ref{csapndx} of the online appendix.} The difference in the non-experimental estimate, obtained from using the doubly weighted estimator, and the experimental estimate provides the first measure of estimated bias associated with the proposed strategy. Combining the experimental control group with the non-experimental comparison group gives a second measure of estimated bias [\citet{hist}]. Much like CS, I report both these estimates across a range of regression specifications for the average returns to training estimates. 

Given the growing importance of estimating distributional impacts of job training programs, I also estimate returns to training at every 10th quantile of the 1979 earnings distribution. The role of double weighting is highlighted for the case of estimating marginal quantiles since covariates, which primarily serve to remove biases arising from nonrandom assignment and missing outcomes, enter the estimating equation only through the two weights. 

\subsection{Results}
First, to evaluate whether women with missing earnings in 1979 were significantly different than those who were observed, Table \ref{tab:miss_ttest} reports the mean and standard deviation of the woman's age, years of schooling, pre-training earnings and other characteristics across the observed and missing samples. In terms of age, the women who were observed in the experimentally treated group of NSW and the PSID-1 sample were, on average, older than those who were missing. The observed women in PSID-1 were also more likely to be married. For the PSID-2 sample, women who were observed had, on average, more kids with higher pre-training earnings. Apart from these minor differences, the observed women did not appear to be systematically different that those who were missing, as measured through observable characteristics.

The presence of non-experimental control groups implies that assignment was nonrandom and therefore an issue in the sample. This is because the comparison groups were drawn from PSID after imposing only a partial version of the full NSW eligibility criteria. Table \ref{tab:stat_var} provides descriptive statistics for the covariates by the treatment status. As can be expected, the treatment and control groups of NSW are not observably different, indicating the strong role that randomization plays in producing comparable groups. In contrast, the women in PSID-1 and PSID-2 groups are statistically different than the treatment group members implying substantial scope for nonrandom assignment. 

Table \ref{tab:ate_est} reports the d-weighted, ps-weighted and unweighted average returns to training estimates which using three different comparison groups; NSW control, PSID-1 and PSID-2. The unweighted (unadjusted and adjusted) experimental estimates given in row 1, are same as the estimates reported by CS in Table 3 of their paper. Overall, one can see that the doubly weighted experimental estimates are more stable than the single weighted or unweighted estimates across the different regression specifications, with a range between \$824-\$828.

For computing the ps-weighted and d-weighted non-experimental estimates, I first trim the sample to ensure common support between the treatment and comparison groups.\footnote{Appendix \ref{csapndx} describes estimation of the two probability weights along with the sample trimming criteria.} This reduces the sample size from 1,248 to 1,016 observations for the PSID-1 estimates and from 782 to 720 observations for the PSID-2 estimates. A pattern that is consistent across the two sets of non-experimental estimates is that weighting gets us much closer to the benchmark relative to not weighting at all. For instance, the unweighted simple difference in means estimate of training, which uses the PSID-1 comparison group, is -\$799 whereas the weighted estimates are \$827 and \$803. For the PSID-2 comparison group, the unweighted estimate which controls for all covariates is \$335 whereas the weighted estimates are \$905 and \$904.

The second panel of Table \ref{tab:ate_est} reports the bias in training estimates from combining the experimental control group with the PSID comparison groups. A similar pattern is seen here with weighted bias estimates being much closer to zero than the unweighted estimates. For instance, the doubly weighted estimate that adjusts for all covariates using the PSID-1 comparison group is -\$21 whereas the unweighted estimates is -\$568. These results suggest that the argument for weighting is strong when using a non-experimental comparison group where nonrandom assignment and missing outcomes are significant problems.\footnote{Note that the large standard errors for the non-experimental estimates can be attributed to the small sample sizes and to the large residual variance of earnings in the PSID-1 and PSID-2 populations.}

Figure \ref{fig:quant_bias} plots the relative bias in UQTE estimates at every 10th quantile of the 1979 earnings distribution. Much like the average training estimates, we see that the weighted estimates consistently lie below the unweighted estimates for most quantiles, irrespective of whether we use the PSID-1 or PSID-2 non-experimental group. Note that I do not plot UQTE estimates for quantiles less than 0.46, since these are all zero.\footnote{There are a lot of women in the experimental and PSID samples with zero real earnings in 1979.} 

This empirical application illustrates the role of proposed estimator in both experimental and observational data contexts. The comparison involving the treatment and control group of NSW demonstrates its use in an experiment with missing outcomes, whereas the non-experimental sample demonstrates its use in the more realistic observational data setting.

[\textit{Table \ref{tab:stat_var} here}]
[\textit{Table \ref{tab:miss_ttest} here}]
[\textit{Table \ref{tab:ate_est} here}]
[\textit{Figure \ref{fig:quant_bias} here}]

\section{Conclusion} \label{end}
In empirical research, the problems of nonrandom assignment and missing outcomes threaten identification of causal parameters. This paper proposes a new class of consistent and asymptotically-normal M-estimators that address these two issues using a double weighting procedure. The method combines propensity score weighting with weighting for missing outcomes in a general M-estimation framework, which can be applied to a range of estimation methods, such as ordinary least squares, quasi maximum likelihood, and quantile regression.
In addition, the proposed class has a \textit{robustness} property which allows us to estimate meaningful causal quantities of interest despite misspecification in either a conditional model of interest or the two weighting functions.

As leading applications, the paper discusses estimation of ATE and QTEs. A Monte Carlo study indicates that the doubly weighted estimates of average and quantile treatment effects have the lowest bias compared to naive alternatives (unweighted or propensity score weighted estimators) under three realistic cases of misspecification. Finally, the estimator is applied to the data on AFDC women from the NSW program compiled by Cal\'{o}nico and Smith (2017). The presence of experimental and non-experimental comparison groups in this application help to quantify the estimated bias in the doubly weighted returns to training estimates as well as the other two estimators. %Results suggest that the argument for weighting is strong whenever nonrandom assignment and (or) missing outcomes are significant concerns. 

Since the severity and magnitude of bias introduced from ignoring either problem cannot be assessed ex-ante, a safe bet from the practitioner's perspective is to report both doubly weighted and unweighted causal effect estimates. 
Practically, the doubly weighted estimator for the ATE is easy to implement. Appendix \ref{atevar} provides an example code that uses Stata's \texttt{gmm} command for implementing it. Computation of analytically correct standard errors, however, requires additional coding and is still a work in progress. Alternatively, one can use bootstrapped standard errors which will provide asymptotically correct inference.

Even though missing outcomes are a common concern in empirical analysis, it is equally common to encounter missing data on the covariates. A particularly important future extension can be to allow for missing data on both. 
In this case, using a generalized method of moments framework which incorporates information on complete and incomplete cases could provide efficiency gains over just using the observed data. A different possibility would be to relax the identifying restrictions to allow for selection on unobservables and possibly explore estimation of local average treatment effect (LATE).

\section*{References}\label{biblio}
\begin{spacing}{1}
\renewcommand{\section}[2]{}%
\bibliographystyle{ecta.bst}
\bibliography{Bibliography}
\end{spacing} 

\appendix
\singlespacing
\numberwithin{equation}{section}
\numberwithin{table}{section}
\numberwithin{figure}{section}
\section{Tables and figures for main text} \label{maintf}
\begin{figure}[H]
	\centering
	\caption{Empirical distribution of estimated ATE for N=5,000}
	\label{fig:ate}
	\begin{minipage}{0.50\columnwidth}
		\caption*{Case 1: Misspecified CEF, correct weights} \vspace{-0.5em}
		\includegraphics[scale=0.45]{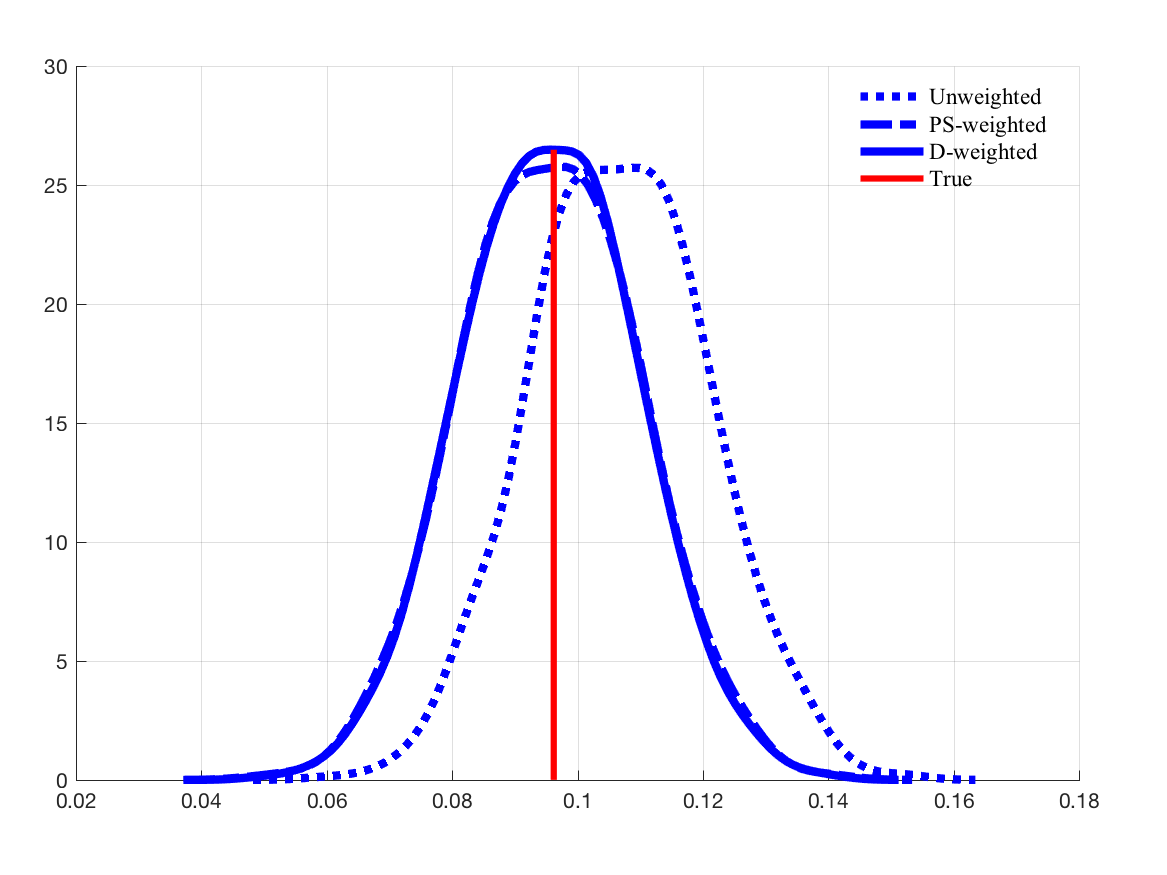}
	\end{minipage}\begin{minipage}{0.50\columnwidth}
		\caption*{Case 2: Misspecified CEF, misspecified weights} \vspace{-0.5em}
		\includegraphics[scale=0.45]{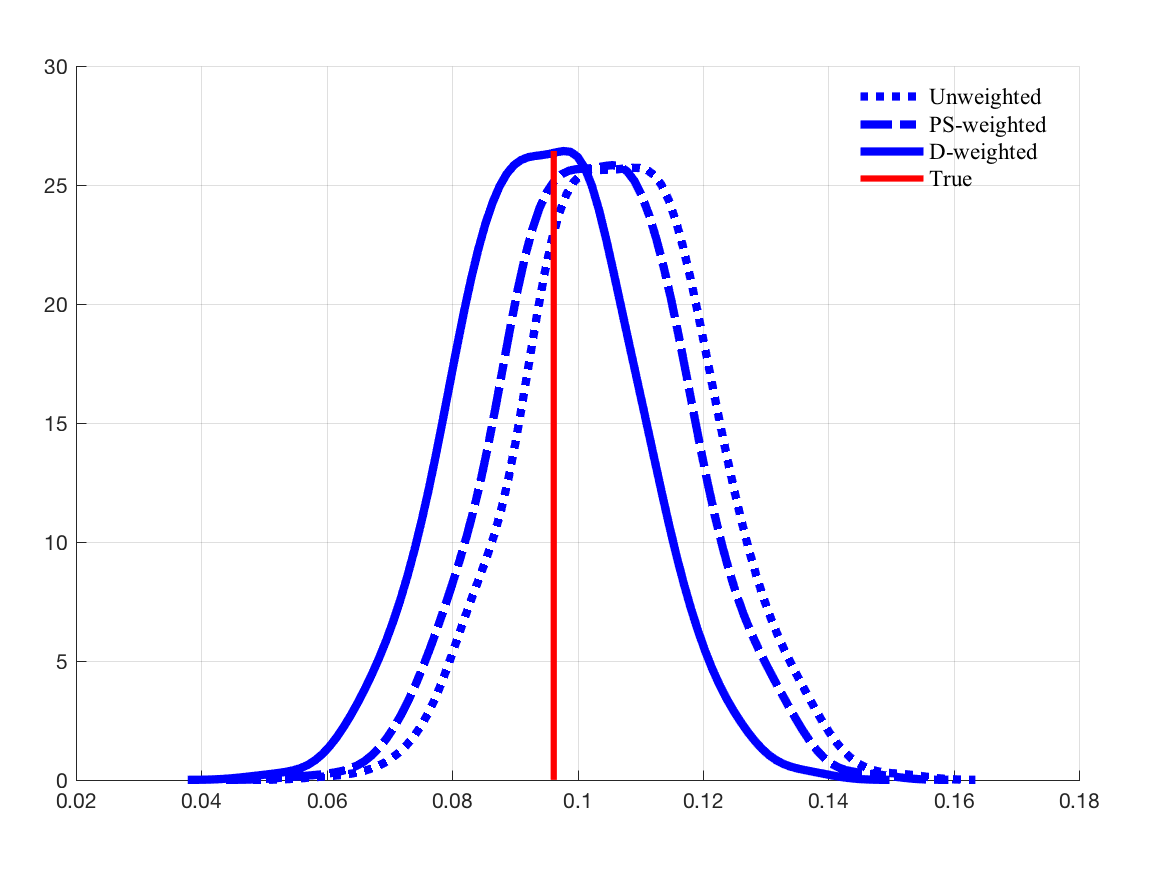}
	\end{minipage}
	\begin{minipage}{\columnwidth}
		\begin{spacing}{0.9}
			{\footnotesize \textit{Notes:} This figure plots the empirical distributions of the unweighted, ps-weighted, and d-weighted ATE estimates using 1,000 Monte Carlo simulation draws of sample size 5,000. The average treated sample size is $N_1 = 5,000\times 0.41\times 0.38 = 779$ and average control sample size is $N_0 = 5,000\times (1-0.41)\times 0.38 = 1,121$. The true ATE = 0.096 and the population is generated using a million observations. The unweighted estimator does not weight the observed data. The ps-weighted estimator weights to correct only for nonrandom assignment and the d-weighted estimator weights by both the treatment and missing outcomes probabilities.\par}
		\end{spacing}
	\end{minipage}

	\begin{minipage}{0.50\columnwidth}
		\caption*{ Case 3: Correct CEF, misspecified weights} \vspace{-0.5em}
		\includegraphics[scale=0.45]{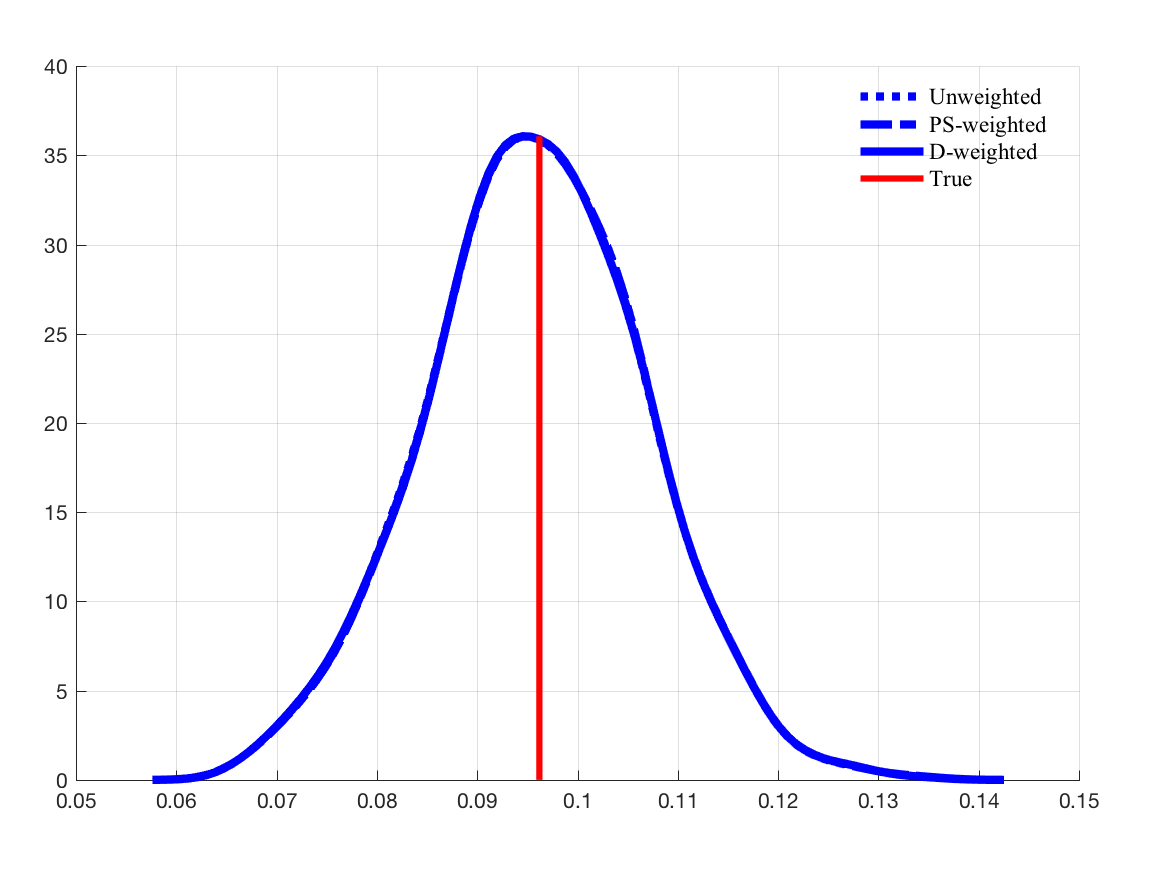}
	\end{minipage}
	\begin{minipage}{\columnwidth}
		\begin{spacing}{0.9}
			{\footnotesize \textit{Notes:} This figure plots the empirical distributions of the unweighted, ps-weighted, and d-weighted ATE estimates using 1,000 Monte Carlo simulation draws of sample size 5,000.  The average treated sample size is $N_1 = 5,000\times 0.41\times 0.38 = 779$ and average control sample size is $N_0 = 5,000\times (1-0.41)\times 0.38 = 1,121$. The true ATE = 0.096 and the population is generated using a million observations. The unweighted estimator does not weight the observed data. The ps-weighted estimator weights to correct only for nonrandom assignment and the d-weighted estimator weights by both the treatment and missing outcomes probabilities. \par}
		\end{spacing}
	\end{minipage}
\end{figure}	

\begin{figure}[H]
	\centering
	\caption{Bias in the estimated LP relative to the true LP to CQTE as a function of $X_1$ for N=5,000} 
	\label{fig:qte}
	\centering {a) $\tau=0.25$} \vspace{1em}
	
	\begin{minipage}{0.50\columnwidth}
		\caption*{Case 1: Misspecified CQF, correct weights} 
		\includegraphics[scale=0.45]{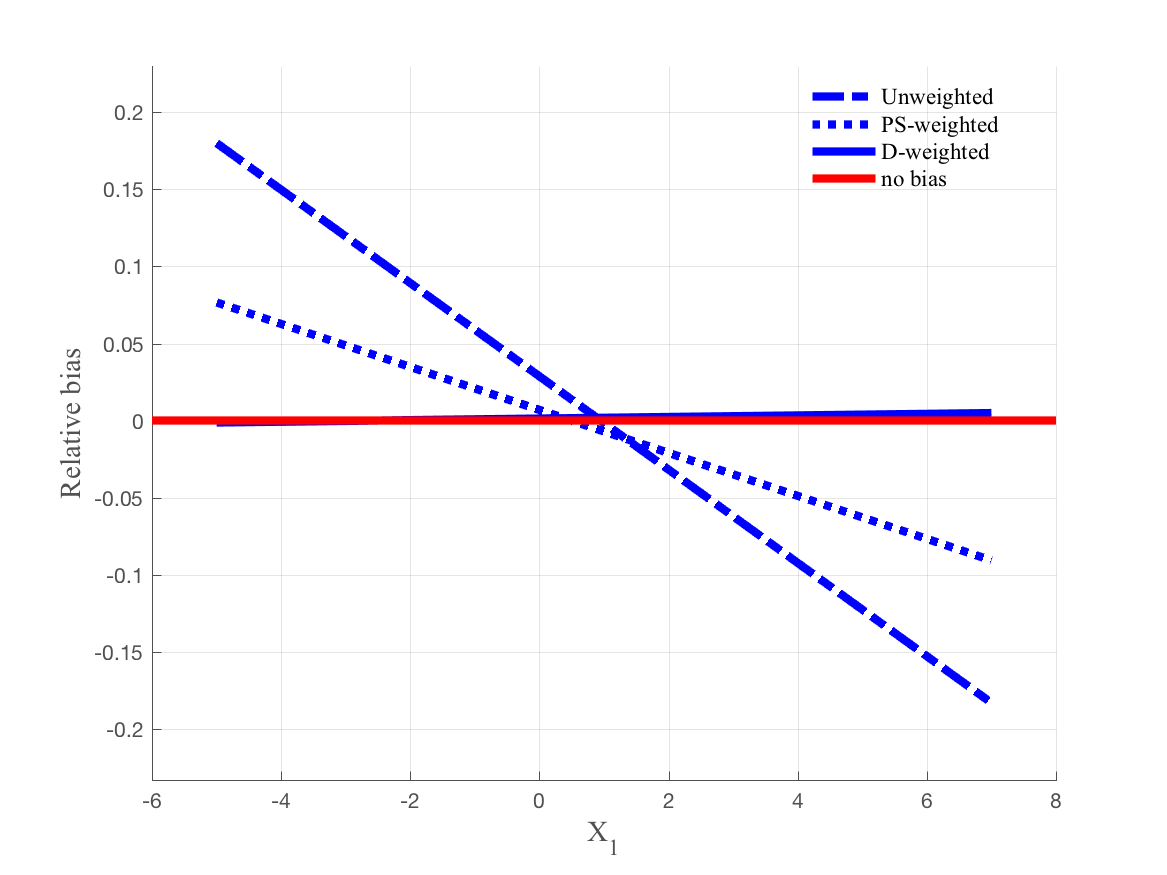}
	\end{minipage}\begin{minipage}{0.50\columnwidth}
		\caption*{Case 2: Misspecified CQF, misspecified weights} 
		\includegraphics[scale=0.45]{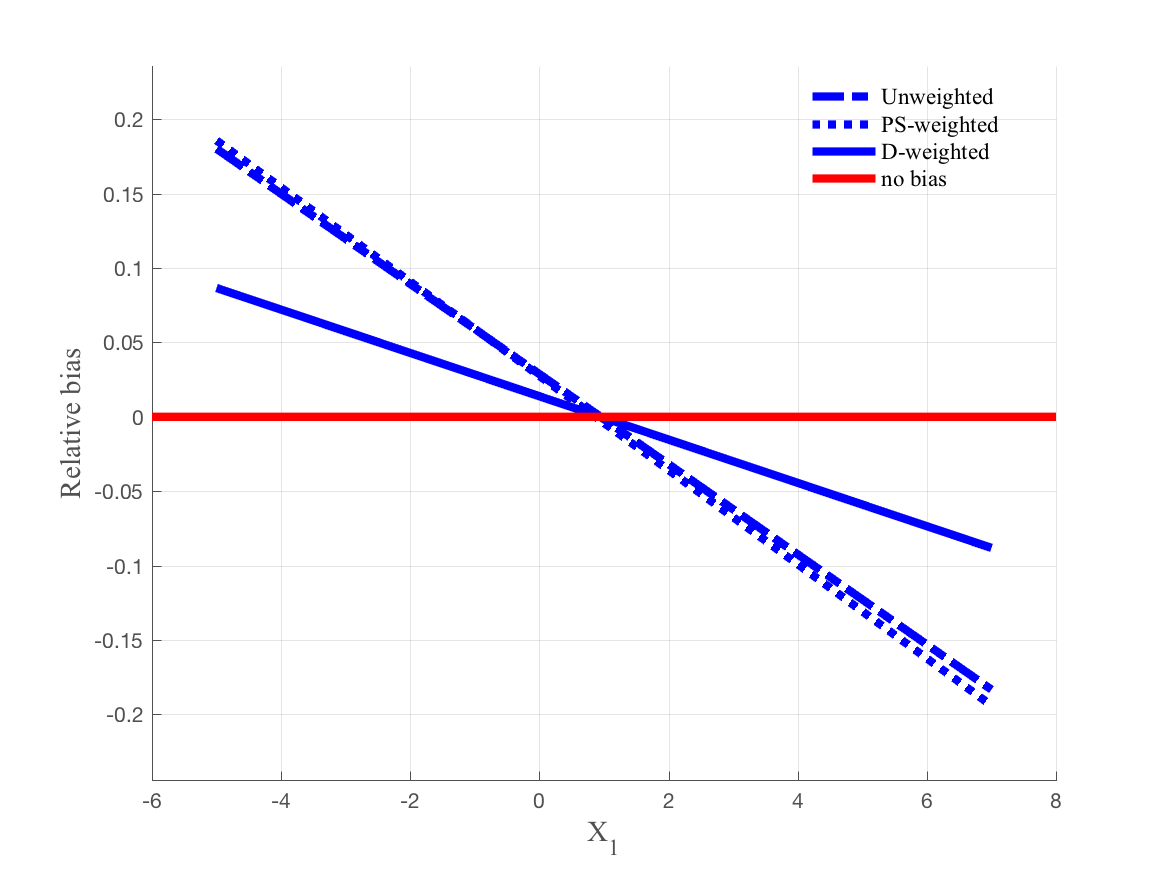}
	\end{minipage}
	\begin{minipage}{\columnwidth}
		\begin{spacing}{0.9}
			{\footnotesize \textit{Notes:} This figure plots the bias in the unweighted, ps-weighted, and d-weighted LPs to CQTE relative to the true population LP for $N=5,000$. The average treated sample size is $N_1 = 5,000\times 0.41\times 0.38 = 779$ and average control sample size is $N_0 = 5,000\times (1-0.41)\times 0.38 = 1,121$. The unweighted estimator does not weight the observed data. The ps-weighted estimator weights to correct only for nonrandom assignment and the d-weighted estimator weights by both the treatment and missing outcomes probabilities. \par}
		\end{spacing}
	\end{minipage}
\end{figure}

\begin{figure}[H]
	\centering
	\caption{Estimated CQTE with true CQTE as a function of $X_1$ for $N=5,000$} 
	\label{fig:cqte}
	\centering {a) $\tau=0.25$}  	\vspace{1em}
	
	\begin{minipage}{0.50\columnwidth}
		\caption*{Case 3: Correct CQF, misspecified weights} 
		\includegraphics[scale=0.45]{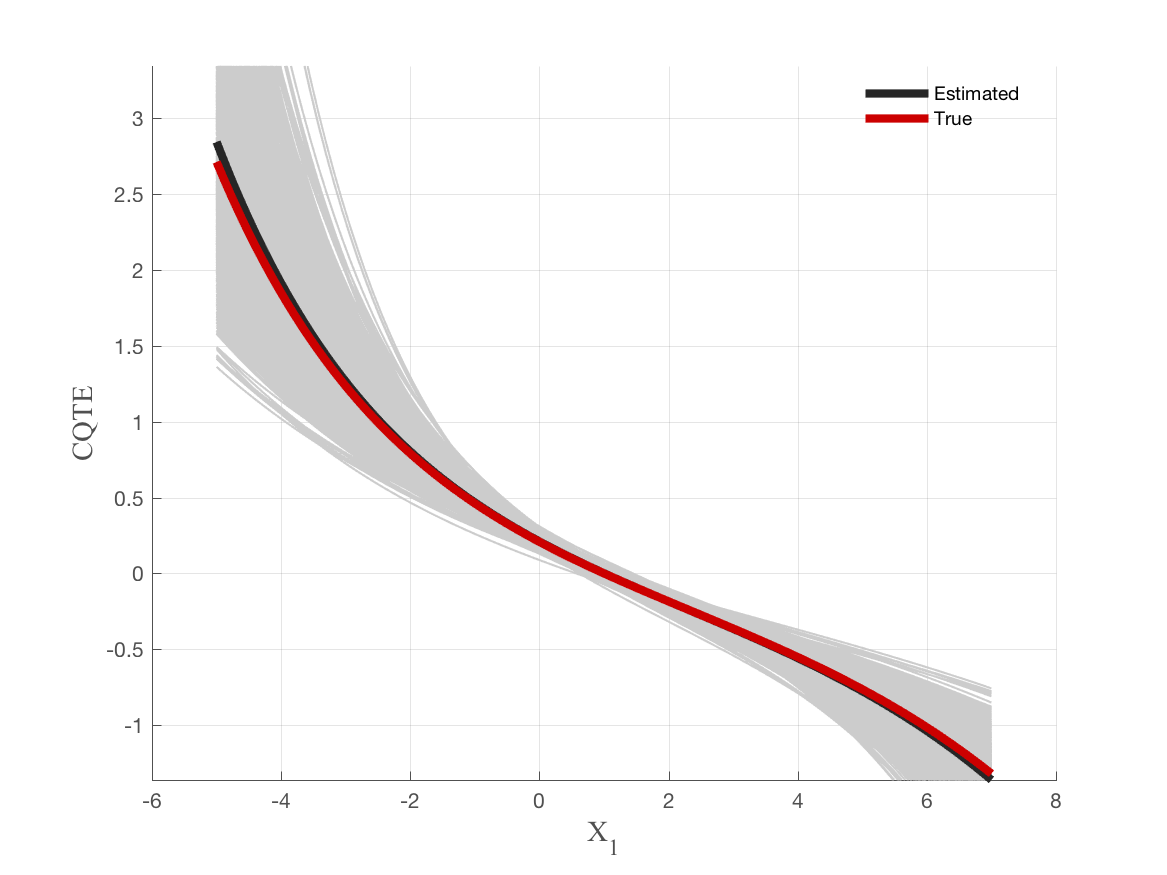}
	\end{minipage}
	\begin{minipage}{\columnwidth}
		\vspace{1em}
		\begin{spacing}{0.9}
			{\footnotesize \textit{Notes:} This figure plots the average d-weighted CQTE function with the true CQTE along $X_1$ for 1,000 Monte Carlo simulation draws of sample size $N=5,000$. Along with these two graphs, the figure also plots the individual function across the 1,000 simulation draws.  The average treated sample is $N_1 = 5,000\times 0.41\times 0.38 = 779$ and average control sample is $N_0 = 5,000\times (1-0.41)\times 0.38 = 1,121$. \par}
		\end{spacing}
	\end{minipage}
\end{figure}	

\begin{figure}[H]
	\caption{Empirical distribution of estimated UQTE for N=5,000}  \vspace{1em}
	\label{fig:uqte}
	\centering {a) $\tau=0.25$}
	\begin{minipage}{0.50\columnwidth}
		\caption*{Case1: Correct weights} \vspace{-1em}
		\includegraphics[scale=0.45]{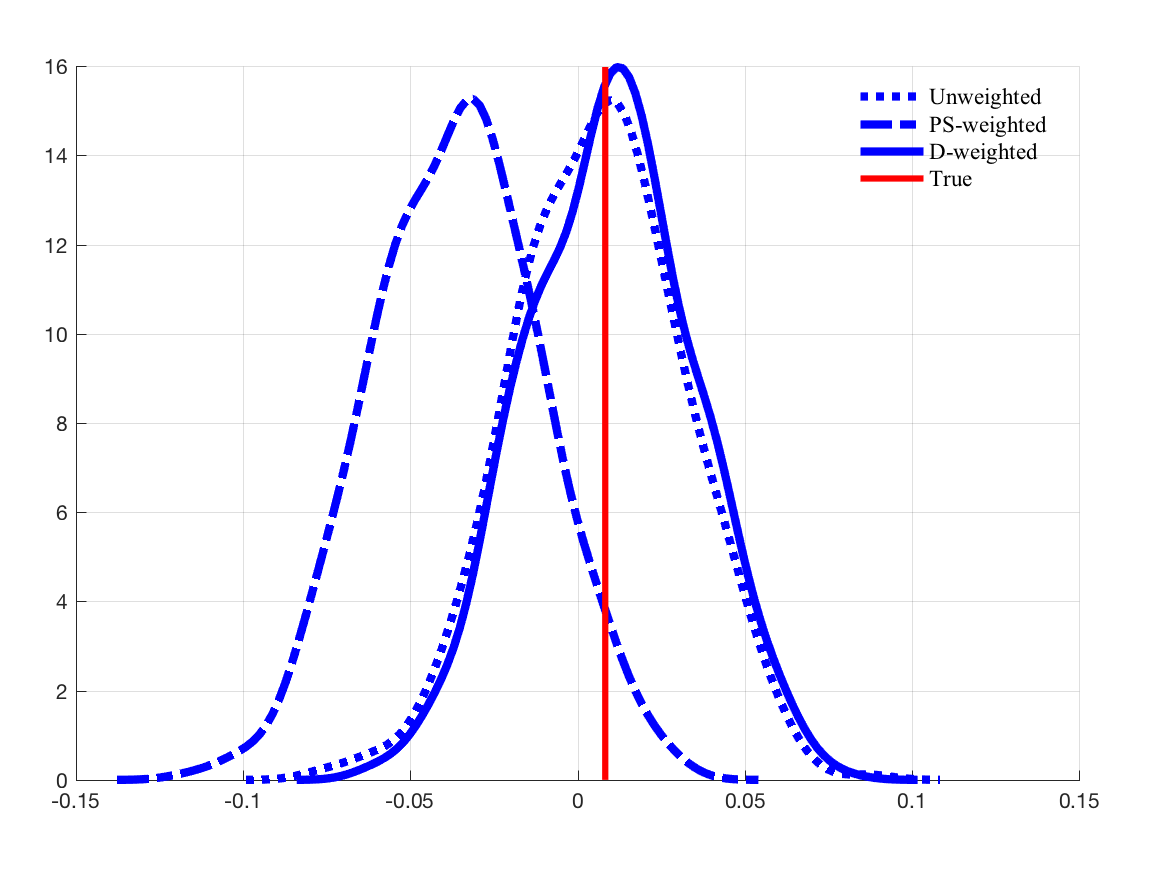}
	\end{minipage}\begin{minipage}{0.50\columnwidth}
		\caption*{Case 2: Misspecified weights} \vspace{-1em}
		\includegraphics[scale=0.45]{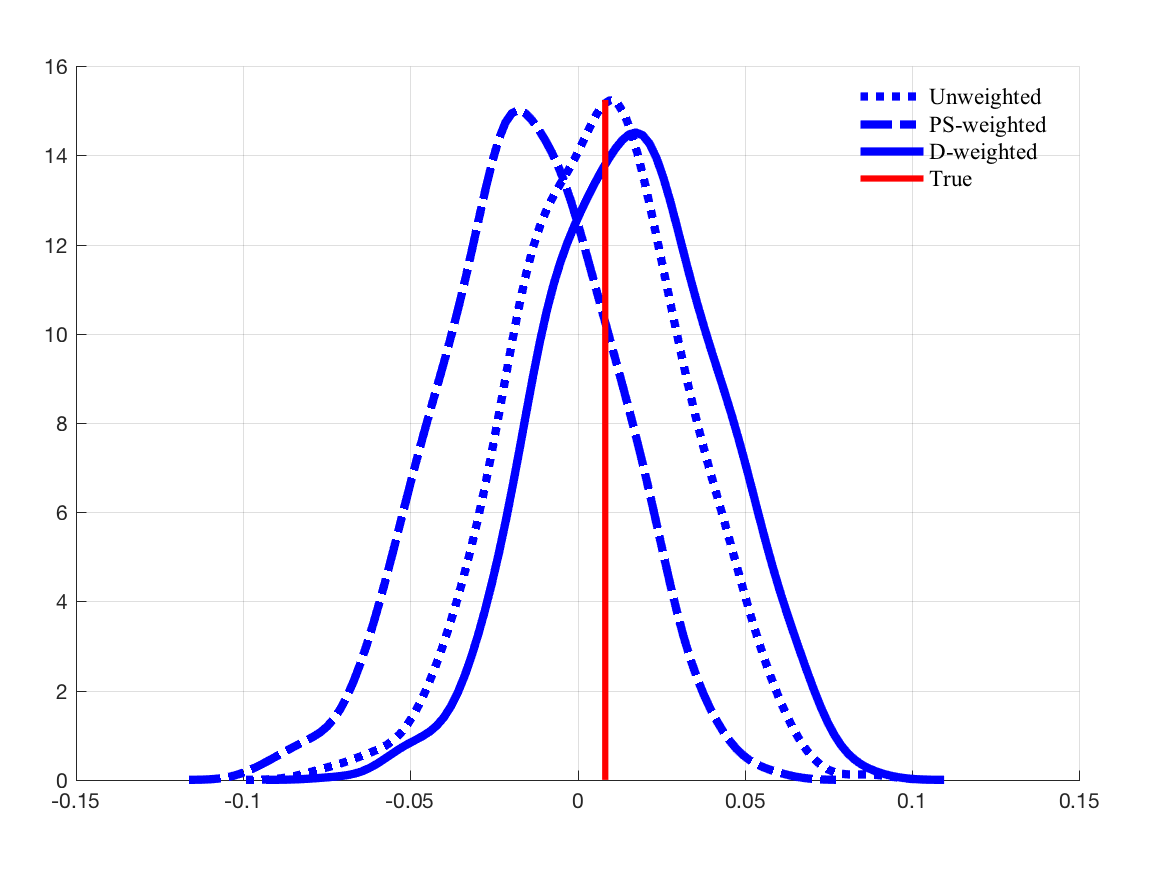}
	\end{minipage}
	\begin{minipage}{\columnwidth}
		\begin{spacing}{0.9}
			{\footnotesize \textit{Notes:} This figure plots the empirical distributions of the unweighted, ps-weighted, and d-weighted UQTE estimates using 1,000 Monte Carlo simulation draws of sample size 5,000.  The average treated sample is $N_1 = 5,000\times 0.41\times 0.38 = 779$ and average control sample is $N_0 = 5,000\times (1-0.41)\times 0.38 = 1,121$. The unweighted estimator does not weight the observed data. The ps-weighted estimator weights to correct only for nonrandom assignment and the d-weighted estimator weights by both the treatment and missing outcomes propensity score models to deal with nonrandom assignment and missing outcome problems. \par}
		\end{spacing}
	\end{minipage}
\end{figure}	

\begin{figure}[H]
	\centering
	\caption{Relative estimated bias in UQTE estimates at different quantiles of the 1979 earnings distribution}
	\begin{minipage}{0.50\columnwidth}
		\caption*{{\small a) PSID-1 control group}}
		\includegraphics[scale=0.32]{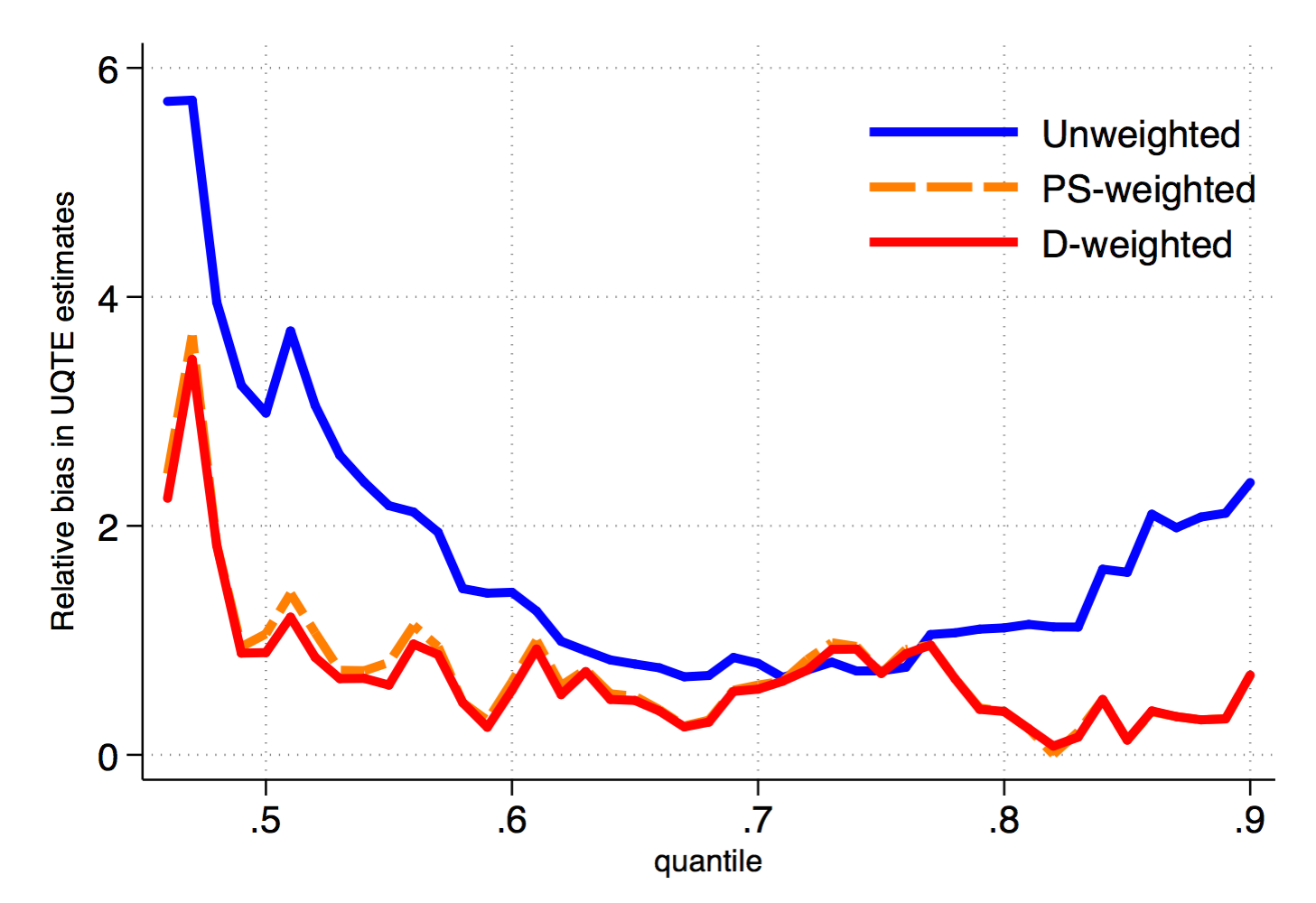}
	\end{minipage}\vspace{-0.3em}\begin{minipage}{0.50\columnwidth}
		\caption*{{\small b) PSID-2 control group}}
		\includegraphics[scale=0.32]{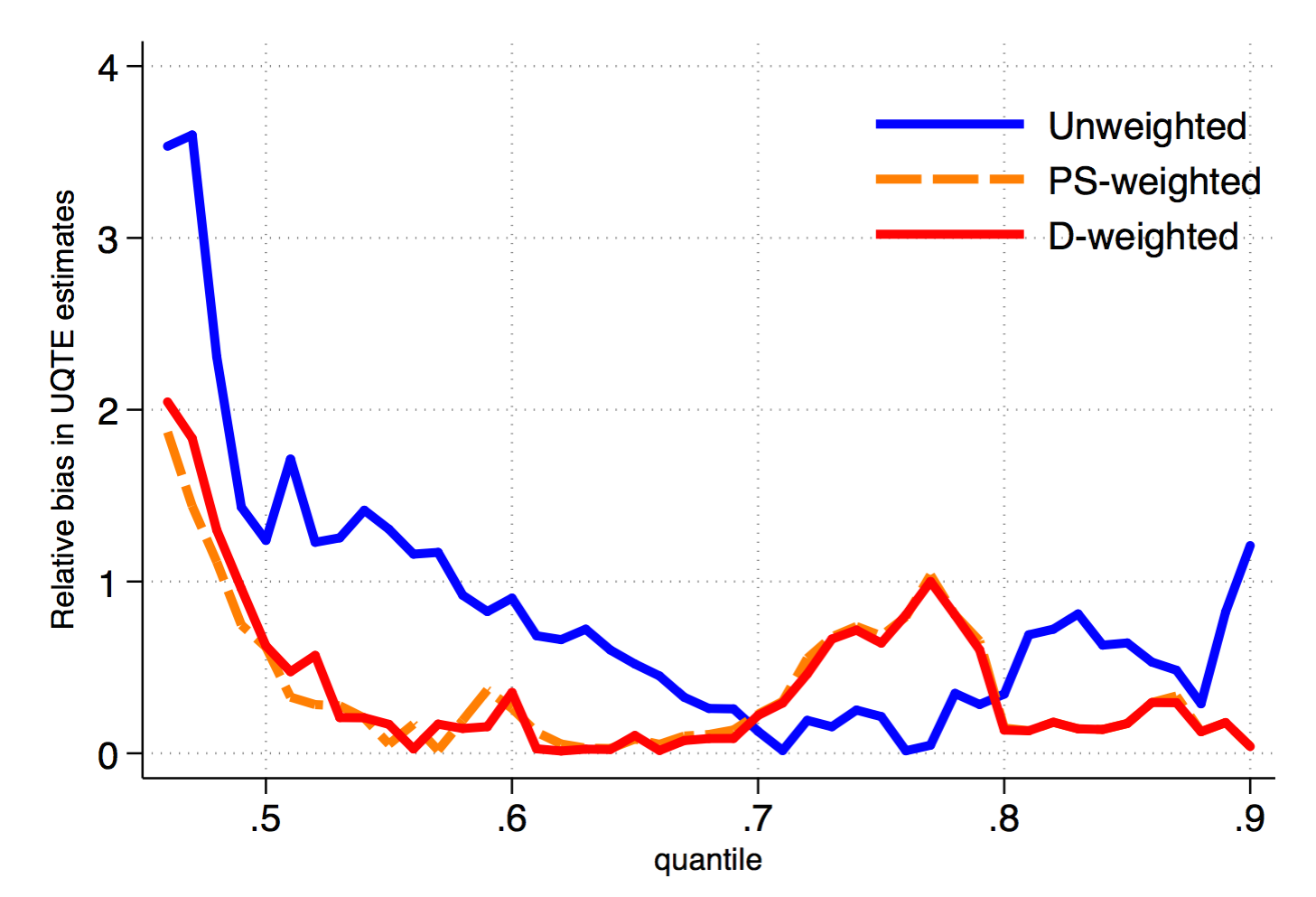}
	\end{minipage}\vspace{1em}
	\begin{minipage}{\columnwidth}
		\begin{spacing}{0.9}{\footnotesize \textit{Notes:} This graph plots the bias in the unweighted, ps-weighted and d-weighted UQTE estimates relative to the true experimental estimates across different quantiles of the 1979 earnings distribution. Panel (a) plots the relative bias estimates using the PSID-1 comparison group and Panel (b) plots the same using the PSID-2 comparison group. The treatment and missing outcome propensity score models have been estimated as flexible logits and the samples used for constructing these estimates have been trimmed to ensure common support across the two groups. The treatment propensity score has been estimated using the full experimental sample along with either PSID-1 or PSID-2 comparison group. The UQTE estimates for $\tau<0.46$ are omitted from the graph since these are zero. \par} \end{spacing}
	\end{minipage} 
	\label{fig:quant_bias}
\end{figure}	

\begin{table}[H]
	\centering
	\caption{Covariate means and p-values from the test of equality of two means, by treatment status}
	\resizebox{\textwidth}{!}{\begin{threeparttable}
			\begin{tabular}{lccc|cc|cc}
				\toprule[1pt]\midrule[0.3pt]
				\textbf{Covariates} & \textbf{Treatment} & \textbf{Control} & \textbf{P}$\mathbf{\left(\abs{T}>\abs{t}\right)}$& \textbf{PSID-1} & \textbf{P}$\mathbf{\left(\abs{T}>\abs{t}\right)}$ & \textbf{PSID-2} & \textbf{P}$\mathbf{\left(\abs{T}>\abs{t}\right)}$ \\
				\midrule
				Age in years & 33.37 & 33.64 & 0.46 & 36.73 & 0.00 & 34.41 & 0.11 \\
				& (7.42)  & (7.19)  &       & (10.60) &       & (9.48)  &  \\
				Years of education & 10.30 & 10.27 & 0.72 & 11.32 & 0.00 & 10.55 & 0.07 \\
				& (1.92)  & (2.00)  &       & (2.71)  &       & (2.09)  &  \\
				Proportion of high school dropouts & 0.70  & 0.69  & 0.73 & 0.45  & 0.00 & 0.59  & 0.00 \\
				& (0.46)  & (0.46)  &       & (0.50)  &       & (0.49)  &  \\
				Proportion Married & 0.02  & 0.04  & 0.03 & 0.02  & 0.05 & 0.01  & 0.08\\
				& (0.15)  & (0.20)  &       & (0.13)  &       & (0.10)  &  \\
				Proportion Black & 0.84  & 0.82  & 0.29 & 0.66  & 0.00 & 0.87  & 0.13 \\
				& (0.37)  & (0.39)  &       & (0.47)  &       & (0.34)  &  \\
				Proportion Hispanic & 0.12  & 0.13  & 0.59 & 0.02  & 0.00 & 0.02  & 0.00 \\
				& (0.32)  & (0.33)  &       & (0.12)  &       & (0.16)  &  \\
				Number of children in 1975 & 2.17  & 2.26  & 0.21 & 1.70  & 0.00 & 2.91  & 0.00 \\
				& (1.30)  & (1.32)  &       & (1.75)  &       & (1.73)  &  \\
				Real earnings in 1975 & 799.88 & 811.19 & 0.91 & 7446.15 & 0.00 & 2069.65 & 0.00 \\
				& (1931.92) & (2041.32) &       & (7515.59) &       & (3474.10) &  \\
				\midrule
				Observations & 796   & 795   &   & 729   &   & 204   &  \\
				\bottomrule
			\end{tabular}%
			\begin{tablenotes}[flushleft]
				\footnotesize
				\item \textit{Notes:} Along with the covariate means and standard deviation (in parentheses), the table also reports p-values from the test of equality for two means. Column 4 tests for differences between the NSW treatment and control groups, column 6 and 8 report the same using PSID-1 and PSID-2 comparison groups respectively. Real earnings in 1975 are expressed in terms of 1982 dollars. 
			\end{tablenotes}
	\end{threeparttable}}%
	\label{tab:stat_var}
\end{table}%

\begin{landscape}
	\begin{table}[H]
		\centering
		\caption{Covariate means and p-values from the test of equality of two means for the observed and missing samples}
		\resizebox{1.4\textwidth}{!}{\begin{threeparttable}
				\begin{tabular}{lccc|ccc|ccc|ccc}
					\toprule[1pt]\midrule[0.3pt]
					&       & \multicolumn{1}{l}{\textbf{Control}} &       &       & \multicolumn{1}{l}{\textbf{Treatment}} &       &       & \multicolumn{1}{l}{\textbf{PSID-1}} &       &       & \multicolumn{1}{l}{\textbf{PSID-2}} &  \\
					\cmidrule{2-13}    \textbf{Covariates} & Missing & Observed & P$\left(\abs{T}>\abs{t}\right)$ & Missing & Observed & P$\left(\abs{T}>\abs{t}\right)$ & Missing & Observed & P$\left(\abs{T}>\abs{t}\right)$ & Missing & Observed & P$\left(\abs{T}>\abs{t}\right)$ \\
					\midrule
					Age   & 33.36 & 33.74 & 0.51  & 32.15 & 33.77 & 0.01  & 34.00 & 37.07 & 0.01  & 33.32 & 34.54 & 0.62 \\
					& (7.30)  & (7.15)  &       & (7.39)  & (7.40)  &       & (10.50) & (10.57) &       & (10.81) & (9.34)  &  \\
					Years of education & 10.29 & 10.26 & 0.85  & 10.29 & 10.31 & 0.89  & 11.44 & 11.30 & 0.60  & 11.05 & 10.49 & 0.18 \\
					& (1.93)  & (2.03)  &       & (2.05)  & (1.88)  &       & (2.17)  & (2.77)  &       & (1.73)  & (2.13)  &  \\
					Proportion of high school dropouts & 0.70  & 0.68  & 0.57  & 0.69  & 0.70  & 0.77  & 0.43  & 0.45  & 0.73  & 0.55  & 0.59  & 0.68 \\
					& (0.46)  & (0.47)  &       & (0.46)  & (0.46)  &       & (0.50)  & (0.50)  &       & (0.51)  & (0.49)  &  \\
					Proportion married & 0.05  & 0.04  & 0.61  & 0.03  & 0.02  & 0.75  & 0.00  & 0.02  & 0.00  & 0.00  & 0.01  & 0.16 \\
					& (0.21)  & (0.19)  &       & (0.16)  & (0.15)  &       & (0.00)  & (0.14)  &       & (0.00)  & (0.10)  &  \\
					Proportion black & 0.81  & 0.82  & 0.81  & 0.83  & 0.84  & 0.87  & 0.74  & 0.65  & 0.10  & 0.91  & 0.86  & 0.50 \\
					& (0.39)  & (0.39)  &       & (0.38)  & (0.37)  &       & (0.44)  & (0.48)  &       & (0.29)  & (0.35)  &  \\
					Proportion hispanic & 0.12  & 0.13  & 0.87  & 0.13  & 0.12  & 0.64  & 0.01  & 0.02  & 0.82  & 0.05  & 0.02  & 0.62 \\
					& (0.33)  & (0.33)  &       & (0.33)  & (0.32)  &       & (0.11)  & (0.12)  &       & (0.21)  & (0.15)  &  \\
					Number of children in 1975 & 2.33  & 2.23  & 0.34  & 2.14  & 2.19  & 0.69  & 1.54  & 1.71  & 0.33  & 2.41  & 2.97  & 0.05 \\
					& (1.29)  & (1.34)  &       & (1.32)  & (1.29)  &       & (1.45)  & (1.78)  &       & (1.14)  & (1.79)  &  \\
					Real earnings in 1975 & 621.54 & 879.28 & 0.12  & 610.77 & 861.65 & 0.11  & 6927.95 & 7510.92 & 0.50  & 896.56 & 2211.45 & 0.02 \\
					& (1,523.00) & (2,194.93) &       & (1,677.36) & (2,005.53) &       & (7,330.74) & (7,541.41) &       & (2,315.12) & (3,567.50) &  \\
					\midrule
					Observations & 795   & 795   &    & 796   & 796   &    & 729   & 729   &    & 204   & 204   &  \\
					\bottomrule
				\end{tabular}%
				\begin{tablenotes}[flushleft]
					\footnotesize
					\item \textit{Notes:} Along with the covariate means and standard deviation (in parentheses), the table also reports p-values from the test of equality for two means between the observed and missing samples. Real earnings in 1975 are expressed in terms of 1982 dollars. 
				\end{tablenotes}
		\end{threeparttable}}%
		\label{tab:miss_ttest}%
	\end{table}%
	
	\begin{table}[H]
		\centering
		\caption{ Unweighted and weighted earnings comparisons and estimated training effects using NSW and PSID comparison groups}
		\resizebox{\columnwidth}{!}{
			\begin{threeparttable} \begin{tabular}{lrrrrrrlll}
					\toprule[1pt]\midrule[0.3pt]
					\multicolumn{1}{c}{\multirow{3}[6]{*}{\textbf{Comparison group}}} &       &       &       & \multicolumn{3}{c}{\textbf{Post-training earnings estimates}} &       &       &  \\
					\cmidrule{4-8}          & \multicolumn{3}{c}{\textbf{Unadjusted}} & \multicolumn{3}{c}{\textbf{Adjusted}} & \multicolumn{3}{c}{\textbf{Adjusted}} \\
					\cmidrule{2-10}          & \multicolumn{1}{c}{Unweighted} & \multicolumn{1}{c}{PS-weighted } & \multicolumn{1}{c|}{D-weighted} & \multicolumn{1}{c}{Unweighted} & \multicolumn{1}{c}{PS-weighted} & \multicolumn{1}{c|}{D-weighted} & \multicolumn{1}{c}{Unweighted} & \multicolumn{1}{c}{PS-weighted} & \multicolumn{1}{c}{D-weighted} \\
					\midrule
					\textbf{NSW} & \multicolumn{1}{c}{821} & \multicolumn{1}{c}{848} & \multicolumn{1}{c}{824} & \multicolumn{1}{c}{845} & \multicolumn{1}{c}{852} & \multicolumn{1}{c}{828} & \multicolumn{1}{c}{864} & \multicolumn{1}{c}{850} & \multicolumn{1}{c}{826} \\
					N=1,185 & \multicolumn{1}{c}{(307.22)} & \multicolumn{1}{c}{(304.04)} & \multicolumn{1}{c}{(304.61)} & \multicolumn{1}{c}{(303.60)} & \multicolumn{1}{c}{(302.94)} & \multicolumn{1}{c}{(303.53)} & \multicolumn{1}{c}{(303.47)} & \multicolumn{1}{c}{(302.96)} & \multicolumn{1}{c}{(303.58)} \\
					&       &       & \multicolumn{1}{r}{} &       &       & \multicolumn{1}{r}{} &       &       &  \\
					\textbf{PSID-1} & \multicolumn{1}{c}{-799} & \multicolumn{1}{c}{827} & \multicolumn{1}{c}{803} & \multicolumn{1}{c}{298} & \multicolumn{1}{c}{909} & \multicolumn{1}{c}{907} & \multicolumn{1}{c}{335} & \multicolumn{1}{c}{905} & \multicolumn{1}{c}{904} \\
					N=1,016 & \multicolumn{1}{c}{(444.84)} & \multicolumn{1}{c}{(503.00)} & \multicolumn{1}{c}{(503.26)} & \multicolumn{1}{c}{(428.60)} & \multicolumn{1}{c}{(497.76)} & \multicolumn{1}{c}{(501.54)} & \multicolumn{1}{c}{(440.18)} & \multicolumn{1}{c}{(518.54)} & \multicolumn{1}{c}{(522.97)} \\
					&       &       & \multicolumn{1}{r}{} &       &       & \multicolumn{1}{r}{} &       &       &  \\
					\textbf{PSID-2} & \multicolumn{1}{c}{-31} & \multicolumn{1}{c}{569} & \multicolumn{1}{c}{566} & \multicolumn{1}{c}{492} & \multicolumn{1}{c}{1,040} & \multicolumn{1}{c}{996} & \multicolumn{1}{c}{698} & \multicolumn{1}{c}{1,082} & \multicolumn{1}{c}{1,049} \\
					N=720 & \multicolumn{1}{c}{(713.88)} & \multicolumn{1}{c}{(1041.81)} & \multicolumn{1}{c}{(1027.12)} & \multicolumn{1}{c}{(664.46)} & \multicolumn{1}{c}{(961.74)} & \multicolumn{1}{c}{(953.80)} & \multicolumn{1}{c}{(784.28)} & \multicolumn{1}{c}{(1264.18)} & \multicolumn{1}{c}{(1217.46)} \\
					\midrule
					&       &       &       & \multicolumn{3}{c}{\textbf{Bias estimates using NSW control }} &       &       &  \\
					\cmidrule{4-8}    \textbf{PSID-1} & \multicolumn{1}{c}{-1,620} & \multicolumn{1}{c}{169} & \multicolumn{1}{c}{156} & \multicolumn{1}{c}{-493} & \multicolumn{1}{c}{-40} & \multicolumn{1}{c}{-21} & \multicolumn{1}{c}{-568} & \multicolumn{1}{c}{-38} & \multicolumn{1}{c}{-21} \\
					N=1,001 & \multicolumn{1}{c}{(431.75)} & \multicolumn{1}{c}{(561.74)} & \multicolumn{1}{c}{(553.07)} & \multicolumn{1}{c}{(427.93)} & \multicolumn{1}{c}{(499.91)} & \multicolumn{1}{c}{(501.44)} & \multicolumn{1}{c}{(434.59)} & \multicolumn{1}{c}{(504.19)} & \multicolumn{1}{c}{(507.02)} \\
					&       &       & \multicolumn{1}{r}{} &       &       & \multicolumn{1}{r}{} &       &       &  \\
					\textbf{PSID-2} & \multicolumn{1}{c}{-853} & \multicolumn{1}{c}{-228} & \multicolumn{1}{c}{-212} & \multicolumn{1}{c}{-109} & \multicolumn{1}{c}{207} & \multicolumn{1}{c}{200} & \multicolumn{1}{c}{-378} & \multicolumn{1}{c}{-17} & \multicolumn{1}{c}{-24} \\
					N=705 & \multicolumn{1}{c}{(707.87)} & \multicolumn{1}{c}{(1041.44)} & \multicolumn{1}{c}{(1025.87)} & \multicolumn{1}{c}{(663.80)} & \multicolumn{1}{c}{(962.85)} & \multicolumn{1}{c}{(954.61)} & \multicolumn{1}{c}{(759.75)} & \multicolumn{1}{c}{(1195.47)} & \multicolumn{1}{c}{(1156.39)} \\
					\midrule
					\textbf{Adjusted covariates} &       &       &       &       &       &       &       &       &  \\
					\cmidrule{1-1}    Pre-training earnings (1975) &       &       &       & \multicolumn{1}{c}{\checkmark} & \multicolumn{1}{c}{\checkmark} & \multicolumn{1}{c}{\checkmark} & \multicolumn{1}{c}{\checkmark} & \multicolumn{1}{c}{\checkmark} & \multicolumn{1}{c}{\checkmark} \\
					Age   &       &       &       & \multicolumn{1}{c}{\checkmark} & \multicolumn{1}{c}{\checkmark} & \multicolumn{1}{c}{\checkmark} & \multicolumn{1}{c}{\checkmark} & \multicolumn{1}{c}{\checkmark} & \multicolumn{1}{c}{\checkmark} \\
					Age$^2$  &       &       &       & \multicolumn{1}{c}{\checkmark} & \multicolumn{1}{c}{\checkmark} & \multicolumn{1}{c}{\checkmark} & \multicolumn{1}{c}{\checkmark} & \multicolumn{1}{c}{\checkmark} & \multicolumn{1}{c}{\checkmark} \\
					Education &       &       &       & \multicolumn{1}{c}{\checkmark} & \multicolumn{1}{c}{\checkmark} & \multicolumn{1}{c}{\checkmark} & \multicolumn{1}{c}{\checkmark} & \multicolumn{1}{c}{\checkmark} & \multicolumn{1}{c}{\checkmark} \\
					High school droput &       &       &       & \multicolumn{1}{c}{\checkmark} & \multicolumn{1}{c}{\checkmark} & \multicolumn{1}{c}{\checkmark} & \multicolumn{1}{c}{\checkmark} & \multicolumn{1}{c}{\checkmark} & \multicolumn{1}{c}{\checkmark} \\
					Black &       &       &       & \multicolumn{1}{c}{\checkmark} & \multicolumn{1}{c}{\checkmark} & \multicolumn{1}{c}{\checkmark} & \multicolumn{1}{c}{\checkmark} & \multicolumn{1}{c}{\checkmark} & \multicolumn{1}{c}{\checkmark} \\
					Hispanic &       &       &       & \multicolumn{1}{c}{\checkmark} & \multicolumn{1}{c}{\checkmark} & \multicolumn{1}{c}{\checkmark} & \multicolumn{1}{c}{\checkmark} & \multicolumn{1}{c}{\checkmark} & \multicolumn{1}{c}{\checkmark} \\
					Marital status &       &       &       & \multicolumn{1}{c}{\checkmark} & \multicolumn{1}{c}{\checkmark} & \multicolumn{1}{c}{\checkmark} & \multicolumn{1}{c}{\checkmark} & \multicolumn{1}{c}{\checkmark} & \multicolumn{1}{c}{\checkmark} \\
					Number of Children (1975) &       &       &       &       &       &       & \multicolumn{1}{c}{\checkmark} & \multicolumn{1}{c}{\checkmark}& \multicolumn{1}{c}{\checkmark} \\
					\bottomrule
				\end{tabular}%
				\begin{tablenotes}[flushleft]
					\footnotesize
					\item \textit{Notes:} This table reports unadjusted and adjusted post-training earnings differences between the NSW treatment and three different comparison groups, namely, NSW control, PSID-1 and PSID-2. The first row reports experimental training estimates which combines the NSW treatment and control group whereas the second and third rows report non-experimental estimates computed from using the PSID-1 and PSID-2 groups respectively. Each of the non-experimental estimates should be compared to the experimental benchmark. The second panel of the table reports bias estimates computed from combining the NSW control with PSID-1 and PSID-2 comparison groups respectively. These represent a second measure of bias which should be compared to zero. Bootstrapped standard errors are given in parentheses and have been constructed  using 10,000 replications. All values are in 1982 dollars. The samples used for estimating the training and bias estimates have been trimmed to ensure common support in the distribution of weights for the treatment and comparison groups. For more detail, see appendix \ref{csapndx}. 
				\end{tablenotes}	
		\end{threeparttable}}
		\label{tab:ate_est}%
	\end{table}%
\end{landscape}

	\title{Online Appendix}
	\maketitle	
	\bigskip
	
		\begin{spacing}{1.0}
		\begin{abstract}
			In this online appendix, section \ref{sims_appndx} provides details of the simulation study. Section \ref{uqterif} discusses an extension of the doubly weighted framework to the case of estimating unconditional quantile treatment effects using recentered influence functions. Section \ref{mvt} provides a simple extension to the case when treatment assumes multiple values. Section \ref{atevar} provides the asymptotic variance expressions for the average treatment effect under the first and second half of asymptotic theory. Section \ref{csapndx} provides some background information on the National supported work demonstration along with augmenting \citet{calonico2017women}'s sample for missing information and trimming rules for the probability weights. Section \ref{prf} contains proofs for results in the main text. Finally sections \ref{tabs} and \ref{figs} provide supplementary tables and figures, respectively.
		\end{abstract}
	\end{spacing}
	
	\appendix
	\numberwithin{equation}{section}
	\numberwithin{table}{section}
	\numberwithin{figure}{section}
	\section{Simulation details}\label{sims_appndx}
	This section outlines details of the simulation study for evaluating the finite sample behavior of unweighted, ps-weighted, and d-weighted (doubly weighted) estimators of ATE and QTE parameters. For each data generating process, the population is generated using a million observations. The empirical distributions of ATE and QTE estimands are simulated from drawing random vectors $\{(Y_i,\mathbf{X}_i,W_i,S_i); \  i=1,2,\ldots,N\}$ of size $N$ a thousand times without replacement from the population. This is done to mimic the setting of "random sampling" from an infinite population. 
	\subsection{Average treatment effect}
	To allow for possible misspecification of the regression functions $\mathbb{E}[Y(g)|\mathbf{X}]$, I simulate two binary potential outcomes generated using a probit as follows
	{\small \begin{spacing}{0.9}\begin{align*}
			Y(g) &=  \ \begin{cases}
			1, \text{ } Y^\ast(g)>0 \\
			0, \text{ } Y^\ast(g)\leq 0 \\
			\end{cases} \\
			Y^\ast(g) &= \ \mathbf{X}\bm{\theta_g^0}+U(g)
			\end{align*} \end{spacing}}
	
	Note that $\mathbf{X}$ here includes an intercept. The linear index, {\small $\mathbf{X}\bm{\theta_g^0}$}, is parameterized to have covariates be only mildly predictive of the potential outcomes with $R_0^2=0.19$ and $R_1^2=0.14$ in the population.\footnote{Here $\bm{\theta_0^0}=(0, 1, 1)^\prime$ and $\bm{\theta_1^0}=(-1,1,1)^\prime$. With cross sectional data, covariates are typically seen to be mildly predictive of the outcome. For example, in the National Supported Work dataset from \citet{calonico2017women}, baseline factors explain about 26-50 percent of the variation in the non-experimental sample and about .04-2 percent in the experimental sample depending upon the included subset of covariates.} The two covariates and the two latent errors are drawn from two independent bivariate normal distributions as follows, \vspace{-1.1em} {\small \begin{spacing}{0.9}\begin{align} \label{covar}
			\begin{pmatrix}
			X_1 \\
			X_2
			\end{pmatrix} \sim N\left(\begin{pmatrix} 1 \\
			2\end{pmatrix}, \begin{pmatrix}
			3 & 0.2 \\
			0.2 & 2
			\end{pmatrix}\right) \text{ and }
			\begin{pmatrix}
			U(0) \\
			U(1)
			\end{pmatrix} \sim N\left(\begin{pmatrix} 0 \\
			0\end{pmatrix}, \begin{pmatrix}
			1 & 0.2 \\
			0.2 & 1
			\end{pmatrix}\right)
			\end{align} \end{spacing}}
	The assignment and missing outcome mechanisms have been simulated to ensure that unconfoundedness and MAR are satisfied \vspace{-0.5em}
	\begin{spacing}{0.9} \begin{align}\label{assign_mech}
		\begin{split}
		W &= \begin{cases}
		1, \text{ } W^\ast>0 \\
		0, \text{ } W^\ast\leq 0\\
		\end{cases} \text{ and} \ \ \ \ S= \begin{cases}
		1, \text{ } S^\ast >0 \\
		0, \text{ } S^\ast \leq 0 \\ 
		\end{cases}
		\end{split}
		\end{align} \end{spacing}
	where \[W^\ast = \mathbf{X}\bm{\gamma_0}+\nu  \hspace{6em}	S^\ast= \mathbf{Z}\bm{\delta_0}+\upsilon  \]
	with the errors $\nu$ and $\upsilon$ drawn from two independent standard logistic distributions.\footnote{This implies that $\mathbb{P}(W=1|\mathbf{X}) \equiv p(\mathbf{X}) = \Lambda(\mathbf{X}\bm{\gamma_0})$ and $\mathbb{P}(S=1|W, \mathbf{X}) \equiv r(\mathbf{X},W) = \Lambda(\mathbf{Z}\bm{\delta_0})$ where $\Lambda(\cdot)$ is the standard logistic CDF.} 
	
	Misspecification in the true assignment and missing outcome distributions is allowed in both the functional form and linear index dimension where for the misspecified cases, I estimate a probit with $X_1$ omitted from the linear index.  For scenarios where the conditional mean is misspecified, I estimate a linear model with a correct index. The parameters, $\bm{\gamma_0}$ and $\bm{\delta_0}$, indexing the assignment and missingness mechanisms have been chosen to ensure average propensity of assignment to be 41\% and average propensity of being observed to be 38\%.\footnote{Here $\bm{\gamma_0} = (0.05, \text{-}0.2, \text{-}0.11)^\prime$, $\bm{\delta_0} = (0.01, 0.03, 0.05, \text{-}0.28)^\prime$ and $\mathbf{Z} =  (1, W, X_1, X_2)$} The missing data have been simulated to imitate empirical settings where a significant portion of the outcomes are missing. The following table gives an estimation summary for the eight different cases of misspecification, 
	
	\begin{table}[H]
		\centering
		\resizebox{0.9\textwidth}{!}{
			\begin{threeparttable}
				\caption{Estimation summary for different cases of misspecification}
				\begin{tabular}{c|cc|cc|cc}
					\toprule[1pt]\midrule[0.3pt]
					\multirow{2}[4]{*}{\textbf{Scenario}} & \multicolumn{2}{c}{\textbf{CEF}} & \multicolumn{2}{c}{$\bm{G(\cdot)}$} & \multicolumn{2}{c}{$\bm{R(\cdot)}$} \\
					\cmidrule{2-7}          & \textbf{Model} & \textbf{Estimation} & \textbf{Model} & \textbf{Estimation} & \textbf{Model} & \textbf{Estimation} \\
					\midrule
					1     & M & $\mathbf{X}\bm{\theta_g}$  & C & $\Lambda(\mathbf{X}\bm{\gamma})$ & C & $\Lambda(\mathbf{Z}\bm{\gamma})$ \\
					2    & M & $\mathbf{X}\bm{\theta_g}$ & M  &  $\Phi(\mathbf{X}^{(1)}\bm{\gamma}^{(1)})$     & M & $\Phi(\mathbf{Z}^{(1)}\bm{\gamma}^{(1)})$ \\
					3     & C & $\Phi(\mathbf{X}\bm{\theta_g})$ & M  &  $\Phi(\mathbf{X}^{(1)}\bm{\gamma}^{(1)})$     & M & $\Phi(\mathbf{Z}^{(1)}\bm{\gamma}^{(1)})$ \\
					\bottomrule
				\end{tabular}%
				\label{tab:ate_cases}%
				\begin{tablenotes}[flushleft]
					\footnotesize 
					\item \textit{Notes}: C and M correspond to whether the estimated model is correctly specified or misspecified. $\mathbf{X}$ and $\mathbf{Z}$ both include an intercept. $\mathbf{X}^{(1)}$ and $\mathbf{Z}^{(1)}$ are the subsets of $\mathbf{X}$ and $\mathbf{Z}$ left after omitting $X_1$. $G(\cdot)$ refers to the propensity score model and $R(\cdot)$ refers to the missing outcomes probability model.
				\end{tablenotes}
		\end{threeparttable}}
	\end{table}%
	
	\subsection{Quantile treatment effects}
	To ensure that the marginal quantiles of the potential outcome distributions are unique with no flat spots, I simulate two continuous non-negative outcomes as follows,
	{\small \begin{align*}
		Y(g) = \text{exp}[\mathbf{X}\bm{\theta_g^0}+U(g)], \text{ for } g=0,1
		\end{align*}}
	\hspace{-0.4em}where {\small $\bm{\theta_1^0} = (0.1, -0.36, -0.1)^\prime$} and {\small $\bm{\theta_0^0} = (0.2, 0.24, -0.45)^\prime$} are parameterized to ensure $R_0^2= 0.15$ and $R_1^2=0.13$ in the population. The two covariates and the two latent errors are drawn from two independent normal distributions following (\ref{covar}). The missing outcomes and the treatment assignment mechanisms are also generated according to eq (\ref{assign_mech}). 
	Since exp$(\cdot)$ is an increasing continuous function, the equivariance property of quantiles implies that \vspace{-0.5em}
	{\small \begin{align*}
		\mathcal{Q}_{\tau}[Y(g)|\mathbf{X}] &= \mathcal{Q}_{\tau}\left[\text{exp}(\mathbf{X}\bm{\theta_g^0}+U(g))|\mathbf{X}\right] \\
		& = \text{exp}\left[\mathcal{Q}_{\tau}(\mathbf{X}\bm{\theta_g^0}+U(g)|\mathbf{X})\right] \\
		& = \text{exp}\left[\mathbf{X}\bm{\theta_g^0}+\mathcal{Q}_{\tau}(U(g)|\mathbf{X})\right] \\
		& = \text{exp}\left[\mathbf{X}\bm{\theta_g^0}+\Phi^{-1}(\tau)\right]
		\end{align*}}
	\hspace{-0.4em}where $\Phi^{-1}(\tau)$ is the inverse standard normal CDF evaluated at $\tau$. This equivariance property helps to characterize and estimate CQTE for cases when the CQF is correct. The three different cases of misspecification are enumerated in Table \ref{tab:quant_cases} below. Case 1 corresponds to the situation for which results are derived in section \ref{theory}, Case 2 allows for misspecification in both conditional quantile function and the probability weights. Even though the theory in this paper does not address that specific case, the simulation results show that the proposed estimator has the lowest bias among all three alternatives. Finally, case 3 relates to situations considered in section \ref{strongid}; correct CQF but misspecified weights.
	
	\begin{table}[H]
		\centering
		\resizebox{0.9\textwidth}{!}{%
			\begin{threeparttable}	
				\caption{Estimation summary for quantile effects under different cases of misspecification}
				\begin{tabular}{c|cc|cc|cc}
					\toprule[1pt]\midrule[0.3pt]
					\multirow{2}[4]{*}{\textbf{Scenario}} & \multicolumn{2}{c}{\textbf{CQF}} & \multicolumn{2}{c}{$\bm{G(\cdot)}$} & \multicolumn{2}{c}{$\bm{R(\cdot)}$} \\
					\cmidrule{2-7}          & \textbf{Model} & \textbf{Estimation} & \textbf{Model} & \textbf{Estimation} & \textbf{Model} & \textbf{Estimation} \\
					\midrule
					\midrule
					%1     & C & $\text{exp}(\mathbf{x}\bm{\theta_g}(\tau))$ & C & $\Lambda(\mathbf{x}\bm{\gamma})$ & C & $\Lambda(\mathbf{z}\bm{\gamma})$ \\
					%2     & C & $\text{exp}(\mathbf{x}\bm{\theta_g}(\tau))$ & C & $\Lambda(\mathbf{x}\bm{\gamma})$ & M & $\Phi(\mathbf{x}^{(1)}\bm{\gamma}^{(1)})$ \\
					%3     & C & $\text{exp}(\mathbf{x}\bm{\theta_g}(\tau))$ & M &   $\Phi(\mathbf{x}^{(1)}\bm{\gamma}^{(1)})$    & C & $\Lambda(\mathbf{z}\bm{\gamma})$\\
					1     & M & $\mathbf{X}\bm{\theta_g}(\tau)$  & C & $\Lambda(\mathbf{X}\bm{\gamma})$ & C & $\Lambda(\mathbf{Z}\bm{\gamma})$ \\
					2     & M & $\mathbf{X}\bm{\theta_g}(\tau)$ & M  &  $\Phi(\mathbf{X}^{(1)}\bm{\gamma}^{(1)})$     & M & $\Phi(\mathbf{X}^{(1)}\bm{\gamma}^{(1)})$ \\
					3    & C & $\text{exp}(\mathbf{X}\bm{\theta_g}(\tau))$ & M  &  $\Phi(\mathbf{X}^{(1)}\bm{\gamma}^{(1)})$     & M & $\Phi(\mathbf{X}^{(1)}\bm{\gamma}^{(1)})$ \\
					\bottomrule
				\end{tabular}%
				\label{tab:quant_cases}
				\begin{tablenotes}[flushleft]
					\footnotesize
					\item \textit{Notes}: C and M denote whether the estimated model is correctly specified or misspecified. $\mathbf{X}$ and $\mathbf{Z}$ both include an intercept. $\mathbf{X}^{(1)}$ and $\mathbf{Z}^{(1)}$ are the subsets of $\mathbf{X}$ and $\mathbf{Z}$ left after omitting $X_1$. Therefore, the probability models have been misspecified in both the functional form and the linear index dimension. $G(\cdot)$ refers to the propensity score model and $R(\cdot)$ refers to the missing outcomes probability model.
				\end{tablenotes}
		\end{threeparttable}}
	\end{table}%

	For plotting the estimated and true CQTE functions, I first collect the estimates that solve the unweighted, ps-weighted, and doubly weighted CQR problem (defined in (\ref{checkfunc})) corresponding to a particular quantile level, $\tau =0.25, 0.50, 0.75$ across 1,000 Monte Carlo simulation draws. I then draw a linearly spaced vector of values for $X_1$  and simulate the CQTE using the 1,000 estimated conditional quantile coefficients. Averaging  these 1,000 functions at each point on the $X_1$ vector gives me the estimated average CQTE function. I plot this along with the 1,000 individual functions and the true CQTE, which is calculated using the population conditional quantile parameters, $\bm{\theta_g^0}$.
	
	\section{Unconditional quantile treatment effect using recentered influence functions} \label{uqterif}
	This section discusses an alternative method of estimating UQTE using \citet{firpo2009unconditional}'s (FFL, thereafter) recentered influence function (RIF) methodology. 
	
	Following FFL, let $v(F)$ be a real valued functional such that $v: \mathcal{F}\rightarrow \mathfrak{R}$ whose domain $\mathcal{F}$ is a class of distribution functions such that $F\in \mathcal{F}$ if $\abs{v(F)}<+\infty$. One may define $v(\cdot)$ to be any distributional statistic of interest like mean, variance, quantiles, inequality indices etc. We can define various treatment effects as the difference in the functionals of the marginal outcome distributions
	\begin{equation}\label{te}
	\Delta_v = v_1-v_0
	\end{equation}
	where $v_g \equiv v(F_{g})$ is the functional of the distribution function for $Y(g)$.\footnote{Note that \citet{firpo2016identification} use the above formulation to consider inequality treatment effects by exclusively considering $v$ to be different inequality measures.} As defined in FFL, the RIF is nothing but the influence function which has been centered at the statistic $v_g$. Formally,
	\begin{equation}\label{rif}
	\text{RIF}(Y(g);v,F_g) = v(F_g)+\text{IF}(Y(g);v, F_g)
	\end{equation} 
	where $\text{IF}(Y(g);v, F_g)$ captures the change in $v_g$ as a result of an  infinitesimal change in the distribution of $\mathbf{X}$. FFL introduce the idea of running a standard regression of RIF on $\mathbf{X}$ with the objective of estimating the function \[\mathbb{E}\left[\text{RIF}(Y(g);v, F_g)|\mathbf{X}\right] = \mathbf{X}\bm{\theta_g^0}\] One can then use the law of iterated expectations to express $v_g$ in terms of the regression function as follows,
	\begin{equation}
		\mathbb{E}[\mathbb{E}(\text{RIF}(Y(g);v,F_g)|\mathbf{X})] = v_g
	\end{equation} 
	
	For $v_g = \mathcal{Q}_{\tau,g}$, equation \ref{rif} defines the UQTE for the $\tau^{th}$ quantile. 
	  We know that the RIF for $\mathcal{Q}_{\tau,g}$ is given as:
	\begin{equation}\label{rif_q}
	\text{RIF}(Y(g);\mathcal{Q}_{\tau}, F_g) = \mathcal{Q}_{\tau,g}+\frac{\tau-\mathbf{1}\{Y(g)\leq \mathcal{Q}_{\tau,g}\}}{f_g(\mathcal{Q}_{\tau,g})} 
	\end{equation}
	where $f_g(\cdot)$ is the density of $Y(g)$.\footnote{Note that \citet{firpo2009unconditional} express the conditional RIF expectation as  $\mathbb{E}\left[\text{RIF}(Y(g);\mathcal{Q}_{\tau}, F_g)|\mathbf{X}\right] = c_{1,\tau,g}\cdot\mathbb{P}[Y(g)>\mathcal{Q}_{\tau,g}|\mathbf{X}]+c_{2,\tau,g}$ where $c_{1,\tau,g} = 1/f_g(\mathcal{Q}_{\tau,g})$ and $c_{2,\tau,g} = \mathcal{Q}_{\tau,g}-c_{1,\tau,g}\cdot(1-\tau)$ for the $\tau^{th}$ quantile of $Y(g)$.}
	Then estimation of doubly weighted UQTE using RIFs involves the following steps:
	\begin{itemize}
		\item[a.] $ \displaystyle \bm{\hat{\theta}_g} = \left(\frac{1}{N}\sum_{i=1}^{N}\widehat{\omega}_{ig}\mathbf{X}_i^\prime\mathbf{X}_i\right)^{-1}\left(\frac{1}{N}\sum_{i=1}^{N}\widehat{\omega}_{ig}\mathbf{X}_i^\prime \cdot\widehat{\text{RIF}}(Y_i;\widehat{\mathcal{Q}}_{\tau}, \widehat{F}_g) \right)$
		\item[b.] $\widehat{\text{RIF}}(Y(g);\widehat{\mathcal{Q}}_{\tau}, \widehat{F}_g) = \widehat{\mathcal{Q}}_{\tau,g}+\displaystyle \frac{\tau-\mathbf{1}\{Y(g)\leq \widehat{\mathcal{Q}}_{\tau,g}\}}{\widehat{f}_g(\widehat{\mathcal{Q}}_{\tau,g})}$ where $\displaystyle \widehat{f}_g(y)$ is the non-parametric kernel density estimator with bandwidth $h_g$. 
		\item[c.] $\displaystyle \widehat{f}_g(\widehat{\mathcal{Q}}_{\tau,g}) = \frac{1}{N}\sum_{i=1}^{N}\widehat{\omega}_{ig} \cdot \frac{1}{h_g} \cdot \mathcal{K}_g\left(\frac{\widehat{\mathcal{Q}}_{\tau,g}}{h_g}\right)$
		\item[d.] $	\widehat{\mathcal{Q}}_{\tau,g} = \underset{\mathcal{Q}_g}{\text{argmin}} \sum_{i=1}^{N} \widehat{\omega}_{ig} \cdot c_{\tau}(Y_i-\mathcal{Q}_g)$
		\item[e.] $\displaystyle \widehat{\omega}_{i1} = \frac{S_i\cdot W_i}{R(\mathbf{X}_i,W_i,\bm{\hat{\delta}})\cdot G(\mathbf{X}_i,\bm{\hat{\gamma}})}$ and $\displaystyle \widehat{\omega}_0= \frac{S_i\cdot (1-W_i)}{R(\mathbf{X}_i,W_i,\bm{\hat{\delta}})\cdot (1-G(\mathbf{X}_i,\bm{\hat{\gamma}}))}$ \\
	\end{itemize}

	where double weighting has to be performed at each stage that uses the observed sample. This implies that for ensuring  consistency of UQTE, the weights would necessarily have to be correctly specified. One may estimate the weights nonparametrically using sieves to sidestep this issue of misspecification. Estimating UQTE in this manner also has the advantage initially put forth in FFL which is that one can directly estimate the effect of covariates on UQTE.

	\section{Multivalued Treatments} \label{mvt}
	One can easily extend the binary treatment case considered here to the case when there are multiple treatment values. Let $Y(g)$ denote the potential outcome for treatment level $g$ where $g=0,1,\ldots, T$ and $W_g$ be a binary indicator for receiving treatment level $g$ such that 
	\begin{equation*}
	\begin{split}
	W_0+W_1+\ldots+W_T&=1 \\
	\mathbb{P}(W_g=1) &\equiv \rho_g > 0
	\end{split}
	\end{equation*} 
	Also, let $\mathbf{W} = \begin{pmatrix} W_0, W_1, \ldots, W_T\end{pmatrix}$ . Then the observed outcome is
	\begin{equation*} 
	Y = W_0\cdot Y(0)+W_1\cdot Y(1)+\ldots+W_T\cdot Y(T)
	\end{equation*}
	Let $\rho_g(\mathbf{x}) \equiv \mathbb{P}(W_g=1|\mathbf{X}=\mathbf{x})$ be the propensity score and $r(\mathbf{x},w) \equiv \mathbb{P}(S=1|\mathbf{X}=\mathbf{x}, W_g=w)$ be the missing outcomes probability for treatment level $g$. One may then consider solving the same population problem, $Q_0(\bm{\theta_0})$ but with true weights given as
	\begin{equation*}
	\omega_{g} = \frac{S\cdot W_{g}}{r(\mathbf{X},W_{g}) \cdot \rho_g(\mathbf{X})}
	\end{equation*}
	
	To construct the doubly weighted estimator, we would assume unconfoundedness and MAR along with assuming parametric models for the two probability weights; $R(\mathbf{X},W_g,\bm{\delta})$ and $G(\mathbf{X},\bm{\gamma_g})$.
	
	\section{Asymptotic variance for ATE} \label{atevar}
	Given $\sqrt{N}$ consistent and  asymptotically normal estimators, $\bm{\hat{\theta}_1}$ and $\bm{\hat{\theta}_0}$, the estimated average treatment effect
	\[\hat{\Delta}_{\text{ate}} = \frac{1}{N}\sum_{i=1}^{N}m(\mathbf{X}_i,\bm{\hat{\theta}_1})-\frac{1}{N}\sum_{i=1}^{N}m(\mathbf{X}_i, \bm{\hat{\theta}_0})\]
	is easily shown to also be $\sqrt{N}$-consistent and asymptotically normal [\citet{wooldridge2010econometric} chapter 21]. Regularity conditions for such an asymptotic result would require that the parametric model, $m(\mathbf{X},\bm{\theta_g})$, is continuously differentiable on the parameter space $\bm{\Theta_g} \subset \mathfrak{R}^{P_g}$ and $\bm{\theta_g^0}$ is in the interior of $\bm{\Theta_g}$. Then, by the continuous mapping theorem and slutsky's theorem,
	\begin{align*}
	\sqrt{N}\left(\hat{\Delta}_{\text{ate}} -\Delta_{\text{ate}} \right)\overset{d}{\rightarrow} N(0, \text{V})
	\end{align*}
	where $\text{V} = \mathbb{E}\left[\psi(\mathbf{X}_i)\psi(\mathbf{X}_i)^\prime\right]$. Let's denote 
	$\mathbb{E}\left[\bm{\nabla_{\theta_g}}m(\mathbf{X}_i,\bm{\theta_g^0})\right]\equiv \mathbf{J}_{\bm{g}}^\mathbf{0}$, then \vspace{-0.5em}
	\begin{align*}
	\psi(\mathbf{X}_i) = \left\{m(\mathbf{X}_i, \bm{\theta_1^0})-m(\mathbf{X}_i, \bm{\theta_0^0})-\Delta_{\text{ate}}\right\}-\mathbf{J_1^0}\cdot \mathbf{H_1^{-1}}\mathbf{u}_{i\mathbf{1}}+\mathbf{J_0^0}\cdot \mathbf{H_0^{-1}}\mathbf{u}_{i\mathbf{0}}
	\end{align*}
	where $\mathbf{H}_{\bm{g}}$ is the Hessian for the treatment group $g$, and $\mathbf{u}_{ig}$ is the residual from the regression of the weighted score on the scores of two probability models. 
	For the case when the conditional mean model is correctly specified, the variance expression simplifies to \vspace{-0.5em}
	\begin{align}\label{ate_var}
	\begin{split}
	\text{V}= \mathbb{E}\left[(m(\mathbf{X}_i, \bm{\theta_1^0})-m(\mathbf{X}_i,\bm{\theta_0^0}))-\Delta_{\text{ate}}\right]^2+\mathbf{J_1^0}\cdot \mathbf{V_1}\cdot \mathbf{J_1^0}^\prime+\mathbf{J_0^0}\cdot \mathbf{V_0}\cdot \mathbf{J_0^0}^\prime 
	\end{split}
	\end{align}
	Here $\mathbf{V_1}$ and $\mathbf{V_0}$ are the asymptotic variances of the doubly weighted estimator that solve the treatment and control group problems, respectively. The above formula makes it clear that it better to use more efficient estimators of $\bm{\hat{\theta}_g}$. But we know from the results in section \ref{strongid} that when the conditional mean model is correctly specified, using estimated weights is as efficient as using known weights. Another alternative in this case is to use unweighted estimators of $\bm{\theta_g^0}$ since under GCIME, unweighted estimators is more efficient than the doubly weighted estimators of $\bm{\theta_g^0}$. 
	
	For the case when the mean model is misspecified, the asymptotic variance of the ATE is given as follows
	\begin{align}
	\begin{split}
	\text{V}& = \mathbb{E}\left[(m(\mathbf{X}_i, \bm{\theta_1^0})-m(\mathbf{X}_i,\bm{\theta_0^0}))-\Delta_{\text{ate}}\right]^2+\mathbf{J_1^0}\cdot \mathbf{V_1}\cdot\mathbf{J_1^0}^\prime + \mathbf{G_0^0}\cdot \mathbf{V_0}\cdot \mathbf{J_0^0}^\prime \\
	&-2\mathbb{E}\left[\{m(\mathbf{X}_i, \bm{\theta_1^0})-m(\mathbf{X}_i,\bm{\theta_0^0})-\Delta_{\text{ate}}\}\mathbf{u}_{i1}^\prime\right]\mathbf{H_1^{-1}}\mathbf{J_1^0}^\prime \\
	&+2\mathbb{E}\left[\{m(\mathbf{X}_i, \bm{\theta_1^0})-m(\mathbf{X}_i,\bm{\theta_0^0})-\Delta_{\text{ate}}\}\mathbf{u}_{i0}^\prime\right]\mathbf{H_0^{-1}}\mathbf{J_0^0}^\prime
	\end{split}
	\end{align}
	In this case, the variance expression is a bit more complicated than the previous case.  Even though it is better to have more efficient estimators of $\bm{\theta_g^0}$ in this case as well, it is not obvious whether that would help obtain a smaller variance for the ATE since we now have cross correlation terms in the variance expression.
	
	\subsection{Proofs}
	\begin{proof}[\textup{\textbf{Asymptotic variance expression for ATE: Correctly specified mean model}}]
			
			Assuming continuous differentiability of $m(\mathbf{X}_i, \bm{\theta_g})$ on $\bm{\Theta_g}$, mean value expansion around $\bm{\theta_g^0}$ gives
			\begin{equation*}
			\frac{1}{N}\sum_{i=1}^{N}m(\mathbf{X}_i, \bm{\hat{\theta}_g}) \approx \frac{1}{N}\sum_{i=1}^{N}m(\mathbf{X}_i, \bm{\theta_g^0})+\frac{1}{N}\sum_{i=1}^{N}\bm{\nabla_{\theta_g}}m(\mathbf{X}_i,\bm{\tilde{\theta}_g})\cdot (\bm{\hat{\theta}_g}-\bm{\theta_g^0}) 
			\end{equation*}
			where $\bm{\tilde{\theta}_g}$ lies between $\bm{\hat{\theta}_g}$ and $\bm{\theta_g^0}$. Since $\bm{\hat{\theta}_g}\overset{p}{\rightarrow}\bm{\theta_g^0}$, so does $\bm{\tilde{\theta}_g}$. Hence, using the weak law of large numbers, we obtain
			{\small\begin{equation*}
				\begin{split}
				\frac{1}{\sqrt{N}}\sum_{i=1}^{N}m(\mathbf{X}_i, \bm{\hat{\theta}_g}) &= \frac{1}{\sqrt{N}}\sum_{i=1}^{N}m(\mathbf{X}_i, \bm{\theta_g^0})+\mathbf{J}_{\bm{g}}^\mathbf{0}\cdot \sqrt{N}(\bm{\hat{\theta}_g}-\bm{\theta_g^0}) + o_p(1) 
				\end{split}
				\end{equation*}}
			Adding and subtracting {\small $\sqrt{N}\cdot \mathbb{E}[m(\mathbf{X}_i,\bm{\theta_g^0})]$} on both sides gives us 
			{\small\begin{equation*}
				\begin{split}
				\frac{1}{\sqrt{N}}\sum_{i=1}^{N}\left\{m(\mathbf{X}_i, \bm{\hat{\theta}_g})-\mathbb{E}[m(\mathbf{X}_i, \bm{\theta_g^0})]\right\} &= \frac{1}{\sqrt{N}}\sum_{i=1}^{N}\left\{m(\mathbf{X}_i, \bm{\theta_g^0})-\mathbb{E}[m(\mathbf{X}_i, \bm{\theta_g^0})]\right\}+\mathbf{J}_{\bm{g}}^\mathbf{0}\cdot \sqrt{N}(\bm{\hat{\theta}_g}-\bm{\theta_g^0}) + o_p(1) 
				\end{split}
				\end{equation*}}
			Then, using the asymptotic results from section \ref{strongid}, where we posit that the conditional feature of interest is correctly specified, we have
			{\small \begin{equation*}
				\begin{split}
				\sqrt{N}\left(\bm{\hat{\theta}_1}-\bm{\theta_1^0}\right) = -\mathbf{H}_{\bm{1}}^{-1}\left\{\frac{1}{\sqrt{N}}\sum_{i=1}^{N}\mathbf{l}_{i1}\right\}+o_p(1) \\
				\sqrt{N}\left(\bm{\hat{\theta}_0}-\bm{\theta_0^0}\right) = -\mathbf{H}_{\bm{0}}^{-1}\left\{\frac{1}{\sqrt{N}}\sum_{i=1}^{N}\mathbf{l}_{i0}\right\}+o_p(1)
				\end{split}
				\end{equation*}}
			Therefore, 
			{\small \begin{equation*}
				\begin{split}
				\sqrt{N}\left(\hat{\Delta}_{\text{ate}}-\Delta_\text{ate}\right)= &\frac{1}{\sqrt{N}}\sum_{i=1}^{N}\Bigg(\{m(\mathbf{X}_i, \bm{\theta_1^0})-m(\mathbf{X}_i, \bm{\theta_0^0})-\Delta_{\text{ate}}\}-\mathbf{J_1^0}\cdot \mathbf{H_1^{-1}}\mathbf{l}_{i1}+\mathbf{J_0^0}\cdot \mathbf{H_0^{-1}}\mathbf{l}_{i0}\Bigg)  + o_p(1)
				\end{split}
				\end{equation*}}
			We may rewrite the above using the influence function representation as
			\begin{equation*}
			\begin{split}
			&\sqrt{N}\left(\hat{\Delta}_{\text{ate}}-\Delta_{\text{ate}}\right)= \frac{1}{\sqrt{N}}\sum_{i=1}^{N}\psi(\mathbf{X}_i)+o_p(1) \text{ where } \mathbb{E}\left[\psi(\mathbf{X}_i)\right]= 0 
			\end{split}
			\end{equation*}
			Then, provided that $\mathbb{E}\left[\psi(\mathbf{X}_i)\psi(\mathbf{X}_i)^\prime\right]$ exists,
			{\small\begin{equation*}
				\begin{split}
				\mathrm{Avar}\left[\sqrt{N}\big(\hat{\Delta}_{\text{ate}}-\Delta_{\text{ate}}\big)\right] = \mathbb{E}\left[\left(m(\mathbf{X}_i, \bm{\theta_1^0})-m(\mathbf{X}_i,\bm{\theta_0^0})\right)-\Delta_{\text{ate}}\right]^2+\mathbf{J_1^0}\cdot \mathbf{V_1}\cdot\mathbf{J_1^0}^\prime
				+\ \mathbf{J_0^0}\cdot \mathbf{V_0}\cdot\mathbf{J_0^0}^\prime
				\end{split}
				\end{equation*}}
			\hspace{-0.3em}Note that the covariance term involving {\small $\mathbf{l}_{ig}$} is zero since they denote scores for the treatment and control group problems. The covariance terms involving 
			{\small $\left\{m(\mathbf{X}_i,\bm{\theta_1^0})-m(\mathbf{X}_i,\bm{\theta_0^0})-\Delta_{\text{ate}}\right\}$} and {\small $\mathbf{l}_{ig}$} will also be zero. This is because  {\small $\bm{\theta_g^0}$} solves the conditional problem. However, using that fact that $\mathbb{E}[\mathbf{h}(Y_i(g),\mathbf{X}_i,\bm{\theta_g^0})|\mathbf{X}_i] = \mathbf{0}$ along with LIE, those covariance terms can be shown to be zero.
			
			\paragraph{Misspecified mean model}
			
			In the case of a misspecified mean model, we still have 
			{\small \begin{equation*}
				\begin{split}
				\frac{1}{\sqrt{N}}\sum_{i=1}^{N}\left\{m(\mathbf{X}_i, \bm{\hat{\theta}_g})-\mathbb{E}\left(m(\mathbf{X}_i, \bm{\theta_g^0})\right)\right\} = &\frac{1}{\sqrt{N}}\sum_{i=1}^{N}\left\{m(\mathbf{X}_i, \bm{\theta_g^0})-\mathbb{E}[m(\mathbf{X}_i, \bm{\theta_g^0})]\right\}+\mathbf{J_g^0}\cdot \\
				&\sqrt{N}(\bm{\hat{\theta}_g}-\bm{\theta_g^0}) + o_p(1) 
				\end{split}
				\end{equation*}}
			Now using results from section \ref{theory}
			{\small \begin{equation*}
				\begin{split}
				\sqrt{N}\left(\bm{\hat{\theta}_1}-\bm{\theta_1^0}\right) &= -\mathbf{H}_{\bm{1}}^{-1}\frac{1}{\sqrt{N}}\sum_{i=1}^{N}\left\{\mathbf{l}_{i1} - \mathbb{E}\left(\mathbf{l}_{i1}\mathbf{b}_i^{\prime}\right)\mathbb{E}\left(\mathbf{b}_i\mathbf{b}_i^\prime\right)^{-1}\mathbf{b}_i - \mathbb{E}(\mathbf{l}_{i1}\mathbf{d}_i^{\prime})\mathbb{E}(\mathbf{d}_i\mathbf{d}_i^{\prime})^{-1}\mathbf{d}_i\right\}+o_p(1) \\
				& = -\mathbf{H}_{\bm{1}}^{-1}\frac{1}{\sqrt{N}}\sum_{i=1}^{N}\mathbf{u}_{i1}+o_p(1) \\
				\sqrt{N}\left(\bm{\hat{\theta}_0}-\bm{\theta_0^0}\right) &= -\mathbf{H}_{\bm{0}}^{-1}\frac{1}{\sqrt{N}}\sum_{i=1}^{N}\left\{\mathbf{l}_{i0} - \mathbb{E}\left(\mathbf{l}_{i0}\mathbf{b}_i^{\prime}\right)\mathbb{E}\left(\mathbf{b}_i\mathbf{b}_i^\prime\right)^{-1}\mathbf{b}_i - \mathbb{E}(\mathbf{l}_{i0}\mathbf{d}_i^{\prime})\mathbb{E}(\mathbf{d}_i\mathbf{d}_i^{\prime})^{-1}\mathbf{d}_i\right\}+o_p(1) \\
				&=-\mathbf{H}_{\bm{0}}^{-1}\frac{1}{\sqrt{N}}\sum_{i=1}^{N}\mathbf{u}_{i0}+o_p(1) 
				\end{split}
				\end{equation*}}
			Then, 
			{\small \begin{equation*}
				\begin{split}
				\sqrt{N}\left(\hat{\Delta}_{\text{ate}}-\Delta_{\text{ate}}\right)= &\frac{1}{\sqrt{N}}\sum_{i=1}^{N}\Bigg(\left\{m(\mathbf{X}_i, \bm{\theta_1^0})-m(\mathbf{X}_i, \bm{\theta_0^0})-\Delta_{\text{ate}}\right\}-\mathbf{J_1^0}\cdot \mathbf{H_1^{-1}}\mathbf{u}_{i\mathbf{1}}+\mathbf{J_0^0}\cdot \mathbf{H_0^{-1}}\mathbf{u}_{i\mathbf{0}}\Bigg)  + o_p(1) \\
				=& \frac{1}{\sqrt{N}}\sum_{i=1}^{N}\psi(\mathbf{X}_i)+o_p(1)
				\end{split}
				\end{equation*}}
			Then, 
			{\small \begin{equation*}
				\begin{split}
				\mathrm{Avar}\left[\sqrt{N}\left(\hat{\Delta}_{\text{ate}}-\Delta_{\text{ate}}\right)\right] & = \mathbb{E}\left[\left(m(\mathbf{X}_i, \bm{\theta_1^0})-m(\mathbf{X}_i,\bm{\theta_0^0})\right)-\Delta_{\text{ate}}\right]^2+\mathbf{J_1^0}\cdot \mathbf{V_1}\cdot\mathbf{J_1^0}^\prime + \mathbf{J_0^0}\cdot \mathbf{V_0}\cdot \mathbf{J_0^0}^\prime \\
				&-2\mathbb{E}\left[\left\{m(\mathbf{X}_i, \bm{\theta_1^0})-m(\mathbf{X}_i,\bm{\theta_0^0})-\Delta_{\text{ate}}\right\}\mathbf{u}_{i1}^\prime\right]\mathbf{H_1^{-1}}\mathbf{J_1^0}^\prime \\
				&+2\mathbb{E}\left[\left\{m(\mathbf{X}_i, \bm{\theta_1^0})-m(\mathbf{X}_i,\bm{\theta_0^0})-\Delta_{\text{ate}}\right\}\mathbf{u}_{i0}^\prime\right]\mathbf{H_0^{-1}}\mathbf{J_0^0}^\prime
				\end{split}
				\end{equation*}}
	\end{proof}
	
	\subsection{Practical advice for obtaining doubly weighted ATE estimates}
	An easy way to obtain the doubly weighted estimates, $\bm{\hat{\theta}_g}$, for estimating ATE, is to combine the treatment and control group problems into a one-step GMM procedure. Essentially, this means that one would stack the moment conditions from the first and second steps, which can then be solved jointly via GMM. Since there are no over-identifying restrictions in the doubly weighted framework, one-step estimation of $\bm{\theta_g^0}$ is equivalent to two-step estimation. Then, suppressing explicit dependence on data, 
	\begin{spacing}{0.5}
		\begin{align*}
		\bar{\mathbf{m}}(\bm{\theta_0, \theta_1, \gamma, \delta})= 
		\frac{1}{N}\sum_{i=1}^{N}\mathbf{m}_i(\bm{\theta_0, \theta_1, \gamma, \delta}) = 
		N^{-1} \begin{pmatrix}
		\frac{N}{N_0}\cdot \sum_{i=1}^{N}\mathbf{m}_{i\mathbf{0}}(\bm{\theta_0, \gamma, \delta}) \\
		\frac{N}{N_1}\cdot\sum_{i=1}^{N}\mathbf{m}_{i\mathbf{1}}(\bm{\theta_1, \gamma, \delta}) \\
		\sum_{i=1}^{N}\mathbf{m}_{i\mathbf{2}}(\bm{\gamma}) \\
		\sum_{i=1}^{N}\mathbf{m}_{i\mathbf{3}}(\bm{\delta}) 
		\end{pmatrix}
		\end{align*}
		where,  \begin{align*}      
		\mathbf{m}_{i\mathbf{0}}(\bm{\theta_0, \gamma, \delta}) = \frac{S_i\cdot (1-W_i)}{R(\mathbf{X}_i, W_i, \bm{\hat{\delta}})\cdot (1-G(\mathbf{X}_i, \bm{\hat{\gamma}}))}\cdot \bm{\nabla}_{\bm{\theta_0}}q(Y_i(0), \mathbf{X}_i, \bm{\theta_0})^\prime \\
		\mathbf{m}_{i\mathbf{1}}(\bm{\theta_1, \gamma, \delta}) = \frac{S_i\cdot W_i}{R(\mathbf{X}_i, W_i, \bm{\hat{\delta}})\cdot G(\mathbf{X}_i, \bm{\hat{\gamma}})}\cdot \bm{\nabla}_{\bm{\theta_1}}q(Y_i(1), \mathbf{X}_i, \bm{\theta_1})^\prime \\
		\mathbf{m}_{i\mathbf{2}}(\bm{\gamma}) = \bm{\nabla}_{\bm{\gamma}}G(\mathbf{X}_i, \bm{\gamma})^\prime \cdot \frac{W_i-G(\mathbf{X}_i, \bm{\gamma})}{G(\mathbf{X}_i,\bm{\gamma})\cdot (1-G(\mathbf{X}_i,\bm{\gamma}))} \\
		\mathbf{m}_{i\mathbf{3}}(\bm{\delta}) = \bm{\nabla}_{\bm{\delta}}R(\mathbf{X}_i,W_i,\bm{\delta})^\prime \cdot \frac{S_i-R(\mathbf{X}_i,W_i, \bm{\delta})}{R(\mathbf{X}_i,W_i,\bm{\delta})\cdot (1-R(\mathbf{X}_i,W_i,\bm{\delta}))}\\
		\end{align*}
	\end{spacing}
	
	The example code below uses STATA's \textbf{\texttt{gmm}} command to estimate the doubly weighted ATE estimate \\
	
	\noindent \textbf{Example code using STATA's \texttt{gmm}} \\
	
	\noindent \begin{minipage}{1.1\columnwidth}
		{\small \noindent\texttt{local Rhat="exp({b31}+{b32}*w+{b33}*x1+{b34}*x2)/(1+exp({b31}+{b32}*w+{b33}*x1+{b34}*x2))"}\\
			\texttt{local Ghat="exp({b21}+{b22}*x1+{b23}*x2)/(1+exp({b21}+{b22}*x1+{b23}*x2))"}\\
			\\
			\texttt{gmm ((-2*s*(1-w)/(`Rhat'*(1-`Ghat')))*(y-{b00}-{b01}*x1-{b02}*x2)*(n/nc)) ///} \\
			\texttt{((-2*s*w/(`Rhat'*`Ghat'))*(y-{b10}-{b11}*x1-{b12}*x2)*(n/nt)) ///} \\
			\texttt{(w-exp({b21}+{b22}*x1+{b23}*x2)/(1+exp({b21}+{b22}*x1+{b23}*x2))) ///} \\
			\texttt{(s-exp({b31}+{b32}*w+{b33}*x1+{b34}*x2)/(1+exp({b31}+{b32}*w+{b33}*x1+{b34}*x2))), ///} \\
			\texttt{instruments(1 2 3: x1 x2) instruments(4: w x1 x2) winitial(identity) ///} \\
			\texttt{nocommonesample onestep from(b00 0.1 b01 0.1 b02 0.1 b10 0.1 b11 0.1 b12 ///} \\
			\texttt{0.1 b21 0.1 b22 0.1 b23 0.1 b31 0.1 b32 0.1 b33 0.1 b34 0.1)} }\\
		
		Then using the GMM estimates, one can estimate the average treatment effect as  \\
		
		{\small \noindent \texttt{gen y0hat = \_b[b00: \_cons]+\_b[b01: \_cons]*x1+\_b[b02: \_cons]*x2} \\
			\texttt{gen y1hat = \_b[b10: \_cons]+\_b[b11: \_cons]*x1+\_b[b12: \_cons]*x2} \\
			\texttt{egen ate = mean(y1hat-y0hat)}}\\
		
	 Since I am estimating the two probability models as logits, the last two moments simplify to	
	\end{minipage}
	
	\begin{align*}
	\mathbf{m}_{i\mathbf{2}}(\bm{\gamma}) = \mathbf{X}_i^\prime\cdot (W_i-\Lambda(\mathbf{X}_i\bm{\gamma}))\\
	\mathbf{m}_{i\mathbf{3}}(\bm{\delta}) =  \mathbf{Z}_i^\prime\cdot (S_i-\Lambda(\mathbf{Z}_i\bm{\delta}))
	\end{align*}
	where $\mathbf{Z}_i \equiv (\mathbf{X}_i,W_i)$. Even though this one-step estimation allows us to obtain variance estimates $\widehat{\textbf{V}}_1$ and $\widehat{\textbf{V}}_0$ for $\bm{\hat{\theta}_1}$ and $\bm{\hat{\theta}_0}$ respectively, obtaining analytically correct standard errors for estimated ATE requires additional work. A command that implements the correct standard errors is still in the works. Meanwhile, one can use bootstrapped standard errors, which provide asymptotically correct inference. 
	
	\section{Appendix to CS (2017) Application} \label{csapndx}
	\subsection{Description of National Supported Work Program} 
	
	The NSW was a transitional and subsidized work experience program that was mainly intended to target four sub-populations; ex-offenders, former drug addicts, women on AFDC welfare and high school dropouts.\footnote{The AFDC program is administered and funded by the federal and state governments and is meant to provide financial assistance to needy families. \textit{Source}: US Census Bureau. Beyond the main eligibility criteria that was applied to all four target populations, the AFDC group was subjected to two additional criteria which were, a) no child below 6 years of age and b) on AFDC welfare for at least 30 of the last 36 months.}  The program became operational in 1975 and continued until 1979 at fifteen locations in the United States. In ten of these sites, the program operated as a randomized experiment where individuals who qualified for the training program were randomly assigned to either the treatment or control group.\footnote{Out of the 10 sites, 7 served AFDC women with random assignment at one or more of these sites in operation from Feb 1976-Aug 1977 (CS (2017)).} At the time of enrollment in April 1975, individuals were given a retrospective baseline survey which was then followed by four follow-up interviews conducted at nine month intervals each. The survey data was collected using these baseline and follow-up interviews over a period of four years. The data included measurement on baseline covariates like age, years of education, number of children in 1975, high school dropout status, marital status, two race indicators for black and Hispanic sub-populations and other demographic and socio-economic information. The main outcome of interest was real earnings for the post-training year of 1979.\\

	\subsection{Augmenting the CS sample to account for missing earnings in 1979} 
	
	I obtain the data from CS's supplementary data files in the Journal of Labor Economics where the authors recreate the experimental sample on AFDC women using the raw public use data files maintained by the Inter-University Consortium for Political and Social Research (ICPSR). Then, I use the PSIDcross file provided by CS along with other supplementary data files to add back the individuals whom CS originally dropped from the analysis for not having valid earnings information between 1975-1979. For this, I apply the same filters applied by CS who use them to match their PSID samples to the ones used by \citet{lalonde1986evaluating}. These filters involve keeping all female household heads continuously from 1975-1979 who were between 20 and 55 years of age in 1975 and were not retired in 1975.\footnote{For the additional filters that CS impose, see their supplementary material provided in JLE.} This constitutes the first non-experimental sample that CS use in their analysis, which they call the PSID-1 sample. The second PSID sample, which they label PSID-2 further restricts the PSID-1 sample to include only those women who received AFDC welfare in 1975.\footnote{Even though the two PSID comparison groups are not perfectly representative of women who would have proven eligible for NSW, there is no clear alternative since the PSID data lacks detailed covariate information that would be needed to impose the full eligibility criteria on the PSID sample.} In order to compare my sample with the original sample used by CS, I first apply all the above mentioned filters and create a dummy variable which I call ``cs''. Next, I remove the filter which requires the women to be continuous household heads and instead only impose that filter for 1975 and 1976. The reason this filter is imposed for both years 1975 and 1976 but not for any other years is because in the PSID datasets, the income information in a particular year corresponds to the previous calendar year. This ensures that merging the cross-file with the separate single-year files for 1975 and 1976 guarantee that only those women are included who do not have any missing earnings information for the pre-training year of 1974 and 1975. This is important since pre-training earnings are treated as any other baseline covariate in this paper, on which I do not allow any missing information.
	
	After merging cross year individual file with the single year family files, I then merge this PSID dataset with the NSW dataset using CS's .do files and generate the various sample dummies essentially in the same manner as they do. After this, I further restrict the sample to include only those women who have valid earnings information in 1975, which is the pre-training year for AFDC women. I also drop the cases where the measured age or education is less than zero. In order to make sure that any observations not used by CS only correspond to the ones that have missing post-program earnings, I also drop observations that do not satisfy the CS criteria but have observed earnings in 1979.  \\
	
	\subsection{Treatment and missing outcome probability specifications and sample trimming}
	
	In this application, I estimate three sets of treatment assignment and missing outcomes probability models depending upon which comparison group is used for obtaining the estimates. For the experimental estimates, I use the experimental treatment and control groups to estimate the propensity score model. For the PSID-1 estimates, I consider the NSW experimental observations to be the treatment group and use PSID-1 as the control group. For estimating the PSID-2 propensity score model, I switch to PSID-2 as being the comparison control group. For estimating the missing outcome probability models, I include the treatment indicator depending upon the comparison group as mentioned above. The probability models are estimated as logits and include the following covariates in their specification. For the treatment probability, I include the real earnings in 1974 and 1975 along with an indicator variable for whether the individual had any zero earnings in 1974 and 1975. Beyond these, I also include Age, Age-squared, Education, High school dropout status, the race indicators of black and Hispanic along as well as the number of children in 1975. CS also add some interaction terms in their propensity score specification which I do not. I noticed that allowing for those terms in my specifications drove the final weights for many women in the sample too close to a 0 or 1. For the missing outcomes probability, I include the treatment indicator along with the same covariates. I kept the specifications to be the same for the three sets of probabilities I estimated. However, my regression specifications include the same covariates as CS to allow for some comparison across the analyses. These comparisons should be made with some caution. Except the estimates that use the NSW control group, all other estimates are obtained using samples that are different than the CS samples. 
	
	The final sample used to obtain estimates for the PSID-1 comparison group is trimmed in order to ensure common support for the weights in the treatment and comparison groups. For the PSID-1 group, this meant dropping observations with final weight either less than 0.03 or greater than 0.8. For the PSID-2 sample, this meant dropping observations with final weight that was either less than 0.1 or greater than 0.86. These final weights are the weights that are specified in the regression commands in Stata and are constructed as follows:
	
	\begin{minipage}{\textwidth}
		\texttt{weight = (w/Ghat+(1-w)/(1-Ghat))*(s/Rhat)}
	\end{minipage}
	
\begin{figure}[H]
		\centering
		\caption{Kernel density plots for the composite probability}
		\begin{minipage}{0.50\columnwidth}
			\caption*{\small a) Experimental treatment and control groups}
			\includegraphics[scale=0.34]{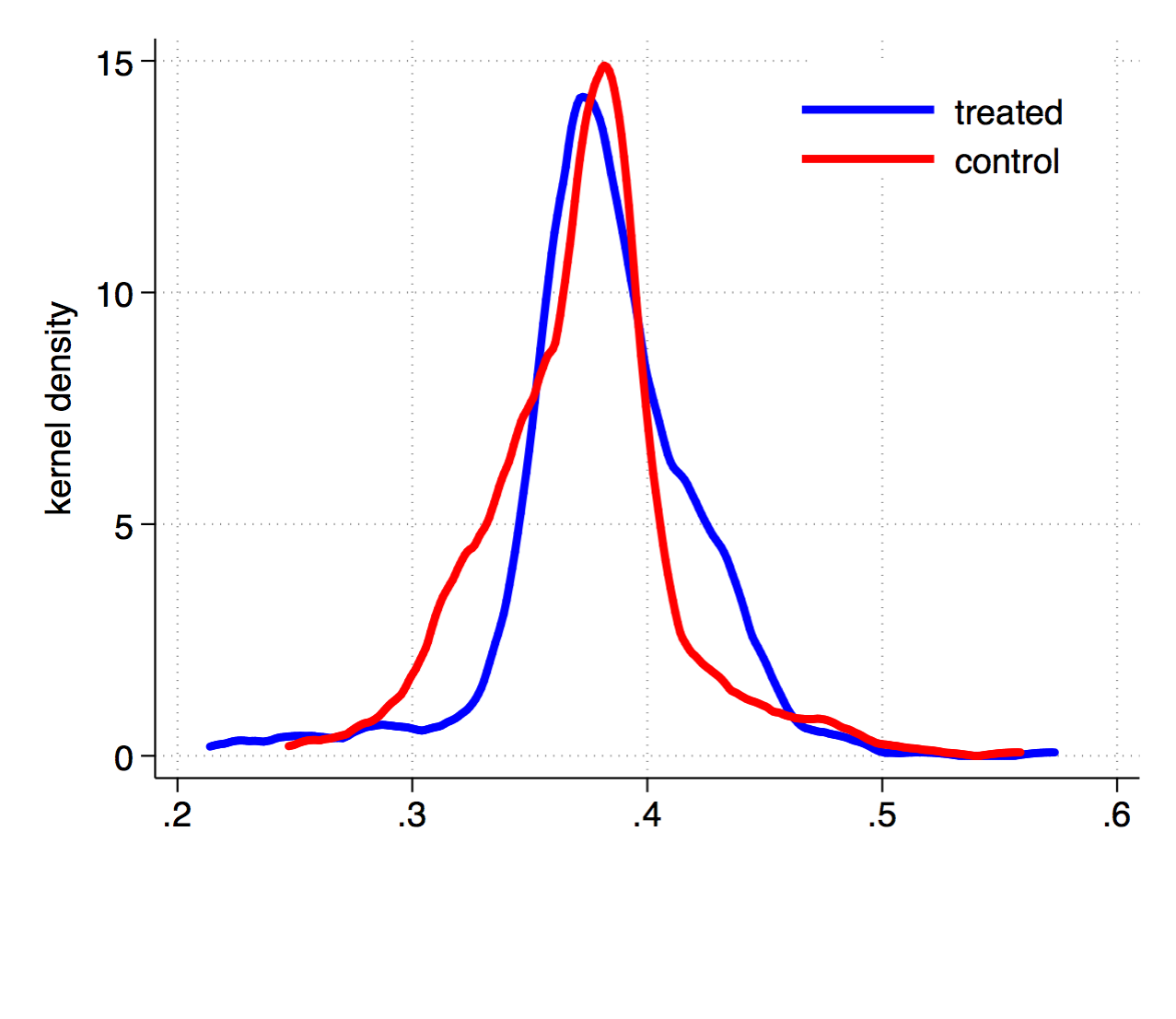}
		\end{minipage}\begin{minipage}{0.50\columnwidth}
			\caption*{\small b) Experimental treatment and PSID-1 group}
			\includegraphics[scale=0.34]{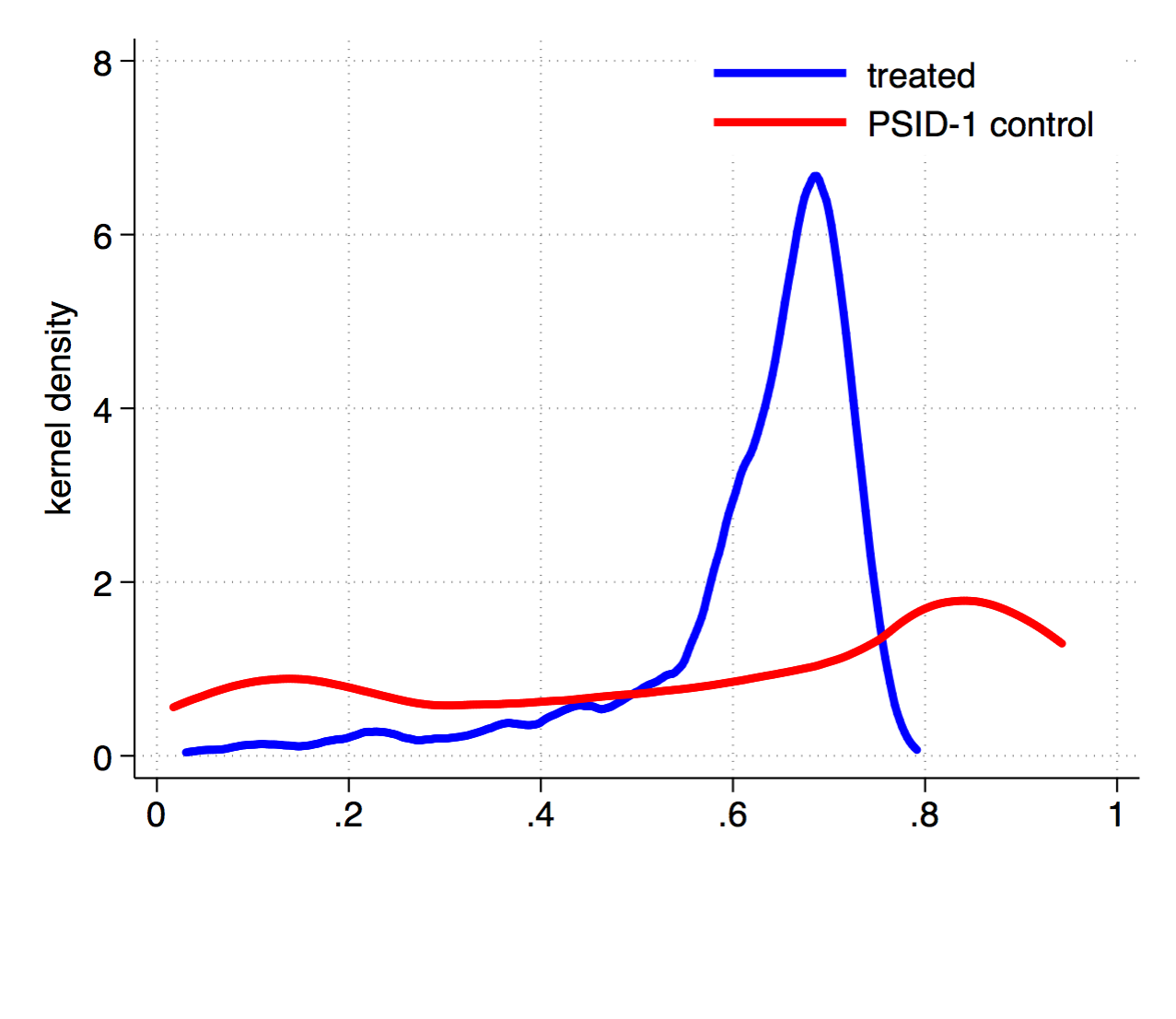}
		\end{minipage} \vspace{-1em}
		\begin{minipage}{0.50\columnwidth}
			\caption*{\small c) Experimental treatment and PSID-2 group}
			\includegraphics[scale=0.34]{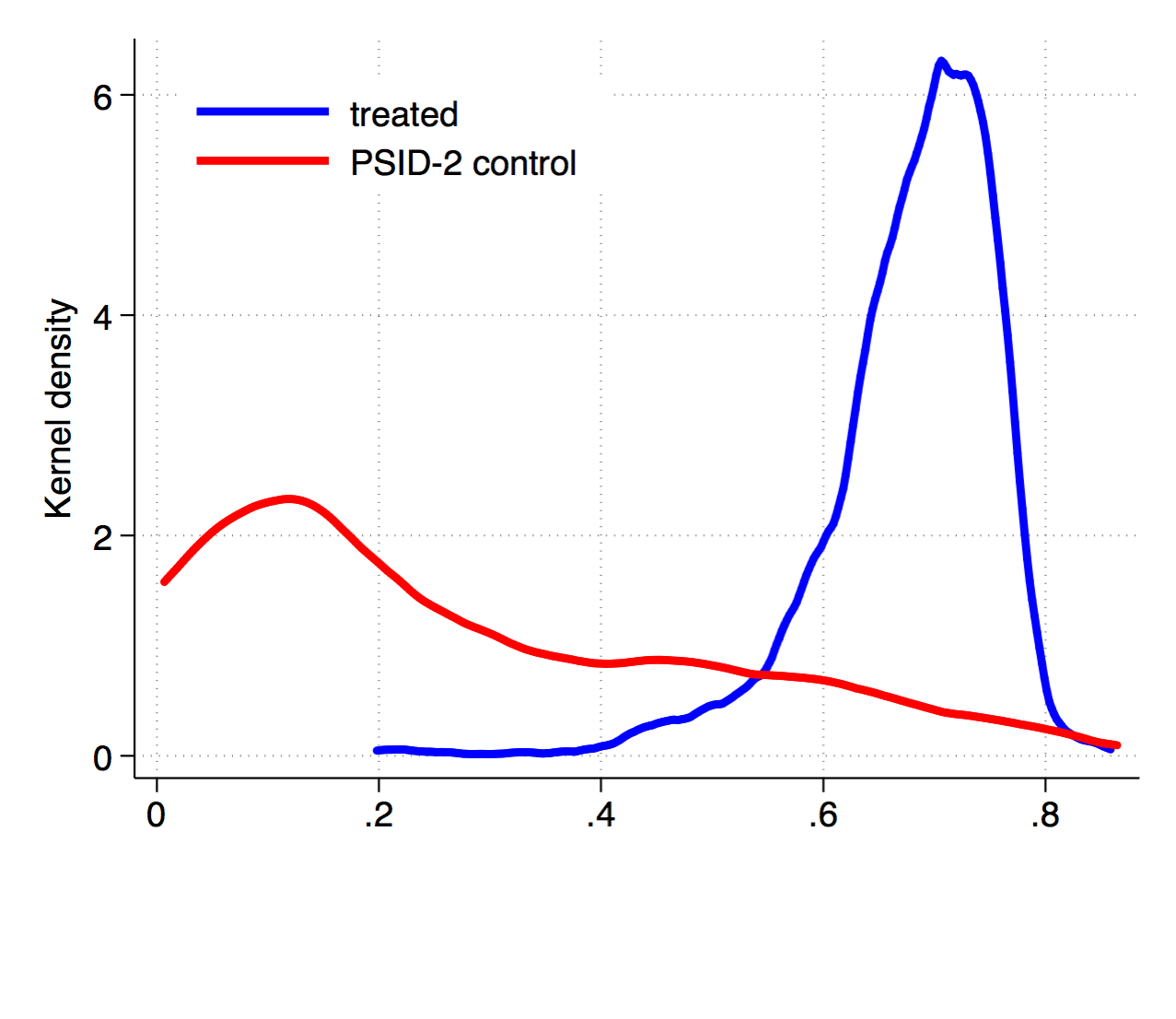}
		\end{minipage}
		\begin{minipage}{\columnwidth}
			{\footnotesize \textit{Notes:} The weights here correspond to the product of the estimated assignment and missing outcomes probabilities.
				Following CS (2017), I exploit the efficiency gain from combining the experimental treatment and control groups for estimating the treatment and missing outcome probability models. For the PSID-1 group, this means using the full experimental group to be the treatment group and the PSID-1 as the control group. Similarly, to construct weights for the PSID-2 group, this means using the full experimental group along with the PSID-2 as the control group. \par}
		\end{minipage}
		\label{double_weight}
	\end{figure}

	The trimming threshold for ps-weighted estimates is kept the same as for computing the doubly weighted estimates since the overlap problem was relatively more severe when using the composite weights than when using propensity scores only.
	The graphs below plot the kernel density for the probabilities \texttt{Rhat*Ghat} for the treatment group and \texttt{Rhat*(1-Ghat)} for the control group. The common support problem due to which the samples were appropriately trimmed can be seen in figure \ref{double_weight}.
	
\begin{figure}[H]
		\centering
		\caption{Kernel density plots for the estimated propensity score}
		\begin{minipage}{0.50\columnwidth}
			\caption*{\small a) Experimental treatment and control groups}
			\includegraphics[scale=0.33]{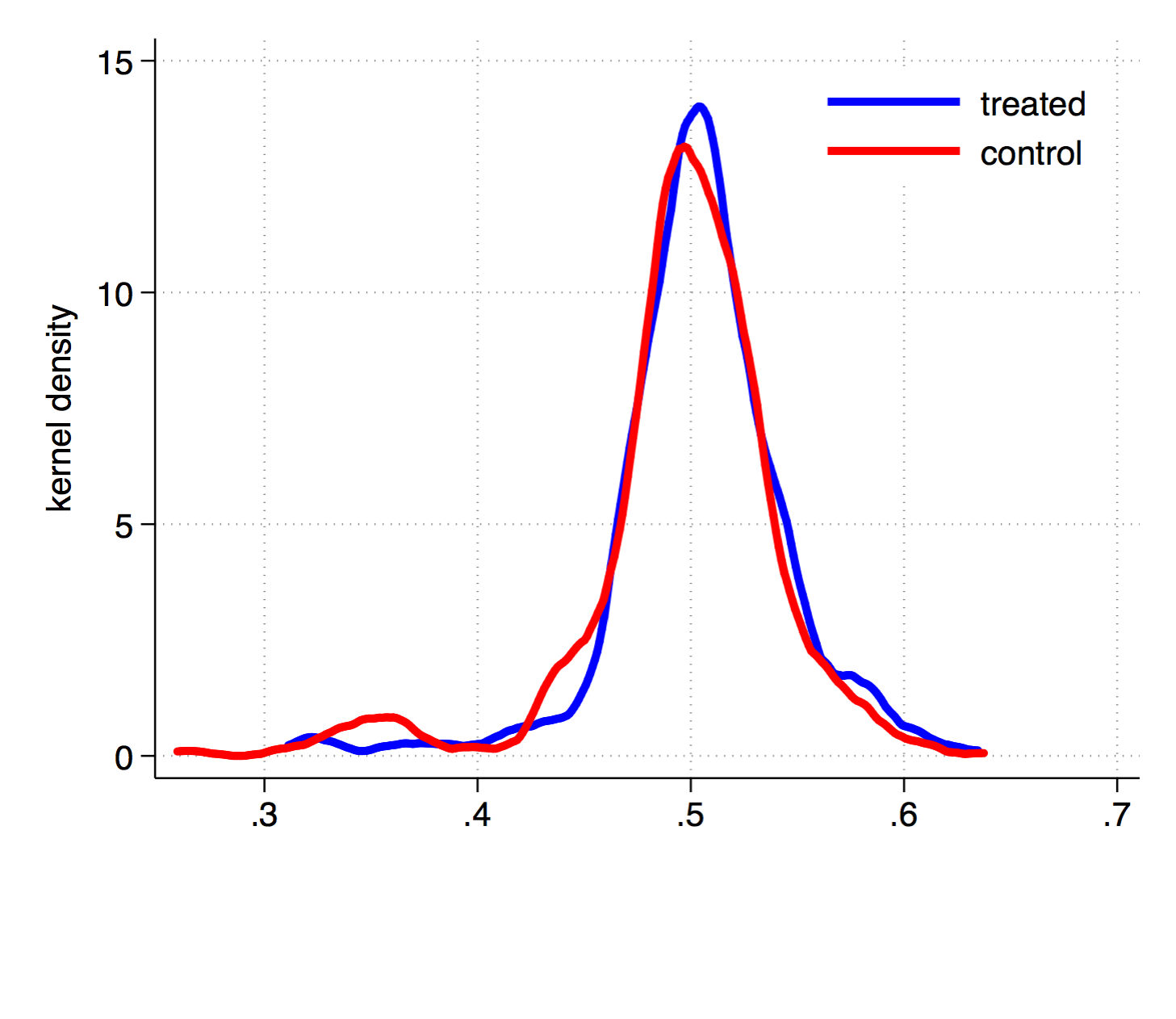}
		\end{minipage}\begin{minipage}{0.50\columnwidth}
			\caption*{\small b) Experimental treatment and PSID-1 group}
			\includegraphics[scale=0.33]{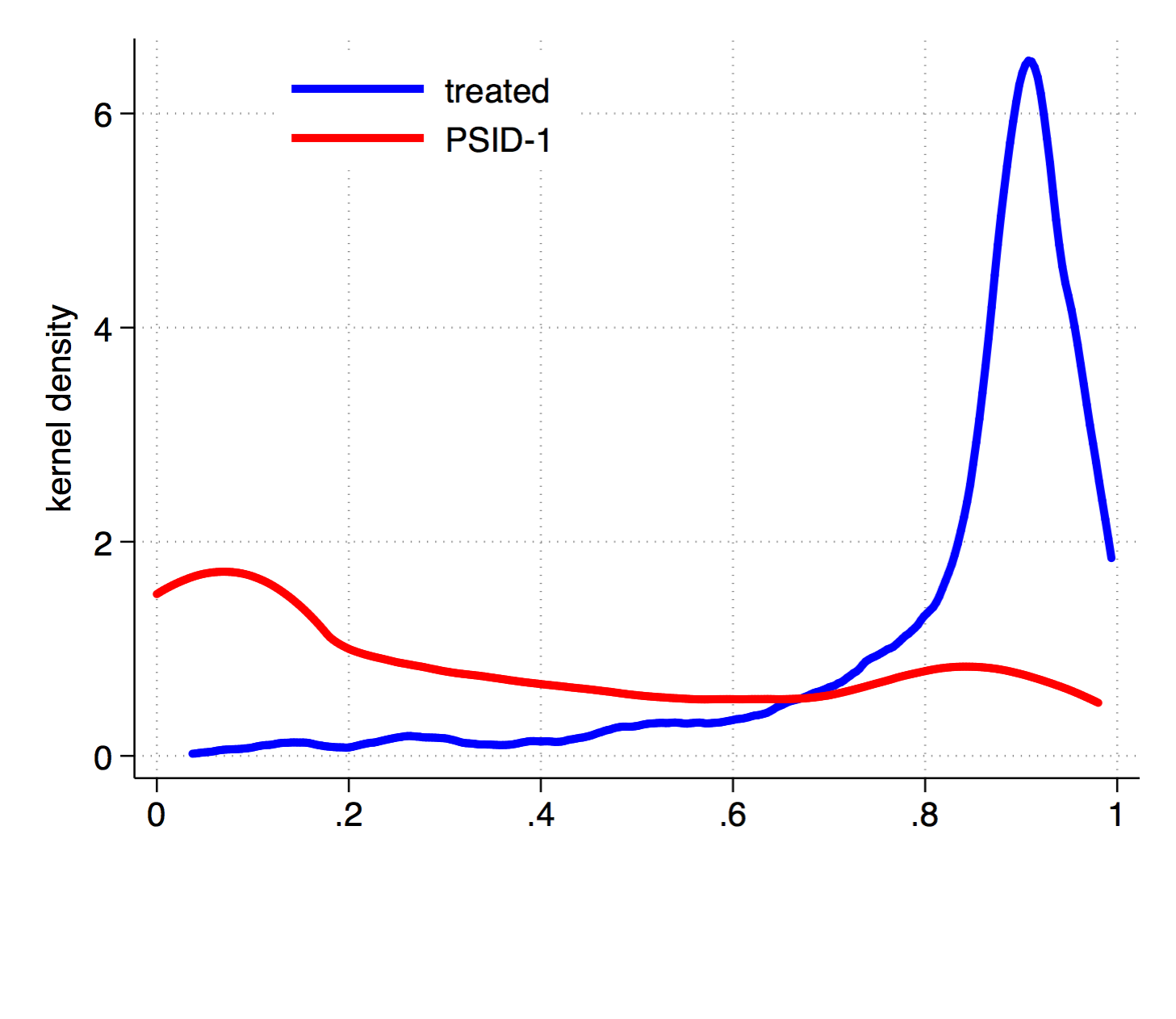}
		\end{minipage} \vspace{-1em}
		\begin{minipage}{0.50\columnwidth}
			\caption*{\small c) Experimental treatment and PSID-2 group}
			\includegraphics[scale=0.33]{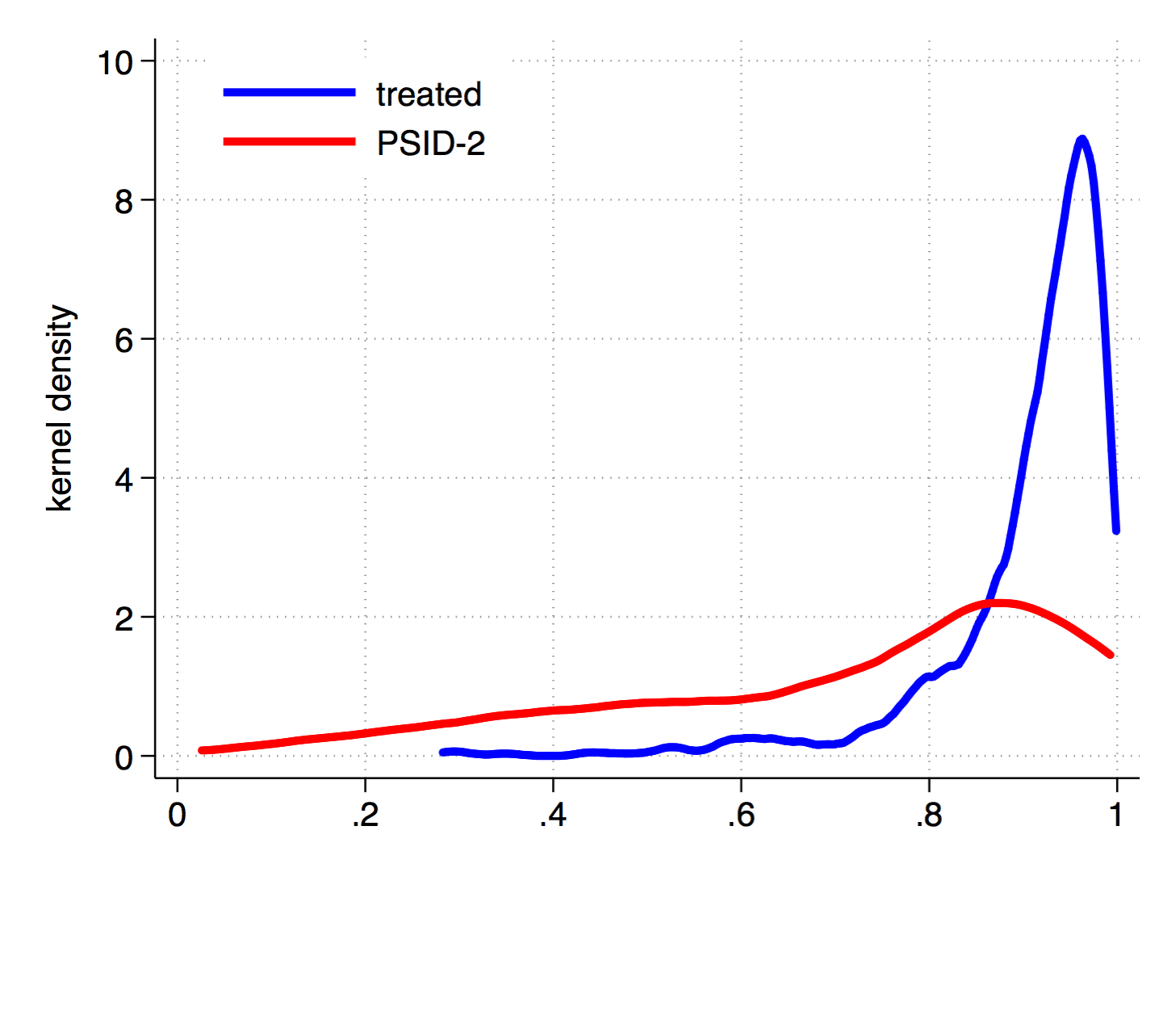}
		\end{minipage}
		\begin{minipage}{\columnwidth}
			{\footnotesize \textit{Notes:} Following CS (2017), I exploit the efficiency gains from combining the experimental treatment and control groups for estimating the propensity scores. For the PSID-1 group, this means using the full experimental group to be the treatment group and the PSID-1 as the control group. Similarly, to construct weights for the PSID-2 group, this means using the full experimental group along with the PSID-2 as the control group. \par}
		\end{minipage}
		\label{ps_weight}
	\end{figure}

\begin{figure}[H]
		\centering
		\caption{Kernel density plots for the estimated missing outcomes probability}
		\begin{minipage}{0.50\columnwidth}
			\caption*{\small a) Experimental treatment and control groups}
			\includegraphics[scale=0.35]{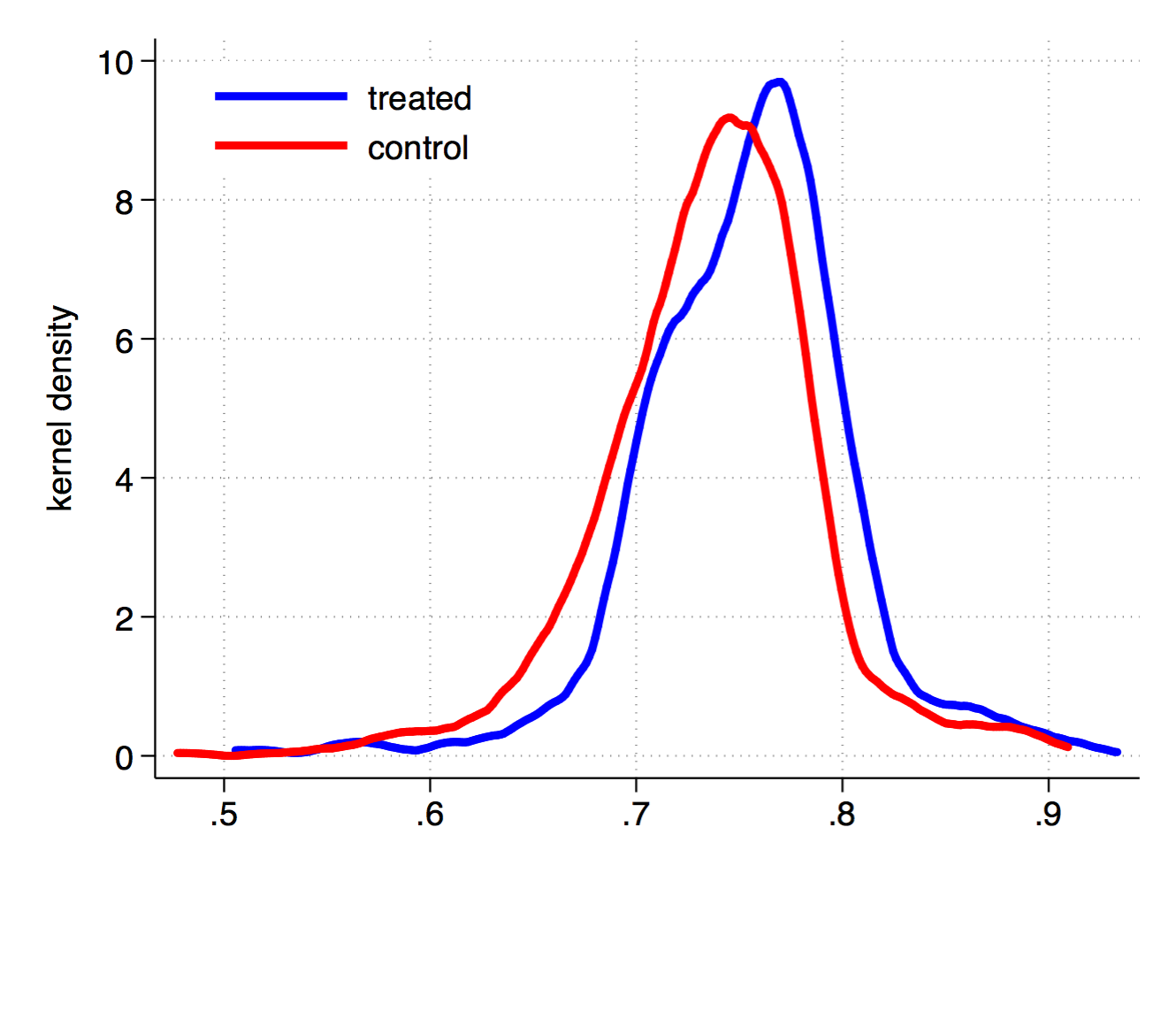}
		\end{minipage}\begin{minipage}{0.50\columnwidth}
			\caption*{\small b) Experimental treatment and PSID-1 group}
			\includegraphics[scale=0.35]{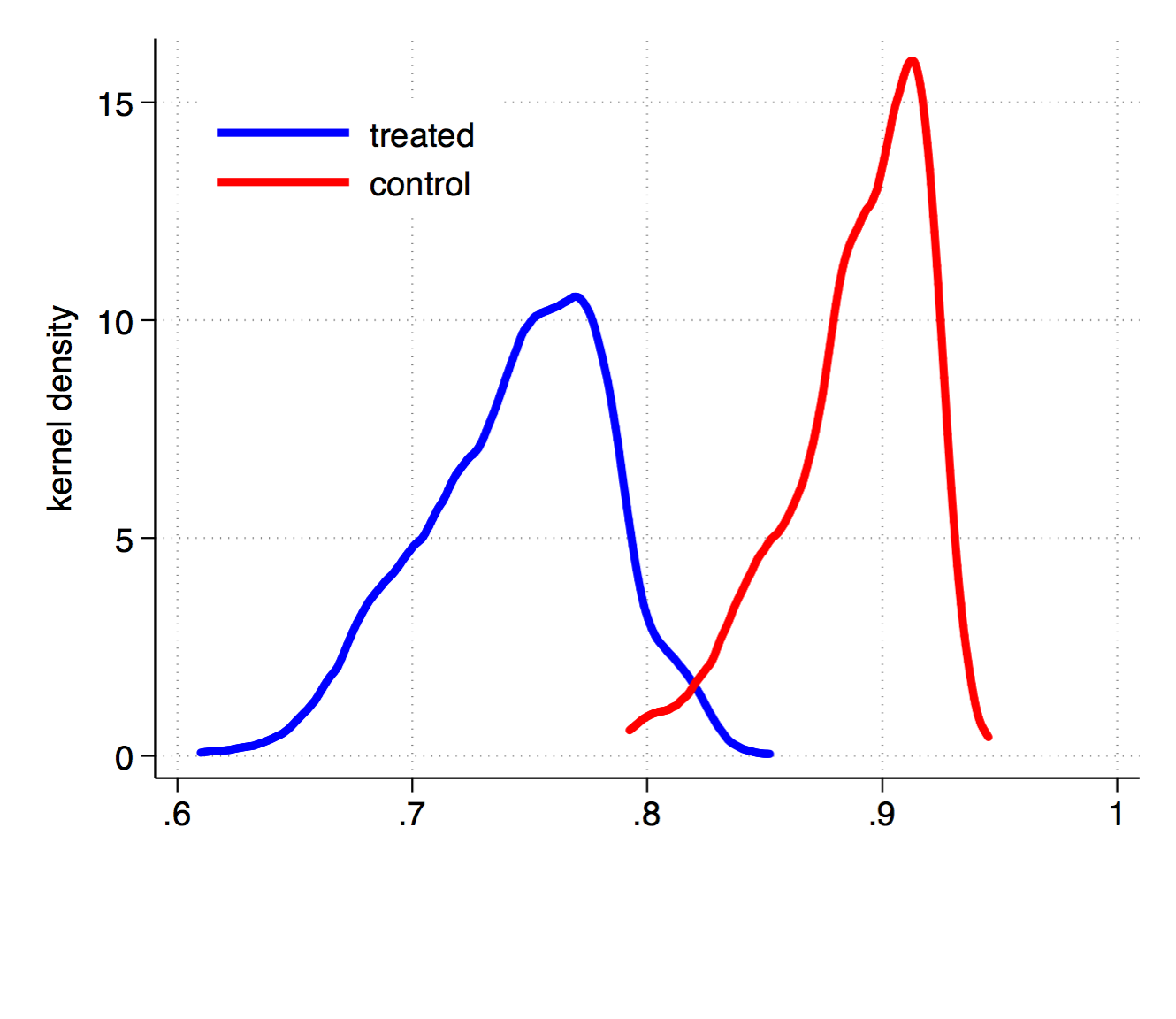}
		\end{minipage} \vspace{-1em}
		\begin{minipage}{0.50\columnwidth}
			\caption*{\small c) Experimental treatment and PSID-2 group}
			\includegraphics[scale=0.35]{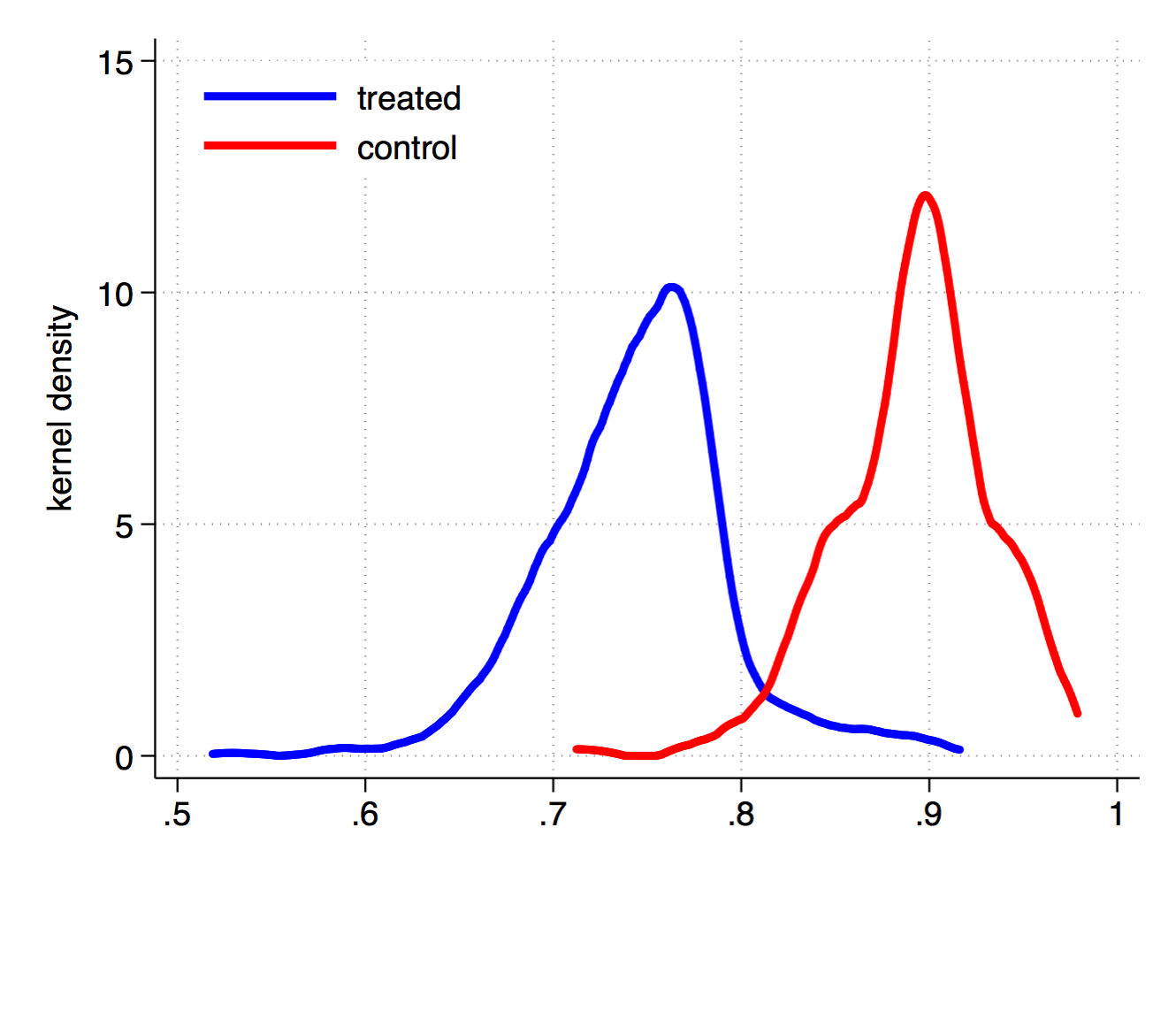}
		\end{minipage}
		\begin{minipage}{\columnwidth}
			{\footnotesize \textit{Notes:} Following CS (2017), I exploit the efficiency gains from combining the experimental treatment and control groups for estimating the missing outcome probability. For the PSID-1 group, this means using the full experimental group to be the treatment group and the PSID-1 as the control group. Similarly, to construct weights for the PSID-2 group, this means using the full experimental group along with the PSID-2 as the control group. \par}
		\end{minipage}
		\label{ps_miss_weight}
	\end{figure}

	Additionally, figures \ref{ps_weight} and \ref{ps_miss_weight} plot the estimated distributions for the propensity score and missing outcomes probability, where panel (a)-(c) display these for the three treatment and comparison group combinations. A couple of points emerge from the estimated graphs. For figure \ref{ps_weight}, panel (a), we see that the treatment and control distributions appear very similar, confirming the strong role of randomization in producing groups that are balanced in terms of covariates. For panel (b), we see that the experimental observations have a relatively high probability of being treated whereas the control group have low probabilities. Note, however, that the common support condition holds quite strongly for the PSID-1 group. In panel (c), while the estimated distribution for the  treated units still has a higher mean, the PSID-2 comparison group distribution is relatively similar than PSID-1 in panel (b). These findings suggest that nonrandom assignment is predicted well by the covariates in the propensity score distributions. The same cannot be said for the estimated missing outcomes probabilities where panel (b) and (c) reveal a strong overlap problem. Moreover, we see that the treated units are less likely to be missing outcomes compared to the comparison groups. 
	
	\section{Proofs} \label{prf}
	\begin{proof}[\textup{\textbf{Proof of Lemma 1}}]
		Let us first consider the argument for $\bm{\theta_1^0}$. By LIE and using the fact that \small{$q(Y, \mathbf{X},\bm{\theta}) = W\cdot q(Y(1), \mathbf{X},\bm{\theta_1})+(1-W)\cdot q(Y(0), \mathbf{X},\bm{\theta_0})$} we can write, 
		\begin{align*}
		\mathbb{E}\left[\omega_1\cdot q\left(Y, \mathbf{X}, \bm{\theta}\right)\right]
		&= \mathbb{E}\left[\mathbb{E}\left(\frac{S}{r(\mathbf{X},W)}\cdot \frac{W}{p(\mathbf{X})}\cdot q\left(Y(1), \mathbf{X}, \bm{\theta_1}\right)\bigg| Y(1),\mathbf{X},W\right) \right] \\
		&=\mathbb{E}\left[\frac{W}{r(\mathbf{X},W)\cdot p(\mathbf{X})}\cdot q\left(Y(1), \mathbf{X}, \bm{\theta_1}\right) \cdot  \mathbb{P}\left(S=1|Y(1),\mathbf{X},W \right)\right]  \\
		&=\mathbb{E}\left[\frac{W}{r(\mathbf{X},W)\cdot p(\mathbf{X})}\cdot q\left(Y(1), \mathbf{X}, \bm{\theta_1}\right) \cdot  \mathbb{P}\left(S=1|\mathbf{X},W \right)\right]  \\
		&=\mathbb{E}\left[\frac{W}{p(\mathbf{X})}\cdot q\left(Y(1), \mathbf{X}, \bm{\theta_1}\right)\right]
		\end{align*}
		Using another application of LIE along with unconfoundedness, we obtain
		\begin{align*}
		\mathbb{E}\left[\frac{W}{p(\mathbf{X})}\cdot q\left(Y(1), \mathbf{X}, \bm{\theta_1}\right)\right] = \mathbb{E}\left[q(Y(1), \mathbf{X}, \bm{\theta_1})\right]
		\end{align*}
		where the third equality follows from MAR and fourth follows from part ii) of Assumption \ref{as:mar}. The proof for $\bm{\theta_0^0}$ follows analogously.
	\end{proof}
	
	\begin{proof}[\textup{\textbf{Proof of Theorem 1}}]
		It has already been established that \[ \mathbb{E}\left[\omega_g\cdot q(Y,\mathbf{X},\bm{\theta})\right] \equiv \mathbb{E}\left[\omega_g\cdot q(Y(g),\mathbf{X},\bm{\theta_g})\right] = \mathbb{E}\left[q(Y(g),\mathbf{X},\bm{\theta_g})\right]\] for both $g=0,1$.
		By iii) $\omega_{g}(\bm{\gamma}, \bm{\delta})$ is continuous in $\bm{\gamma}$ and $\bm{\delta}$ and is bounded in absolute value by Assumptions \ref{as:correctps} and \ref{as:correctsp}. Moreover, $\omega_{g}(\cdot, \bm{\gamma}, \bm{\delta})q(\cdot, \bm{\theta})$ is continuous with probability one. Then, along with v), DCT, and boundedness of $\omega_{g}(\cdot, \cdot)$ we obtain, 
		{\small \begin{align}
			\underset{\left(\bm{\theta_g}, \bm{\gamma}, \bm{\delta}\right)\in\left(\bm{\Theta_g}, \bm{\tilde{\Gamma}}, \bm{\tilde{\Delta}}\right)}{\mathrm{sup}}  \Bigg\lvert \frac{1}{N} \sum_{i=1}^{N}\omega_{ig}(\bm{\gamma}, \bm{\delta})\cdot q(Y_i(g),\mathbf{X}_i,\bm{\theta_g})-\mathbb{E}\left[\omega_g(\bm{\gamma}, \bm{\delta}) \cdot q(Y(g),\mathbf{X},\bm{\theta_g})\right]\Bigg\rvert \overset{p}{\rightarrow} 0
			\end{align}}
		by Lemma 2.4 in \citet{newey1994large}.\footnote{$\bm{\tilde{\Gamma}}$ and $\bm{\tilde{\Delta}}$ are compact neighborhoods around $\bm{\gamma_0}$ and $\bm{\delta_0}$.} Then, by triangle inequality, 
		{\small	\begin{align}
			&\underset{\bm{\theta_g}\in\bm{\Theta_g}}{\mathrm{sup}}  \Bigg\lvert \frac{1}{N} \sum_{i=1}^{N}\widehat{\omega}_{ig}\cdot q(Y_i(g),\mathbf{X}_i,\bm{\theta_g})-\mathbb{E}\left[\omega_g \cdot q(Y(g),\mathbf{X},\bm{\theta_g})\right]\Bigg\rvert\notag \\ \leq 		&\underset{\bm{\theta_g}\in\bm{\Theta_g}}{\mathrm{sup}}  \Bigg\lvert \frac{1}{N} \sum_{i=1}^{N_g}\widehat{\omega}_{ig}\cdot q(Y_i(g),\mathbf{X}_i,\bm{\theta_g})-\mathbb{E}\left[\widehat{\omega}_g \cdot q(Y(g),\mathbf{X},\bm{\theta_g})\right]\Bigg\rvert \\ &+\underset{\bm{\theta_g}\in\bm{\Theta_g}}{\mathrm{sup}}  \Bigg\lvert \mathbb{E}\left[\widehat{\omega}_g \cdot q(Y(g),\mathbf{X},\bm{\theta_g})\right]-\mathbb{E}\left[\omega_g \cdot q(Y(g),\mathbf{X},\bm{\theta_g})\right]\Bigg\rvert
			\end{align} }
		(A.2) is $o_p(1)$ because of (A.1). (A.3) is $o_p(1)$ due to $\bm{\hat{\gamma}}\overset{p}{\rightarrow} \bm{\gamma_0}$, $\bm{\hat{\delta}}\overset{p}{\rightarrow} \bm{\delta_0}$ and uniform continuity of $\mathbb{E}\left[\omega_g\cdot q(Y(g), \mathbf{X},\bm{\delta_g})\right]$ on $\bm{\Theta_g} \times \bm{\tilde{\Gamma}} \times\bm{\tilde{\Delta}}$.
		Then consistency of $\bm{\hat{\theta}_g}$ for $\bm{\theta_g^0}$ follows from Theorem 2.1 of \citet{newey1994large}. 
	\end{proof}
	
	\begin{proof}[\textup{\textbf{Proof of Theorem 2}}]
		Explicit dependence on data is suppressed for notational simplicity. Then expanding $\widehat{\omega}_{ig}$ around $\omega_{ig}$,
		\begin{equation*}
		\widehat{\omega}_{ig} \approx \omega_{ig}-\widetilde{\omega}_{ig}\mathbf{b}_i^\prime(\bm{\tilde{\delta}})\cdot (\bm{\hat{\delta}}-\bm{\delta_0})-\widetilde{\omega}_{ig}\mathbf{d}_i^\prime(\bm{\tilde{\gamma}}) \cdot  (\bm{\hat{\gamma}}-\bm{\gamma_0})
		\end{equation*}
		where $\bm{\tilde{\delta}}$ lies between $\bm{\hat{\delta}}$ and $\bm{\delta_0}$ and $\bm{\tilde{\gamma}}$ lies between $\bm{\hat{\gamma}}$ and $\bm{\gamma_0}$.
		Then, consider
		\begin{align*}
		&N^{-1/2}\sum_{i=1}^{N}\widehat{\omega}_{ig}\cdot \mathbf{h}_{ig} \\
		=&N^{-1/2}\sum_{i=1}^{N}\left\{\omega_{ig}\mathbf{h}_{ig} -\widetilde{\omega}_{ig} \mathbf{h}_{ig}\cdot \mathbf{b}_i^\prime(\bm{\tilde{\delta}})\cdot (\bm{\hat{\delta}}-\bm{\delta_0})-\widetilde{\omega}_{ig} \mathbf{h}_{ig}\cdot \mathbf{d}_i^\prime(\bm{\tilde{\gamma}}) \cdot  (\bm{\hat{\gamma}}-\bm{\gamma_0})\right\} \\
		=&N^{-1/2}\sum_{i=1}^{N}\omega_{ig}\mathbf{h}_{ig}  -N^{-1}\sum_{i=1}^{N}\widetilde{\omega}_{ig} \mathbf{h}_{ig} \mathbf{b}_i^\prime(\bm{\tilde{\delta}})\cdot\sqrt{N}(\bm{\hat{\delta}}-\bm{\delta_0})-N^{-1}\sum_{i=1}^{N}\widetilde{\omega}_{ig} \mathbf{h}_{ig}\mathbf{d}_i^\prime(\bm{\tilde{\gamma}})\cdot
		\sqrt{N}(\bm{\hat{\gamma}}-\bm{\gamma_0})
		\end{align*}
		Now let, $(\bm{\theta_g}^\ast, \bm{\delta}^\ast) =  \underset{\bm{\theta_g}\in\bm{\Theta_g}, \bm{\delta}\in \bm{\Delta}}{\text{arg sup}} \lVert \mathbf{h}(\bm{\theta_g})\cdot \mathbf{b}^\prime(\bm{\delta})\rVert$. Then, 
		{\small \begin{equation}\label{holders}
			(\mathbb{E}\big[\lVert \mathbf{h}(\bm{\theta_g}^\ast)\mathbf{b}^{\prime}(\bm{\delta}^\ast)\rVert\big])^2\leq \mathbb{E}\big[\lVert \mathbf{h}(\bm{\theta_g}^\ast)\rVert^2\big] \mathbb{E}\big[\lVert \mathbf{b}^{\prime}(\bm{\delta}^\ast)\rVert^2\big] \leq \mathbb{E}\bigg[{\small \underset{\bm{\theta_g}\in\bm{\Theta_g}}{\text{sup}}}\lVert \mathbf{h}(\bm{\theta_g})\rVert^2\bigg] \mathbb{E}\bigg[{\small \underset{\bm{\theta_g}\in\bm{\Theta_g}}{\text{sup}}}\lVert \mathbf{b}^{\prime}(\bm{\delta})\rVert^2\bigg] <\infty
			\end{equation}} where first inequality holds by cauchy-schwartz, second holds due to the definition of supremums, and third by conditions iv) and vi). Then, 
		\begin{equation*}
		\mathbb{E}\bigg[\underset{\bm{\theta_g}\in\bm{\Theta_g}, \bm{\delta}\in \bm{\Delta}}{\text{sup}}\lVert \mathbf{h}(\bm{\theta_g})\mathbf{b}^{\prime}(\bm{\delta})\rVert\bigg] \leq  \left(\mathbb{E}\bigg[\underset{\bm{\theta_g}\in\bm{\Theta_g}, \bm{\delta}\in \bm{\Delta}}{\text{sup}}\lVert \mathbf{h}(\bm{\theta_g})\mathbf{b}^{\prime}(\bm{\delta})\rVert\bigg]\right)^2 <\infty
		\end{equation*}
		where the first inequality holds trivially and second inequality holds because of (\ref{holders}). An analogous argument may be made for showing {\small $\mathbb{E}\bigg[\underset{\bm{\theta_g}\in\bm{\Theta_g}, \bm{\gamma}\in \bm{\Gamma}}{\text{sup}}\lVert \mathbf{h}(\bm{\theta_g})\mathbf{d}^{\prime}(\bm{\gamma})\rVert\bigg] <\infty$}. Using the fact that $\omega_{g}(\bm{\gamma},\bm{\delta})$ is continuous and bounded along with continuity of $\mathbf{l}(\bm{\theta_g})$ (condition ii)), $\mathbf{b}(\bm{\delta})$, $\mathbf{d}(\bm{\gamma})$ (condition iii) of theorem \ref{theorem:consistency}), we obtain
		\begin{equation} \label{lln}
		\begin{split}
		\frac{1}{N}\sum_{i=1}^{N}\widetilde{\omega}_{ig} \mathbf{h}_{ig}\mathbf{b}_i^\prime(\bm{\tilde{\delta}})  &= \mathbb{E}\left[\omega_{ig} \mathbf{h}_{ig}\mathbf{b}_i^\prime\right]+o_p(1) \\
		\frac{1}{N}\sum_{i=1}^{N}\widetilde{\omega}_{ig} \mathbf{h}_{ig}\mathbf{d}_i^\prime(\bm{\tilde{\gamma}})  &= \mathbb{E}\left[\omega_{ig} \mathbf{h}_{ig}\mathbf{d}_i^\prime\right]+o_p(1) \\
		\end{split}	
		\end{equation} using Lemma 4.3 in \citet{newey1994large} as $\bm{\tilde{\gamma}}\rightarrow_p \bm{\gamma_0}$ and $\bm{\tilde{\delta}} \rightarrow_p \bm{\delta_0}$. Rewriting (7) using influence function representations for $\bm{\hat{\gamma}}$ and $\bm{\hat{\delta}}$ along with (\ref{lln})
		\begin{align}\label{wfoc}
		N^{-1/2}\sum_{i=1}^{N}\widehat{\omega}_{ig}\mathbf{h}_{ig}  &= 
		N^{-1/2}\sum_{i=1}^{N}\left\{\mathbf{l}_{ig} -\mathbb{E}\left[ \mathbf{l}_{ig}\mathbf{b}_i^\prime\right]\cdot \mathbb{E}\left[\mathbf{b}_i\mathbf{b}_i^\prime\right]^{-1}\mathbf{b}_i-\mathbb{E}\left[\mathbf{l}_{ig}\mathbf{d}_i^\prime\right]\cdot \mathbb{E}\left[\mathbf{d}_i\mathbf{d}_i^\prime\right]^{-1}\mathbf{d}_i\right\}+o_p(1) \nonumber \\
		&\equiv N^{-1/2}\sum_{i=1}^{N}\mathbf{u}_{ig}+o_p(1)  \nonumber \\
		& \overset{d}{\rightarrow} N(\mathbf{0}, \mathbf{\Omega_g})
		\end{align}
		where $\mathbf{u}_{ig} \equiv \mathbf{l}_{ig} -\mathbb{E}\left[ \mathbf{l}_{ig}\mathbf{b}_i^\prime\right]\cdot \mathbb{E}\left[\mathbf{b}_i\mathbf{b}_i^\prime\right]^{-1}\mathbf{b}_i-\mathbb{E}\left[\mathbf{l}_{ig}\mathbf{d}_i^\prime\right]\cdot \mathbb{E}\left[\mathbf{d}_i\mathbf{d}_i^\prime\right]^{-1}\mathbf{d}_i$. Since $\mathbb{E}(\mathbf{u}_{ig})=\mathbf{0}$,
		\begin{align*}
		\mathbf{\Omega_g} &= \mathbb{E}\left(\mathbf{l}_{ig}\mathbf{l}_{ig}^{\prime}\right) - \mathbb{E}\left(\mathbf{l}_{ig}\mathbf{b}_i^{\prime}\right)\mathbb{E}\left(\mathbf{b}_i\mathbf{b}_i^{\prime}\right)^{-1}\mathbb{E}\left(\mathbf{b}_i\mathbf{l}_{ig}^{\prime}\right)-\mathbb{E}\left(\mathbf{l}_{ig}\mathbf{d}_i^{\prime}\right)\mathbb{E}\left(\mathbf{d}_i\mathbf{d}_i^{\prime}\right)^{-1}\mathbb{E}\left(\mathbf{d}_i\mathbf{l}_{ig}^{\prime}\right) 
		\end{align*} 
		Next part of the proof uses the theory of empirical processes for obtaining asymptotic normality of the doubly weighted estimator. Using the definition in (\ref{ep}) along with the fact that $\mathbb{E}[\widehat{\omega}_{ig}\mathbf{h}_i(\bm{\theta_g})]\overset{p}{\rightarrow}\mathbb{E}[\omega_{ig}\mathbf{h}_i(\bm{\theta_g})]$ (by continuity of $\omega(\bm{\gamma},\bm{\delta})\mathbf{h}(\bm{\theta_g})$, condition iv) and DCT as $(\bm{\hat{\gamma}}, \bm{\hat{\delta}}) \overset{p}{\rightarrow} (\bm{\gamma_0}, \bm{\delta_0})$), rewrite
		\begin{align} \label{emp}
		\bm{v}_N(\bm{\theta_g})  = \bm{v}_N^\ast(\bm{\theta_g})+ o_p(1) 
		\end{align}	
		where $\bm{v}_N^\ast(\bm{\theta_g}) \equiv \frac{1}{N}\sum_{i=1}^{N}\left\{\widehat{\omega}_{ig}\mathbf{h}_{i}(\bm{\theta_g})-\mathbb{E}\big[\omega_{ig}\mathbf{h}_{i}(\bm{\theta_g})\big]\right\}$. Let
		\begin{equation*}
		\begin{split}
		\bm{\bar{m}}_N(\bm{\theta_g}) &= \frac{1}{N}\sum_{i=1}^{N}\widehat{\omega}_{ig}\mathbf{h}_i(\bm{\theta_g}) \\ \bm{m^\ast}_N(\bm{\theta_g}) &= \mathbb{E}\big[\omega_{ig}\mathbf{h}_i(\bm{\theta_g}) \big]
		\end{split}
		\end{equation*} Then performing element by element mean value expansions of $\bm{m^\ast}_N(\bm{\hat{\theta}_g})$ around $\bm{\theta_g^0}$, we obtain
		\begin{equation*}
		\mathbf{0} = \sqrt{N}\bm{m^\ast}_N(\bm{\theta_g^0}) = \sqrt{N}\bm{m^\ast}_N(\bm{\hat{\theta}_g}) -\nabla_{\bm\theta_g}\bm{m^\ast}_N(\bm{\tilde{\theta}_g})^\prime\cdot \sqrt{N} (\bm{\hat{\theta}_g}-\bm{\theta_g^0}) 
		\end{equation*}
		where $\bm{\tilde{\theta}_g}$ lies between $\bm{\hat{\theta}_g}$ and $\bm{\theta_g^0}$. Since the population first order condition is zero at the truth 
		\begin{equation*}
		\begin{split}
		\mathbf{0} &= \nabla_{\bm{\theta_g}}\mathbb{E}\left[\omega_g\cdot q(Y(g), \mathbf{X}, \bm{\theta_g^0})\right] \\
		&= \mathbb{E}\left[\omega_g\cdot \mathbf{h}(Y(g),\mathbf{X},\bm{\theta_g^0})\right] \equiv \bm{m^\ast}_N(\bm{\theta_g^0})
		\end{split}
		\end{equation*} The second equality follows from dominance condition iv) and application of Lemma 3.6 in \citet{newey1994large}. Then, by the continuity of $\nabla_{\bm\theta_g}\mathbb{E}\big[\omega_{ig}\mathbf{h}_i(\bm{\theta_g})\big]$ (condition vi))
		\begin{equation*}
		\nabla_{\bm\theta_g}\bm{m^\ast}_N(\bm{\tilde{\theta}_g})\overset{p}{\rightarrow} \mathbf{H_g}
		\end{equation*}
		By continuous mapping theorem and condition viii), 
		\begin{equation}\label{tylrexp}
		\sqrt{N} (\bm{\hat{\theta}_g}-\bm{\theta_g^0}) = (\mathbf{H}^{-1}_{\mathbf{g}}+o_p(1))\cdot \sqrt{N}\bm{m^\ast}_N(\bm{\hat{\theta}_g})
		\end{equation}
		Consider,  
		\begin{align*}
		-\sqrt{N}\bm{m^\ast}_N(\bm{\hat{\theta}_g}) &= \bm{v^\ast}_N(\bm{\hat{\theta}_g})-\sqrt{N}\bm{\bar{m}}_N(\bm{\hat{\theta}_g}) \\
		& = \bm{v^\ast}_N(\bm{\hat{\theta}_g})-\bm{v^\ast}_N(\bm{\theta_g^0})+\bm{v^\ast}_N(\bm{\theta_g^0})-\sqrt{N}\bm{\bar{m}}_N(\bm{\hat{\theta}_g}) \\
		& = \bm{v^\ast}_N(\bm{\theta_g^0})+o_p(1)
		\end{align*}
		since $\bm{v^\ast}_N(\bm{\hat{\theta}_g})-\bm{v^\ast}_N(\bm{\theta_g^0})= o_p(1)$ by asymptotic equivalence in (\ref{emp}) and stochastic equicontinuity by condition ix). Moreover, $\sqrt{N}\bm{\bar{m}}_N(\bm{\hat{\theta}_g})=o_p(1)$ by condition iii). Therefore, 
		\begin{equation*}
		\bm{v^\ast}_N(\bm{\theta_g^0}) = \frac{1}{N}\sum_{i=1}^{N}\widehat{\omega}_{ig}\mathbf{h}_{ig}\overset{d}{\rightarrow} N(\mathbf{0}, \mathbf{\Omega_g})  
		\end{equation*}
		by (\ref{wfoc}). Then using (\ref{tylrexp}) along with slutsky's theorem, $\sqrt{N}(\bm{\hat{\theta}_g}-\bm{\theta_g^0})\overset{d}{\rightarrow} N\left(\bm{0}, \mathbf{H}_{\mathbf{g}}^{-1}\mathbf{\Omega_g}\mathbf{H}_{\mathbf{g}}^{-1}\right)$.
	\end{proof}
	
	\begin{proof}[\textup{\textbf{Proof of Corollary 1}}]
		Consider, 
		\begin{align*}
		\bm{\Sigma_g}-\bm{\Omega_g}&= \mathbb{E}\big(\mathbf{l}_{ig}\mathbf{l}_{ig}^{\prime}\big)  - \{\mathbb{E}\big(\mathbf{l}_{ig}\mathbf{l}_{ig}^{\prime}\big) - \mathbb{E}\left(\mathbf{l}_{ig}\mathbf{b}_i^{\prime}\right)\mathbb{E}\left(\mathbf{b}_i\mathbf{b}_i^\prime\right)^{-1}\mathbb{E}(\mathbf{b}_i\mathbf{l}_{ig}^{\prime})-\mathbb{E}\left(\mathbf{l}_{ig}\mathbf{d}_i^{\prime}\right)\mathbb{E}\left(\mathbf{d}_i\mathbf{d}_i^{\prime}\right)^{-1}\mathbb{E}(\mathbf{d}_i\mathbf{l}_{ig}^{\prime}) \} \\
		&=\mathbb{E}\left(\mathbf{l}_{ig}\mathbf{b}_i^\prime\right)\mathbb{E}\left(\mathbf{b}_i\mathbf{b}_i^\prime\right)^{-1}\mathbb{E}(\mathbf{b}_i\mathbf{l}_{ig}^{\prime})+\mathbb{E}\left(\mathbf{l}_{ig}\mathbf{d}_i^{\prime}\right)\mathbb{E}\left(\mathbf{d}_i\mathbf{d}_i^{\prime}\right)^{-1}\mathbb{E}(\mathbf{d}_i\mathbf{l}_{ig}^{\prime})
		\end{align*}
		since each component matrix in the above expression is positive semi-definite, therefore the sum of the two matrices is also positive semi-definite. 
	\end{proof}
	
	\begin{proof}[\textup{\textbf{Proof of Theorem 3}}]
		It has already been established that $\bm{\theta_g^0}$ solves \[ \mathbb{E}\left[\omega_g^\ast\cdot q(Y(g),\mathbf{X},\bm{\theta_g})\right]\] 
		The proof of uniform convergence follows similar to the proof of theorem \ref{theorem:consistency} where we replace $\omega_g$ by $\omega_g^\ast$. Then, consistency of $\bm{\hat{\theta}_g}$ for $\bm{\theta_g^0}$ follows from Theorem 2.1 in \citet{newey1994large}.
	\end{proof}
	
	\begin{proof}[\textup{\textbf{Proof of Theorem 4}}] 
		The proof follows in the manner of Theorem \ref{theorem:normality} where we replace $\omega_g$ by $\omega_g^\ast$. Also, $\bm{\Omega_g}$ now denotes the variance of the score of the objective function, $\mathbf{l}_{ig}$, without the first stage adjustment for the estimated weights. This is because, $\mathbb{E}(\mathbf{l}_{ig}\mathbf{b}_i^\prime) = \mathbb{E}(\mathbf{l}_{ig}\mathbf{d}_i^\prime) = \mathbf{0}$ because the conditional score of $\mathbf{l}_{ig}$, $\mathbb{E}[\mathbf{h}(Y(g),\mathbf{X},\bm{\theta_g^0})|\mathbf{X}] = \mathbf{0}$ due to strong identification of $\bm{\theta_g^0}$.
	\end{proof}
	
	\begin{proof}[\textup{\textbf{Proof of corollary 2}}] This proof follows from the proof of theorem \ref{theorem:normality_2}, and the asymptotic variance of the estimator that uses known weights which is 
		\[\mathrm{Avar}\left[\sqrt{N}\big(\bm{\tilde{\theta}_g-\bm{\theta_g^0}}\big)\right] = \mathbf{H}^{-1}_\mathbf{g}\mathbf{\Omega_g} \mathbf{H}^{-1}_\mathbf{g}\]
		where $\mathbf{\Omega_g} = \mathbb{E}\left(\mathbf{l}_{ig}\mathbf{l}_{ig}^\prime\right)$. The result follows immediately.
	\end{proof}
	
	\begin{proof}[\textup{\textbf{Proof of Corollary 3} (Efficiency gain with unweighted estimator under GCIME)}]
		Using two applications of LIE and invoking MAR and unconfoundedness, I can rewrite
		{\small\begin{align*}	
			\mathbb{E}\left[\frac{S_i\cdot W_i}{R(\mathbf{X}_i, W_i, \bm{\delta^\ast})\cdot G(\mathbf{X}_i, \bm{\gamma^\ast})}\cdot q(Y_i(1), \mathbf{X}_i, \bm{\theta_1^0})\right] 
			=\mathbb{E}\left[\frac{r(\mathbf{X}_i, 1)}{R(\mathbf{X}_i, 1, \bm{\delta^\ast})}\cdot \frac{p(\mathbf{X}_i)}{G(\mathbf{X}_i,\bm{\gamma^\ast})}\cdot q(Y_i(1), \mathbf{X}_i, \bm{\theta_1^0})\right] 
			\end{align*}}
		Using another application of LIE, I can rewrite the above as
		\begin{align*}
		=\mathbb{E}&\left[\frac{r(\mathbf{X}_i, 1)}{R(\mathbf{X}_i, 1, \bm{\delta^\ast})}\cdot \frac{p(\mathbf{X}_i)}{G(\mathbf{X}_i,\bm{\gamma^\ast})}\cdot \mathbb{E}\left\{q(Y_i(1), \mathbf{X}_i, \bm{\theta_1^0})|\mathbf{X}_i\right\}\right] 
		\end{align*}
		Then, 
		\begin{align*}
		\mathbf{H_1}&=\mathbb{E}\left[\frac{r(\mathbf{X}_i, 1)}{R(\mathbf{X}_i, 1, \bm{\delta^\ast})}\cdot \frac{p(\mathbf{X}_i)}{G(\mathbf{X}_i,\bm{\gamma^\ast})}\cdot \nabla_{\bm{\theta_1}}\mathbb{E}\left\{\mathbf{h}(Y_i(1), \mathbf{X}_i, \bm{\theta_1^0})|\mathbf{X}_i\right\}\right] \\
		&=\mathbb{E}\left[\frac{r(\mathbf{X}_i, 1)}{R(\mathbf{X}_i, 1, \bm{\delta^\ast})}\cdot \frac{p(\mathbf{X}_i)}{G(\mathbf{X}_i,\bm{\gamma^\ast})}\cdot \mathbf{A}(\mathbf{X}_i, \bm{\theta_1^0})\right]
		\end{align*}
		Similarly, I use LIE to express $\bm{\Omega_1}$ as
		\begin{align*}
		\bm{\Omega_1} &= \mathbb{E}\bigg[\frac{r(\mathbf{X}_i, 1)}{R^2(\mathbf{X}_i, 1, \bm{\delta^\ast})}\cdot \frac{p(\mathbf{X}_i)}{G^2(\mathbf{X}_i,\bm{\gamma^\ast})}\cdot \mathbb{E}\big\{\mathbf{h}(Y_i(1), \mathbf{X}_i, \bm{\theta_1^0})\mathbf{h}(Y_i(1), \mathbf{X}_i, \bm{\theta_1^0})^\prime\big|\mathbf{X}_i\big\}\bigg] \\
		&= \sigma_{01}^2\cdot \mathbb{E}\left[\frac{r(\mathbf{X}_i, 1)}{R^2(\mathbf{X}_i, 1, \bm{\delta^\ast})}\cdot \frac{p(\mathbf{X}_i)}{G^2(\mathbf{X}_i,\bm{\gamma^\ast})}\cdot \mathbf{A}(\mathbf{X}_i, \bm{\theta_1^0})\right]
		\end{align*}
		
		For the unweighted estimator, the variance simplifies, and this happens precisely due to the GCIME. To see this, consider $\mathbf{H_1^u}$. Then using LIE, I can rewrite  
		\begin{align*}	
		\mathbf{H_1^u}&= \mathbb{E}\left[r(\mathbf{X}_i, 1)\cdot p(\mathbf{X}_i)\cdot \nabla_{\bm{\theta_1}}\mathbb{E}\left\{\mathbf{h}(Y_i(1), \mathbf{X}_i, \bm{\theta_1^0})|\mathbf{X}_i\right\}\right]   \\
		& =\mathbb{E}\left[r(\mathbf{X}_i, 1)\cdot p(\mathbf{X}_i)\cdot  \mathbf{A}(\mathbf{X}_i, \bm{\theta_1^0})\right]  
		\end{align*}
		and similarly we can rewrite $\bm{\Omega}_{\mathbf{1}}^{\mathbf{u}}$ using LIE as
		\begin{align*}
		\mathbf{\Omega}_{\mathbf{1}}^{\mathbf{u}}& = \mathbb{E}\left[r(\mathbf{X}_i, 1)\cdot p(\mathbf{X}_i)\cdot \mathbb{E}\left\{\mathbf{h}(Y_i(1), \mathbf{X}_i, \bm{\theta_1^0})\mathbf{h}(Y_i(1), \mathbf{X}_i, \bm{\theta_1^0})^\prime|\mathbf{X}_i\right\}\right]   \\
		& = \sigma_{01}^2\cdot\mathbb{E}\left[r(\mathbf{X}_i, 1)\cdot p(\mathbf{X}_i)\cdot  \mathbf{A}(\mathbf{X}_i, \bm{\theta_1^0})\right]  
		\end{align*}
		Therefore, the asymptotic variance simplifies to simply 
		\begin{align*}
		\mathrm{Avar}\left[\sqrt{N}\left(\bm{\hat{\theta}_1^u}-\bm{\theta_1^0}\right)\right] = \sigma_{01}^2\cdot \left(\mathbb{E}\left[r(\mathbf{X}_i, 1)\cdot p(\mathbf{X}_i)\cdot  \mathbf{A}(\mathbf{X}_i, \bm{\theta_1^0})\right]\right)^{-1} 
		\end{align*}
		For showing that the two variances are positive semi-definite consider the following
		\begin{align*}
		&\left[\mathrm{Avar}\left\{\sqrt{N}\left(\bm{\hat{\theta}_1^u}-\bm{\theta_1^0}\right)\right\}\right]^{-1}- \left[\mathrm{Avar}\left\{ \sqrt{N}\left(\bm{\hat{\theta}_1}-\bm{\theta_1^0}\right)\right\}\right]^{-1} \\
		=&\frac{1}{\sigma_{01}^2}\cdot \left\{\mathbb{E}\left(r_{i1}\cdot p_i\cdot  \mathbf{A}_i\right) - \mathbb{E}\left(\frac{r_{i1}\cdot p_i}{R_{i1}\cdot G_i}\cdot \mathbf{A}_i\right)\cdot \mathbb{E}\left(\frac{r_{i1}\cdot p_i}{R^2_{i1}\cdot G^2_i}\cdot \mathbf{A}_i\right)^{-1}\cdot \mathbb{E}\left(\frac{r_{i1}\cdot p_i}{R_{i1}\cdot G_i}\cdot \mathbf{A}_i\right)\right\} \\
		&\text{ Let } \mathbf{B}_i = r_{i1}^{1/2}\cdot p_i^{1/2}\cdot \mathbf{A}_i^{1/2} \text{ and } \mathbf{D}_i = \left(r_{i1}^{1/2}/R_{i1}\right) \cdot \left(p_i^{1/2}/G_i\right) \cdot \mathbf{A}_i^{1/2} \\
		=&\frac{1}{\sigma_{01}^2}\left\{\mathbb{E}\left(\mathbf{B}_i^\prime\mathbf{B}_i\right) - \mathbb{E}\left(\mathbf{B}_i^\prime\mathbf{D}_i\right)\cdot \mathbb{E}\left(\mathbf{D}_i^\prime\mathbf{D}_i\right)^{-1}\cdot \mathbb{E}\left(\mathbf{D}_i^\prime\mathbf{B}_i\right) \right\}
		\end{align*}
		where the quantity inside the brackets is nothing but the variance of the residuals from the population regression of $\mathbf{B}_i$ on $\mathbf{D}_i$. Hence, the difference is positive semi-definite. The results for $g=0$ can be proven analogously.
	\end{proof}
	
	\subsection{Identification of ATE using pooled and separate slopes mean functions under second half of DR}
	\begin{proof}[\textup{\textbf{Pooled slopes}}]
		Let us assume that $m(\mathbf{X},\bm{\theta_g}) = h(\mathbf{X}\bm{\theta}+\eta W)$ is the chosen mean function for $\mathbb{E}[Y(g)|\mathbf{X}]$. Then, in the presence of nonrandom sampling, we have the following first order conditions
		\begin{align*}
		\sum_{i=1}^{N}S_i\cdot \left(\frac{W_i}{\hat{R}_i\cdot \hat{G}_i}+\frac{(1-W_i)}{\hat{R}_i\cdot (1-\hat{G}_i)}\right)\cdot \left[Y_i-h(\mathbf{X}_i\bm{\hat{\theta}}+\hat{\eta}W_i)\right] &=0\\
		\sum_{i=1}^{N} \frac{S_i\cdot W_i}{\hat{R}_i\cdot \hat{G}_i}\cdot \left[Y_i-h(\mathbf{X}_i\bm{\hat{\theta}}+\hat{\eta}W_i)\right] &=0 \\
		\sum_{i=1}^{N} S_i\cdot \left(\frac{W_i}{\hat{R}_i\cdot \hat{G}_i}+\frac{(1-W_i)}{\hat{R}_i\cdot (1-\hat{G}_i)}\right)\cdot \mathbf{X}_i^{\prime}\left[Y_i-h(\mathbf{X}_i\bm{\hat{\theta}}+\hat{\eta}W_i)\right] &=0 
		\end{align*}
		where $\hat{R} = R(\mathbf{X},W,\bm{\hat{\delta}})$ and $\hat{G} = G(\mathbf{X}, \bm{\hat{\gamma}})$. Ignoring the last set of moment conditions, the population counterpart to the FOCs above are:
		\begin{align}
		\mathbb{E}\left[S\cdot \left(\frac{W}{R\cdot G}+\frac{(1-W)}{R\cdot (1-G)}\right)\cdot \left[Y-h(\mathbf{X}\bm{\theta}^{\ast}+\eta^{\ast} W)\right]\right] &=0 \label{eq:poolfoc}\\
		\mathbb{E} \left[\frac{S\cdot W}{R\cdot G}\cdot \left[Y-h(\mathbf{X}\bm{\theta}^{\ast}+\eta^{\ast} W)\right]\right] &=0 \label{eq:treatfoc}
		\end{align}
		where $\bm{\theta}^{\ast}$ and $\eta^\ast$ are the probability limits of QMLE estimators $\bm{\hat{\theta}}$ and $\hat{\eta}$. Rearranging (\ref{eq:poolfoc}) and (\ref{eq:treatfoc}) gives us
		\begin{align}
		\mathbb{E}\left[\frac{S}{R}\cdot \left(\frac{W}{G}+\frac{(1-W)}{(1-G)}\right) \cdot Y\right] &=\mathbb{E}\left[\frac{S}{R}\cdot \left(\frac{W}{G}+\frac{(1-W)}{(1-G)}\right)\cdot  h(\mathbf{X}\bm{\theta}^{\ast}+\eta^{\ast} W)\right] \label{eq:poolfocnew} \\
		\mathbb{E} \left[\frac{S\cdot W}{R\cdot G}\cdot Y\right]&=\mathbb{E}\left[\frac{S\cdot W}{R\cdot G} \cdot h(\mathbf{X}\bm{\theta}^{\ast}+\eta^{\ast} W)\right] \label{eq:treatfocnew}
		\end{align}
		Now, $Y=Y(1)\cdot W+Y(0)\cdot (1-W)$ which implies that we can replace $Y$ in the above two equations to obtain the LHS of (\ref{eq:poolfocnew}) equal to 
		\begin{equation*}
		\mathbb{E}\left[\frac{S}{R}\cdot \left\{\frac{W}{G}\cdot Y(1)+\frac{(1-W)}{(1-G)}\cdot Y(0)\right\} \right] 
		\end{equation*}
		By using iterated expectations we can rewrite the above equation as
		\begin{equation*}
		\mathbb{E}\left[\frac{W}{G \cdot R}\cdot \mathbb{E}(S \cdot Y(1)|\mathbf{X}, W)+\frac{(1-W)}{ (1-G)\cdot R}\cdot \mathbb{E}(S\cdot Y(0)|\mathbf{X}, W)\right] 
		\end{equation*}
		Due to MAR, we can split the conditional expectation into parts. 
		\begin{equation*}
		\mathbb{E}\left[\frac{W}{G \cdot R}\cdot \mathbb{E}(S|\mathbf{X}, W) \cdot \mathbb{E}(Y(1)|\mathbf{X}, W)+\frac{(1-W)}{ (1-G)\cdot R}\cdot \mathbb{E}(S|\mathbf{X}, W) \cdot \mathbb{E}(Y(0)|\mathbf{X}, W)\right] 
		\end{equation*} 
		Note that, $W\cdot \mathbb{E}(S|\mathbf{X},W) = W\cdot R$. similarly, $(1-W)\cdot \mathbb{E}(S|\mathbf{X},W) =(1-W)\cdot R$ and due to unconfoundedness we have, $\mathbb{E}\left[Y(1)|\mathbf{X},W\right] =\mathbb{E}\left[Y(1)|\mathbf{X}\right]$ and $\mathbb{E}\left[Y(0)|\mathbf{X},W\right] =\mathbb{E}\left[Y(0)|\mathbf{X}\right]$. Therefore, we can simplify the above expression into
		\begin{equation*}
		\mathbb{E}\left[\frac{W\cdot R}{G \cdot R}\cdot \mathbb{E}(Y(1)|\mathbf{X})+\frac{(1-W)\cdot R}{ (1-G)\cdot R} \cdot \mathbb{E}(Y(0)|\mathbf{X})\right] 
		\end{equation*} 
		Another application of iterated expectation gives us 
		\begin{equation*}
		\begin{split}
		&\mathbb{E}\left[\frac{\mathbb{E}(Y(1)|\mathbf{X})}{G}\cdot \mathbb{E}[W|\mathbf{X}]+\frac{\mathbb{E}(Y(0)|\mathbf{X})}{ (1-G)}\cdot \mathbb{E}[(1-W)|\mathbf{X}]\right]  \\
		=& \  \mathbb{E}\left[\mathbb{E}(Y(1)|\mathbf{X})+\mathbb{E}(Y(0)|\mathbf{X})\right] \\
		=& \ \mathbb{E}[Y(1)]+\mathbb{E}[Y(0)]
		\end{split}
		\end{equation*} 
		Manipulating the RHS of (\ref{eq:poolfocnew}) using iterated expectations gives us
		\begin{equation*}
		\begin{split}
		&\mathbb{E}\left[ h(\mathbf{X}\bm{\theta}^{\ast}+\eta^{\ast} W) \cdot \left\{ \frac{W_1}{G}\cdot \frac{1}{R}\cdot \mathbb{E}(S|\mathbf{X},W_1)+\frac{(1-W)}{(1-G)}\cdot \frac{1}{R}\cdot \mathbb{E}(S|\mathbf{X},W) \right\}\right] \\
		=& \ \mathbb{E}\left[ h(\mathbf{X}\bm{\theta}^{\ast}+\eta^{\ast} W) \cdot \left\{ \frac{W}{G}+\frac{(1-W)}{(1-G)}\right\}\right]  \\
		=&  \ \mathbb{E}\left[ h(\mathbf{X}\bm{\theta}^{\ast}+\eta^{\ast} W) \cdot \frac{W}{G}\right]+\mathbb{E}\left[ h(\mathbf{X}\bm{\theta}^{\ast}+\eta^{\ast} W) \cdot \frac{(1-W)}{(1-G)}\right]
		\end{split}
		\end{equation*}
		Therefore, combining the LHS and RHS give the result
		\begin{equation}
		\mathbb{E}[Y(1)]+\mathbb{E}[Y(0)] = \mathbb{E}\left[ h(\mathbf{X}\bm{\theta}^{\ast}+\eta^{\ast} W) \cdot \frac{W}{G}\right]+\mathbb{E}\left[ h(\mathbf{X}\bm{\theta}^{\ast}+\eta^{\ast} W) \cdot \frac{(1-W)}{(1-G)}\right] \label{eq:mu1+mu0}
		\end{equation}
		Now, consider the LHS of \ref{eq:treatfocnew}.
		\begin{align*}
		\mathbb{E} \left[\frac{S\cdot W}{R\cdot G}\cdot Y\right] &= \mathbb{E} \left[\frac{S\cdot W}{R\cdot G}\cdot Y(1)\right] \\
		& =  \mathbb{E}[Y(1)] \tag*{\text{ (by LIE)}}
		\end{align*}
		Similarly using LIE, the RHS of \ref{eq:treatfocnew} can be re-written as 
		\begin{equation*}
		\begin{split}
		\mathbb{E}\left[\frac{S\cdot W}{R\cdot G} \cdot h(\mathbf{X}\bm{\theta}^{\ast}+\eta^{\ast} W)\right] &= \mathbb{E}\left[h(\mathbf{X}\bm{\theta}^{\ast}+\eta^{\ast} W) \cdot \frac{W}{G}\cdot \frac{1}{R}\cdot \mathbb{E}(S|\mathbf{X},W)\right] \\
		& = \mathbb{E}\left[h(\mathbf{X}\bm{\theta}^{\ast}+\eta^{\ast} W) \cdot \frac{W}{G}\right]
		\end{split}
		\end{equation*}	
		Therefore combining the LHS and RHS give us 
		\begin{equation}
		\mathbb{E}[Y(1)] =   \mathbb{E}\left[h(\mathbf{X}\bm{\theta}^{\ast}+\eta^{\ast} W) \cdot \frac{W}{G}\right] \label{eq:mu1}
		\end{equation}
		Then using \ref{eq:mu1} along with \ref{eq:mu1+mu0} implies that 
		\begin{equation}
		\begin{split}
		\mathbb{E}[Y(0)] &=   \mathbb{E}\left[h(\mathbf{X}\bm{\theta}^{\ast}+\eta^{\ast} W) \cdot \frac{(1-W)}{(1-G)}\right]  \label{eq:mu0}
		\end{split}
		\end{equation}
		Consider 
		\begin{equation*}
		\begin{split}
		&\mathbb{E}\left[h(\mathbf{X}\bm{\theta}^{\ast}+\eta^{\ast} W)\cdot W|\mathbf{X}\right]\\
		=& \ \mathbb{E}\left[h(\mathbf{X}\bm{\theta}^{\ast}+\eta^{\ast})\right]\cdot P(W=1|\mathbf{X}) 
		\end{split}
		\end{equation*}
		Therefore, $\mathbb{E}\left[h(\mathbf{X}\bm{\theta}^{\ast}+\eta^{\ast} W) \cdot \displaystyle \frac{W}{G}\right] = \mathbb{E}\left[h(\mathbf{X}\bm{\theta}^{\ast}+\eta^{\ast})\right]$. Similarly, we can also show that
		 \[\mathbb{E}\left[h(\mathbf{X}\bm{\theta}^{\ast}+\eta^{\ast} W) \cdot \frac{(1-W)}{(1-G)}\right] = \mathbb{E}\left[h(\mathbf{X}\bm{\theta}^{\ast})\right]\]
		
		Hence, the pooled regression adjustment estimator can be written as
		\begin{equation*}
		\Delta_{\text{ate}}^{\text{P}} = \mathbb{E}\left[h(\mathbf{X}\bm{\theta}^{\ast}+\eta^{\ast})\right]-\mathbb{E}\left[h(\mathbf{X}\bm{\theta}^{\ast})\right]
		\end{equation*}
		so a consistent estimator of the QMLE pooled regression adjustment estimator can be obtained by replacing the population expectation by the sample average in the above expression and weighting by the appropriate probabilities to recover the balance of the random sample which gives us
		\begin{equation*}
		\hat{\Delta}_{\text{ate}}^{\text{P}} = \frac{1}{N}\sum_{i=1}^{N} h(\mathbf{X}_i\bm{\hat{\theta}}+\hat{\eta})-\frac{1}{N}\sum_{i=1}^{N} h(\mathbf{X}_i\bm{\hat{\theta}})
		\end{equation*}
	\end{proof}
	
	\begin{proof}[\textup{\textbf{Separate slopes}}]
		Let us assume that $m(\mathbf{X},\bm{\theta_g}) = h(\mathbf{X}\bm{\theta_g})$ is the chosen mean function for $\mathbb{E}\left[Y(g)|\mathbf{X}\right]$. Then the population FOCs are 
		\begin{align}
		\mathbb{E}\left[\frac{S\cdot W}{R\cdot G}\cdot \left[Y-h(\mathbf{X}\bm{\theta_1}^{\ast})\right]\right] &= 0 \label{eq:treatfocsep} \\
		\mathbb{E}\left[\frac{S\cdot (1-W)}{R\cdot (1- G)}\cdot \left[Y-h(\mathbf{X}\bm{\theta_0}^\ast)\right]\right] &= 0 \label{eq:controlfocsep} 
		\end{align}
		where $\bm{\theta_g}^{\ast}$ are the probability limits of QMLE estimators $\bm{\hat{\theta}_g}$. Rearranging \ref{eq:treatfocsep} and \ref{eq:controlfocsep} just like in the pooled case gives us the following equalities. 
		\begin{equation*}
		\begin{split}
		\mathbb{E}\left[\frac{S\cdot W}{R\cdot G}\cdot Y\right]&=\mathbb{E}\left[\frac{S\cdot W}{R\cdot G}\cdot h(\mathbf{X}\bm{\theta_1}^{\ast})\right] \\
		\mathbb{E}\left[\frac{S\cdot (1-W)}{R\cdot (1- G)}\cdot Y\right]&=\mathbb{E}\left[\frac{S\cdot (1-W)}{R\cdot (1-G)} \cdot h(\mathbf{X}\bm{\theta_0}^\ast)\right]
		\end{split}
		\end{equation*}
		Proceeding with the above two equations in the same way as in the pooled case gives us the results 
		\begin{equation*}
		\begin{split}
		\mathbb{E}[Y(1)] &= \mathbb{E}\left[h(\mathbf{X}\bm{\theta_1}^{\ast})\right] \\
		\mathbb{E}[Y(0)] &=  \mathbb{E}\left[h(\mathbf{X}\bm{\theta_0}^{\ast})\right] 
		\end{split}
		\end{equation*}
		Therefore, 
		$\Delta_{\text{ate}}^{\text{F}} = \mathbb{E}\left[h(\mathbf{X}\bm{\theta_1}^{\ast})\right]-\mathbb{E}\left[h(\mathbf{X}\bm{\theta_0}^{\ast})\right]$ and a consistent estimator of the QMLE separate regression adjustment estimator can be obtained as
		\begin{equation*}
		\hat{\Delta}_{\text{ate}}^{\text{F}}  = \frac{1}{N}\sum_{i=1}^{N} h(\mathbf{X}_i\bm{\hat{\theta}_1})-\frac{1}{N}\sum_{i=1}^{N} h(\mathbf{X}_i\bm{\hat{\theta}_0})
		\end{equation*}
	\end{proof}

		\section{Supplementary Tables} \label{tabs}
		
		{\small \begin{table}[h!]
				\centering
				\caption{Proportion of missing earnings in the experimental sample}
				\begin{tabular}{l|ccc}
					\toprule
					\textbf{Earnings in 1979} & \textbf{Treated} & \textbf{Control} & \textbf{Total} \\
					\midrule
					Missing & 196   & 210   & 406 \\
					Observed & 600   & 585   & 1185 \\
					\midrule
					\textbf{Total} & 796   & 795   & 1591 \\
					\bottomrule
				\end{tabular}%
				\label{tab:miss_es}%
		\end{table}}%
		
		\begin{table}[h!]
			\centering
			\caption{Proportion of missing data in the PSID samples}
			\begin{tabular}{l|cc}
				\toprule
				\textbf{Earnings in 1979} & \multicolumn{1}{c}{\textbf{PSID-1}} & \multicolumn{1}{c}{\textbf{PSID-2}} \\
				\midrule
				Missing & 81& 22  \\
				Observed & 648 & 182 \\
				\midrule
				\textbf{Total} & 729 & 204 \\
				\bottomrule
			\end{tabular}%
			\label{tab:miss_psid}%
		\end{table}%
	
		\begin{landscape}
		\begin{table}[H]
			\centering
			\caption{Unweighted and weighted earnings comparisons and estimated training effects using NSW and PSID comparison groups}
			\resizebox{0.66\columnwidth}{!}{
				\begin{threeparttable}  \begin{tabular}{lrrrrrr}
						\toprule[1pt]\midrule[0.3pt]
						\multicolumn{1}{c}{\multirow{3}[6]{*}{\textbf{Comparison group}}} &       &       & \multicolumn{2}{c}{\textbf{Pre-training estimates}} &       &  \\
						\cmidrule{4-5}          &       & \multicolumn{1}{c}{\textbf{Unadjusted}} &       &       & \multicolumn{1}{c}{\textbf{Adjusted}} &  \\
						\cmidrule{2-7}          & \multicolumn{1}{c}{Unweighted} & \multicolumn{1}{c}{PS-weighted} & \multicolumn{1}{c}{D-weighted} & \multicolumn{1}{c}{Unweighted} & \multicolumn{1}{c}{PS-weighted} & \multicolumn{1}{c}{D-weighted} \\
						\midrule
						\textbf{NSW} & \multicolumn{1}{c}{-18} & \multicolumn{1}{c}{-9} & \multicolumn{1}{c}{1} & \multicolumn{1}{c}{-22} & \multicolumn{1}{c}{-10} & \multicolumn{1}{c}{-1} \\
						N=1,185 & \multicolumn{1}{c}{(123.45)} & \multicolumn{1}{c}{(51.07)} & \multicolumn{1}{c}{(48.76)} & \multicolumn{1}{c}{(124.70)} & \multicolumn{1}{c}{(51.34)} & \multicolumn{1}{c}{(48.97)} \\
						&       &       &       &       &       &  \\
						\textbf{PSID-1} & \multicolumn{1}{c}{-2,534} & \multicolumn{1}{c}{-222} & \multicolumn{1}{c}{-255} & \multicolumn{1}{c}{-2,804} & \multicolumn{1}{c}{-199} & \multicolumn{1}{c}{-222} \\
						N=1,016 & \multicolumn{1}{c}{(283.95)} & \multicolumn{1}{c}{(213.57)} & \multicolumn{1}{c}{(205.59)} & \multicolumn{1}{c}{(281.49)} & \multicolumn{1}{c}{(212.55)} & \multicolumn{1}{c}{(205.45)} \\
						&       &       &       &       &       &  \\
						\textbf{PSID-2} & \multicolumn{1}{c}{-2,080} & \multicolumn{1}{c}{-1,371} & \multicolumn{1}{c}{-1,357} & \multicolumn{1}{c}{-2,181} & \multicolumn{1}{c}{-1,505} & \multicolumn{1}{c}{-1,467} \\
						N=720 & \multicolumn{1}{c}{(411.23)} & \multicolumn{1}{c}{(331.41)} & \multicolumn{1}{c}{(317.41)} & \multicolumn{1}{c}{(427.24)} & \multicolumn{1}{c}{(359.98)} & \multicolumn{1}{c}{(342.16)} \\
						\midrule
						&       &       & \multicolumn{2}{c}{\textbf{Bias using NSW control}} &       &  \\
						\cmidrule{4-5}    \textbf{PSID-1} & \multicolumn{1}{c}{-2,517} & \multicolumn{1}{c}{289} & \multicolumn{1}{c}{236} & \multicolumn{1}{c}{-2,760} & \multicolumn{1}{c}{334} & \multicolumn{1}{c}{287} \\
						N=1,001 & \multicolumn{1}{c}{(279.38)} & \multicolumn{1}{c}{(256.93)} & \multicolumn{1}{c}{(247.18)} & \multicolumn{1}{c}{(283.09)} & \multicolumn{1}{c}{(257.50)} & \multicolumn{1}{c}{(248.20)} \\
						&       &       &       &       &       &  \\
						\textbf{PSID-2} & \multicolumn{1}{c}{-2,063} & \multicolumn{1}{c}{-1,249} & \multicolumn{1}{c}{-1,255} & \multicolumn{1}{c}{-2,144} & \multicolumn{1}{c}{-1,306} & \multicolumn{1}{c}{-1,297} \\
						N=705 & \multicolumn{1}{c}{(416.53)} & \multicolumn{1}{c}{(323.36)} & \multicolumn{1}{c}{(310.59)} & \multicolumn{1}{c}{(435.74)} & \multicolumn{1}{c}{(354.12)} & \multicolumn{1}{c}{(337.68)} \\
						\midrule
						\textbf{Adjusted covariates} &       &       &       &       &       &  \\
						\cmidrule{1-1}    Pre-training earnings (1975) &       &       &       & \multicolumn{1}{c}{\checkmark} & \multicolumn{1}{c}{\checkmark} & \multicolumn{1}{c}{\checkmark} \\
						Age   &       &       &       & \multicolumn{1}{c}{\checkmark} & \multicolumn{1}{c}{\checkmark} & \multicolumn{1}{c}{\checkmark} \\
						Age2  &       &       &       & \multicolumn{1}{c}{\checkmark} & \multicolumn{1}{c}{\checkmark} & \multicolumn{1}{c}{\checkmark} \\
						Education &       &       &       & \multicolumn{1}{c}{\checkmark} & \multicolumn{1}{c}{\checkmark} & \multicolumn{1}{c}{\checkmark} \\
						High school droput &       &       &       & \multicolumn{1}{c}{\checkmark} & \multicolumn{1}{c}{\checkmark} & \multicolumn{1}{c}{\checkmark} \\
						Black &       &       &       & \multicolumn{1}{c}{\checkmark} & \multicolumn{1}{c}{\checkmark} & \multicolumn{1}{c}{\checkmark} \\
						Hispanic &       &       &       & \multicolumn{1}{c}{\checkmark} & \multicolumn{1}{c}{\checkmark} & \multicolumn{1}{c}{\checkmark} \\
						Marital status &       &       &       & \multicolumn{1}{c}{\checkmark} & \multicolumn{1}{c}{\checkmark} & \multicolumn{1}{c}{\checkmark} \\
						Number of Children (1975) &       &       &       &       &       &  \\
						\bottomrule
					\end{tabular}%
					\begin{tablenotes}[flushleft]
						\footnotesize
						\item \textit{Notes:} This table reports unadjusted and adjusted pre-training earnings differences where the first row reports the experimental estimates which combines the NSW treatment and control groups. The second and third row reports non-experimental earnings estimates computed from using the PSID-1 and PSID-2 comparison groups respectively. The second panel of the table reports bias estimates computed from combining the NSW control and PSID-1 and PSID-2 comparison groups respectively. Both the pre-training estimates and the bias estimates should be compared to zero. Bootstrapped standard errors are given in parentheses and have been constructed from using 10,000 replications. All values are in 1982 dollars. The samples used for estimating the training and bias estimates using PSID-1 and PSID-2 comparison groups have been trimmed to ensure common support in the distribution of weights for the NSW-treatment and comparison groups. For more detail, see appendix \ref{csapndx}. 
					\end{tablenotes}	
			\end{threeparttable}}
		\end{table}%
	\end{landscape}
	
	\begin{table}[H]
		\centering
		\caption{Unconditional quantile treatment effect (UQTE) using PSID-1 comparison group}
		\begin{threeparttable}
			\begin{tabular}{crrrr}
				\toprule[1pt]\midrule[0.3pt]
				\multicolumn{1}{l}{Quantile} & \multicolumn{1}{l}{Experimental} & \multicolumn{1}{l}{Unweighted} & \multicolumn{1}{l}{PS-weighted} & \multicolumn{1}{l}{D-weighted} \\
				\midrule
				0.1   & 0     & 0     & 0     & 0 \\
				& (0)   & (0)   & (0)   & (0) \\
				0.2   & 0     & 0     & 0     & 0 \\
				& (0)   & (0)   & (0)   & (0) \\
				0.3   & 0     & 0     & 0     & 0 \\
				& (0)   & (12.91) & (0)   & (0) \\
				0.4   & 0     & -1124.61 & 0     & 0 \\
				& (11.17) & (552.97) & (207.14) & (174.89) \\
				0.5   & 993.52 & -2227.26 & 2076.58 & 1847.04 \\
				& (695.93) & (983.43) & (851.09) & (829.42) \\
				0.6   & 2004.40 & -860.55 & 3602.76 & 3535.85 \\
				& (1112.82) & (964.97) & (1299.08) & (1284.64) \\
				0.7   & 2129.93 & 428.01 & 3415.47 & 3340.84 \\
				& (716.04) & (728.22) & (988.24) & (992.95) \\
				0.8   & 1753.27 & -190.60 & 2019.44 & 2019.44 \\
				& (372.37) & (519.63) & (984.59) & (999.47) \\
				0.9   & 1134.21 & -1563.27 & -385.45 & -385.45 \\
				& (449.86) & (952.85) & (1059.43) & (1056.09) \\
				\bottomrule
			\end{tabular}%
			\begin{tablenotes}[flushleft]
				\footnotesize
				\item \textit{Notes:} This table reports unweighted, ps-weighted and d-weighted UQTE estimates for three different comparison groups, namely, NSW control, PSID-1 and PSID-2. The estimates are reported at every 10th quantile of the 1979 earnings distribution. The experimental and PSID-1 estimates have been constructed using N=1,185 and N=1,016 observations respectively. Bootstrapped standard errors are given in parentheses and have been constructed using 1,000 replications. All values are in 1982 dollars. The samples used for constructing these estimates have been trimmed to ensure common support across the treatment and comparison groups. 
			\end{tablenotes}	
		\end{threeparttable}
		\label{tab:quant_psid1}%
	\end{table}%
	
	\begin{table}[H]
		\centering
		\caption{Unconditional quantile treatment effect (UQTE) using PSID-2 comparison group}
		\begin{threeparttable}
			\begin{tabular}{crrrr}
				\toprule[1pt]\midrule[0.3pt]
				\multicolumn{1}{l}{Quantile} & \multicolumn{1}{l}{Experimental} & \multicolumn{1}{l}{Unweighted} & \multicolumn{1}{l}{PS-weighted} & \multicolumn{1}{l}{D-weighted} \\
				\midrule
				0.1   & 0     & 0     & 0     & 0 \\
				& (0)   & (0)   & (0)   & (0) \\
				0.2   & 0     & 0     & 0     & 0 \\
				& (0)   & (0)   & (10.07) & (10.07) \\
				0.3   & 0     & 0     & 0     & 0 \\
				& (0)   & (111.74) & (136.31) & (129.77) \\
				0.4   & 0     & -795.71 & 0     & 0 \\
				& (13.25) & (672.87) & (573.22) & (546.78) \\
				0.5   & 993.52 & -237.98 & 378.98 & 372.07 \\
				& (693.73) & (1232.63) & (1312.93) & (1267.28) \\
				0.6   & 2004.40 & 193.77 & 1480.47 & 1294.77 \\
				& (1114.65) & (1426.40) & (1647.31) & (1659.69) \\
				0.7   & 2129.93 & 1857.64 & 2616.22 & 2599.73 \\
				& (710.26) & (943.38) & (1217.80) & (1209.60) \\
				0.8   & 1753.27 & 1148.85 & 2010.87 & 1990.37 \\
				& (371.73) & (1152.92) & (1541.14) & (1553.67) \\
				0.9   & 1134.21 & -237.08 & 1089.10 & 1089.10 \\
				& (452.08) & (1888.06) & (3321.56) & (3246.78) \\
				\bottomrule
			\end{tabular}%
			\begin{tablenotes}[flushleft]
				\footnotesize
				\item \textit{Notes:} This table reports unweighted, ps-weighted and d-weighted UQTE estimates for three different comparison groups, namely, NSW control, PSID-1 and PSID-2. The estimates are reported at every 10th quantile of the 1979 earnings distribution. The experimental and PSID-2 estimates have been computed using N=1,185 and N=720 observations respectively. Bootstrapped standard errors are given in parentheses and have been constructed using 1,000 replications. All values are in 1982 dollars. The samples used for constructing these estimates have been trimmed to ensure common support across the treatment and comparison groups. 
			\end{tablenotes}	
		\end{threeparttable}
		\label{tab:quant_psid2}%
	\end{table}%
	
	\clearpage
	\section{Supplementary Figures} \label{figs}
	\begin{figure}[H]
		\centering
		\caption{Estimated CQTE with true CQTE as a function of $X_1$ for N=5,000} 
		\centering{Case 3: Correct CQF, misspecified weights} \vspace{1em} 
		
		\begin{minipage}{0.50\columnwidth}
			\caption*{{\small b) $\tau=0.50$}} \vspace{-1em}
			\includegraphics[scale=0.46]{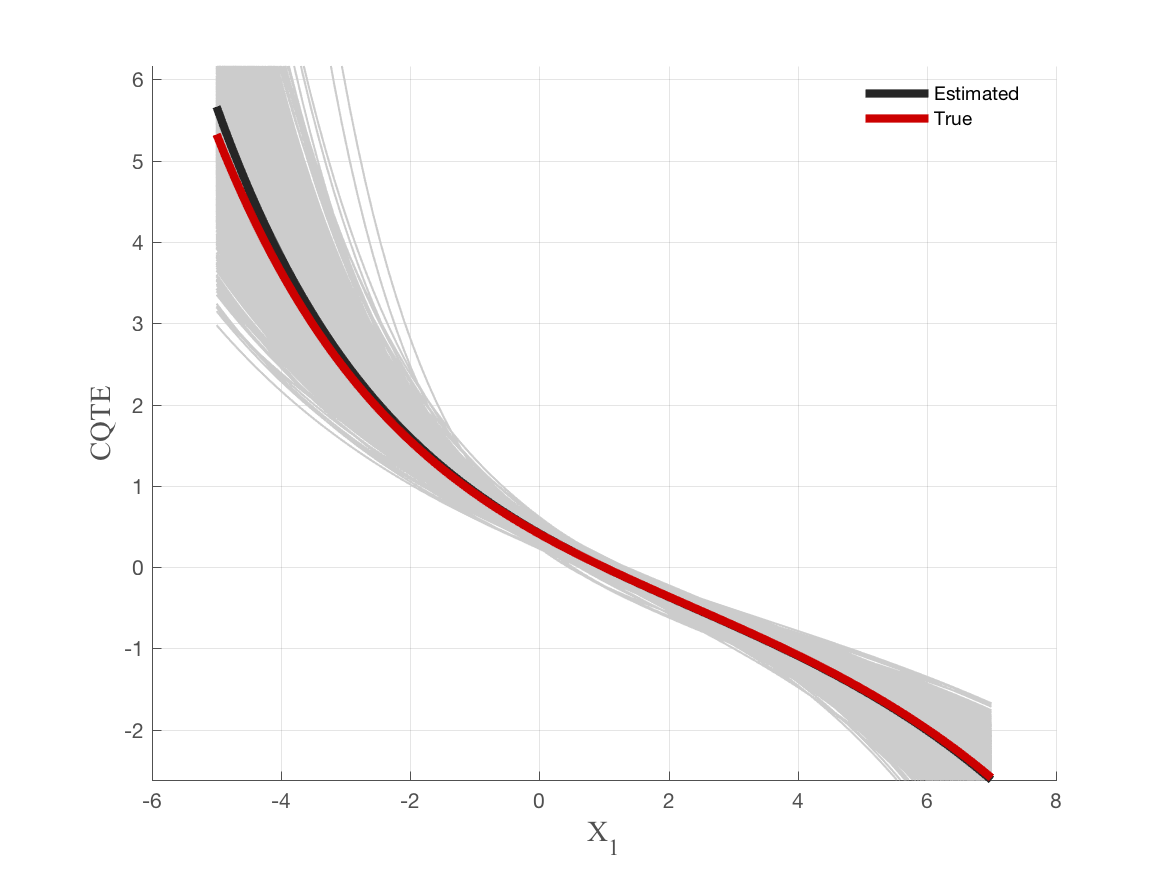}
		\end{minipage}\begin{minipage}{0.50\columnwidth}
			\caption*{{\small c) $\tau=0.75$}} \vspace{-1em}
			\includegraphics[scale=0.46]{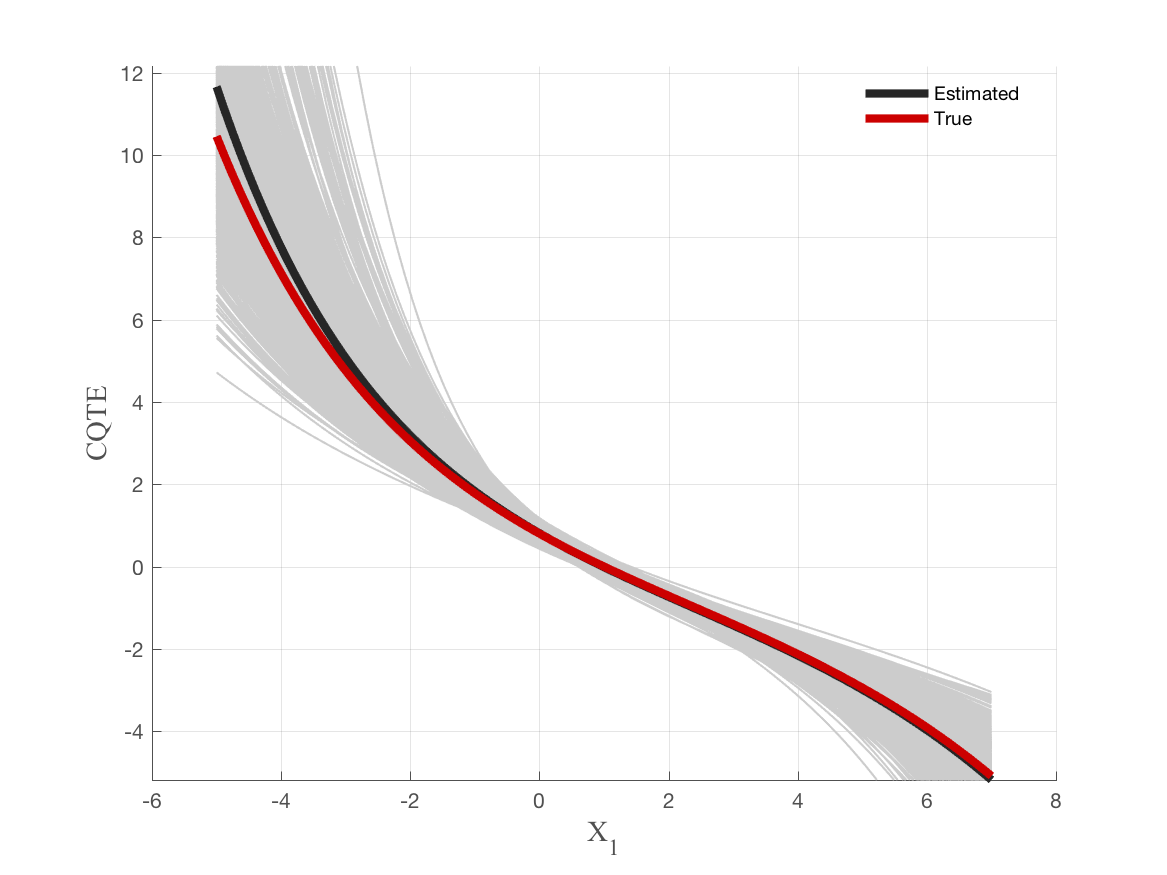}
		\end{minipage}
		\begin{minipage}{\columnwidth}
			\begin{spacing}{0.9}
				{\footnotesize \textit{Notes:} This figure plots the average d-weighted CQTE function with the true CQTE along $X_1$ for 1,000 Monte Carlo simulation draws of sample size $N=5,000$. Along with these two graphs, the figure also plots the individual function across the 1,000 simulation draws. The average treated sample is $N_1 = 5,000\times 0.41\times 0.38 = 779$ and average control sample is $N_0 = 5,000\times (1-0.41)\times 0.38 = 1,121$. \par}
			\end{spacing}
		\end{minipage}
	\end{figure}	
	
	\clearpage
	\begin{figure}[H]
		\centering
		\caption{Bias in the estimated LP relative to the true LP of CQTE as a function of $X_1$ for N=5,000}
		\centering {Case 1: Misspecified CQF, correct weights} \vspace{1em}
		
		\begin{minipage}{0.50\columnwidth}
			\caption*{{\small b) $\tau=0.50$}} \vspace{-1em}
			\includegraphics[scale=0.46]{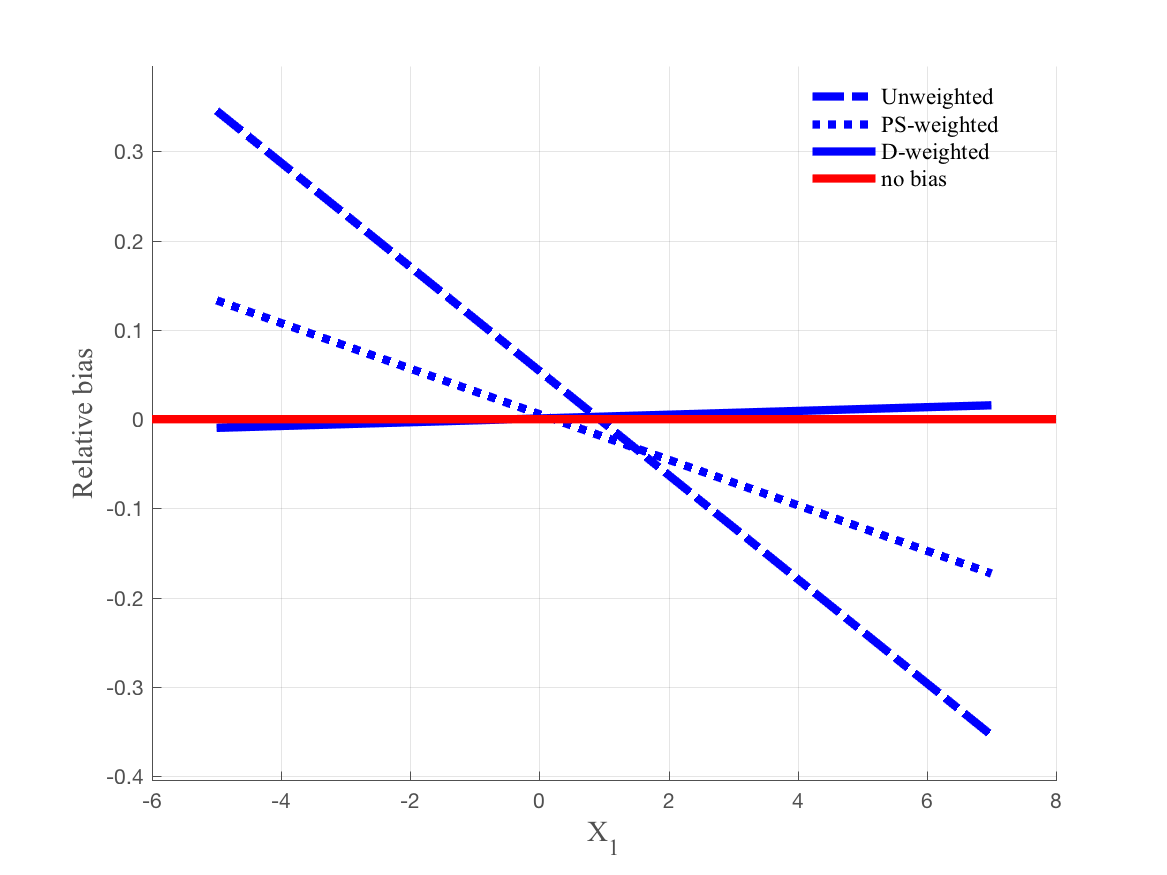}
		\end{minipage}\begin{minipage}{0.50\columnwidth}
			\caption*{{\small c) $\tau=0.75$}} \vspace{-1em}
			\includegraphics[scale=0.46]{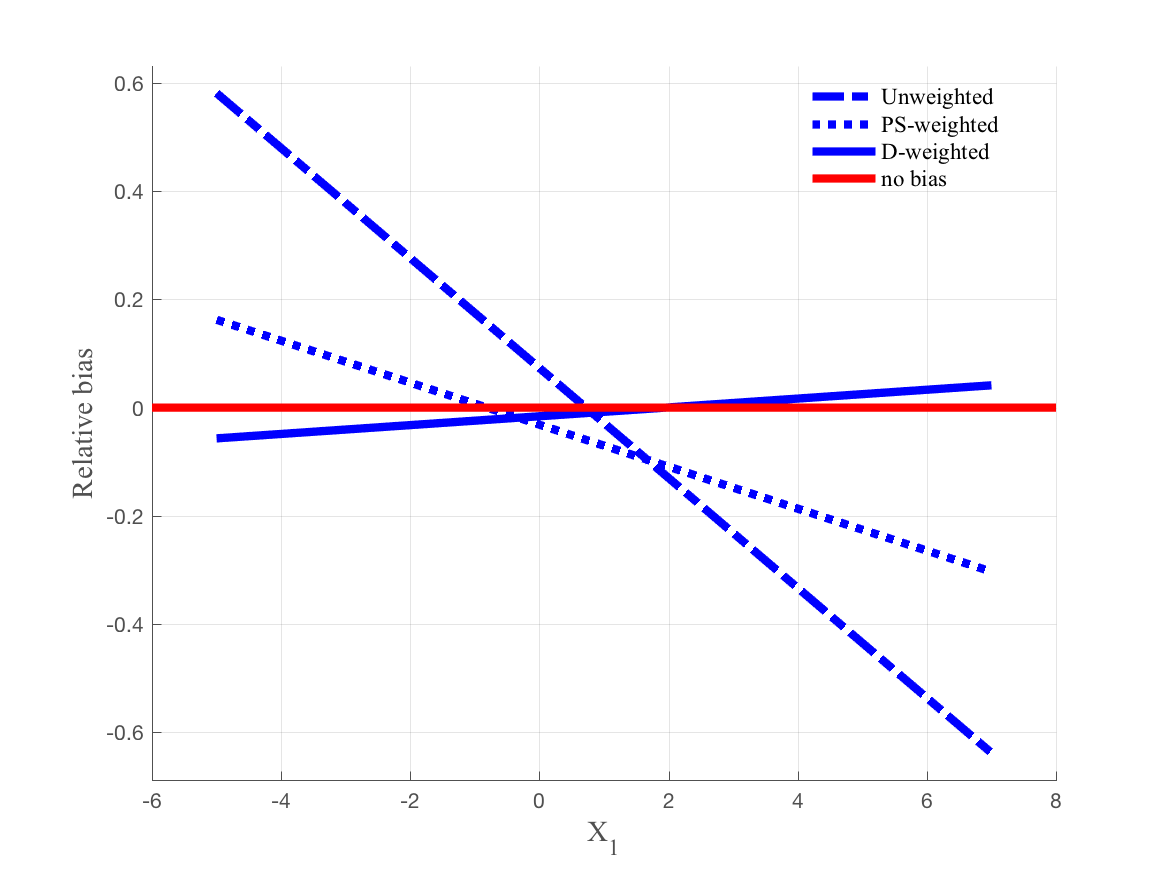}
		\end{minipage}
	\begin{minipage}{\columnwidth}
		\begin{spacing}{0.9}
			{\footnotesize \textit{Notes:} This figure plots the bias in the unweighted, ps-weighted, and d-weighted LP of the true CQTE relative to the true population LP of CQTE. The average treated sample is $N_1 = 5,000\times 0.41\times 0.38 = 779$ and average control sample is $N_0 = 5,000\times (1-0.41)\times 0.38 = 1,121$. The unweighted estimator does not weight the observed data. The ps-weighted estimator weights to correct only for nonrandom assignment and the d-weighted estimator weights by both the treatment and missing outcomes propensity score models to deal with nonrandom assignment and missing outcome problems. \par}
		\end{spacing}
	\end{minipage}
	\end{figure}	
	
	\begin{figure}[H]
		\centering
		\caption{Bias in the estimated linear projection relative to the true linear projection for N=5,000}
		\centering {Case 3: Misspecified CQF, misspecified weights}  \vspace{1em}
		
		\begin{minipage}{0.50\columnwidth}
			\caption*{{\small b) $\tau=0.50$}} \vspace{-1em}
			\includegraphics[scale=0.45]{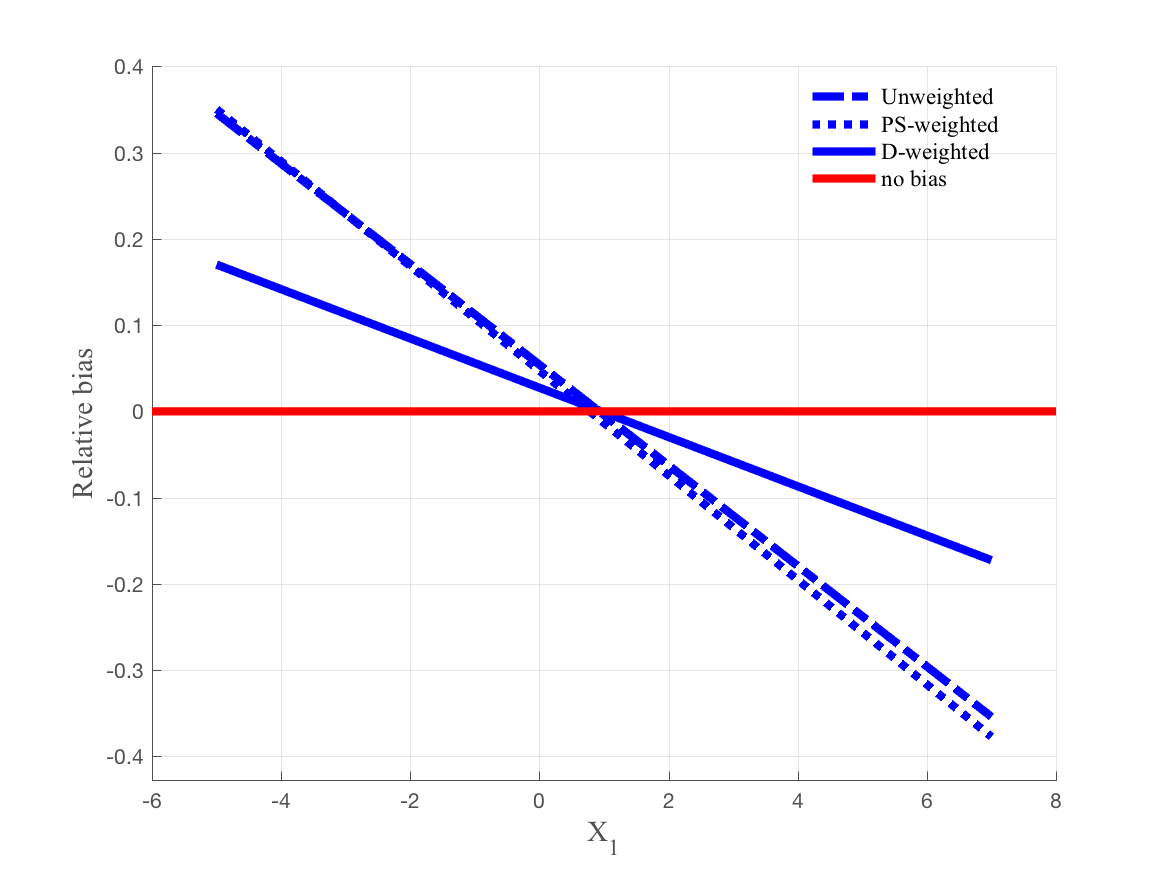}
		\end{minipage}\begin{minipage}{0.50\columnwidth}
			\caption*{{\small c) $\tau=0.75$}} \vspace{-1em}
			\includegraphics[scale=0.45]{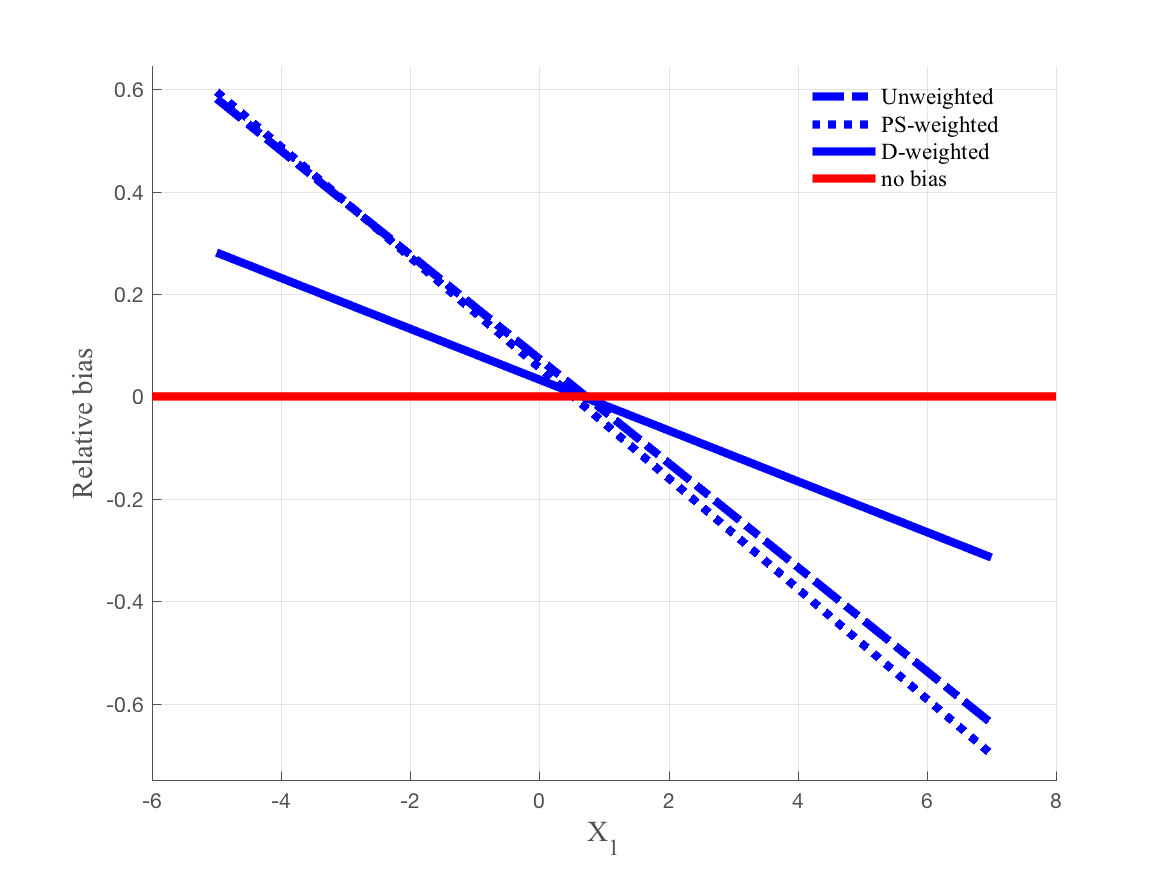}
		\end{minipage}
		\begin{minipage}{\columnwidth}
			\begin{spacing}{0.9}
				{\footnotesize \textit{Notes:} This figure plots the bias in the unweighted, ps-weighted, and d-weighted LP of the true CQTE relative to the true population LP of CQTE. The average treated sample is $N_1 = 5,000\times 0.41\times 0.38 = 779$ and average control sample is $N_0 = 5,000\times (1-0.41)\times 0.38 = 1,121$. The unweighted estimator does not weight the observed data. The ps-weighted estimator weights to correct only for nonrandom assignment and the d-weighted estimator weights by both the treatment and missing outcomes propensity score models to deal with nonrandom assignment and missing outcome problems. \par}
			\end{spacing}
		\end{minipage}
	\end{figure}	
	
	\clearpage
	\begin{figure}[H]
		\centering
		\caption{Empirical distribution of estimated UQTE for N=5,000 when weights are wrong}
		\begin{minipage}{0.50\columnwidth}
			\caption*{{\small b) $\tau=0.50$}} \vspace{-1em}
			\includegraphics[scale=0.45]{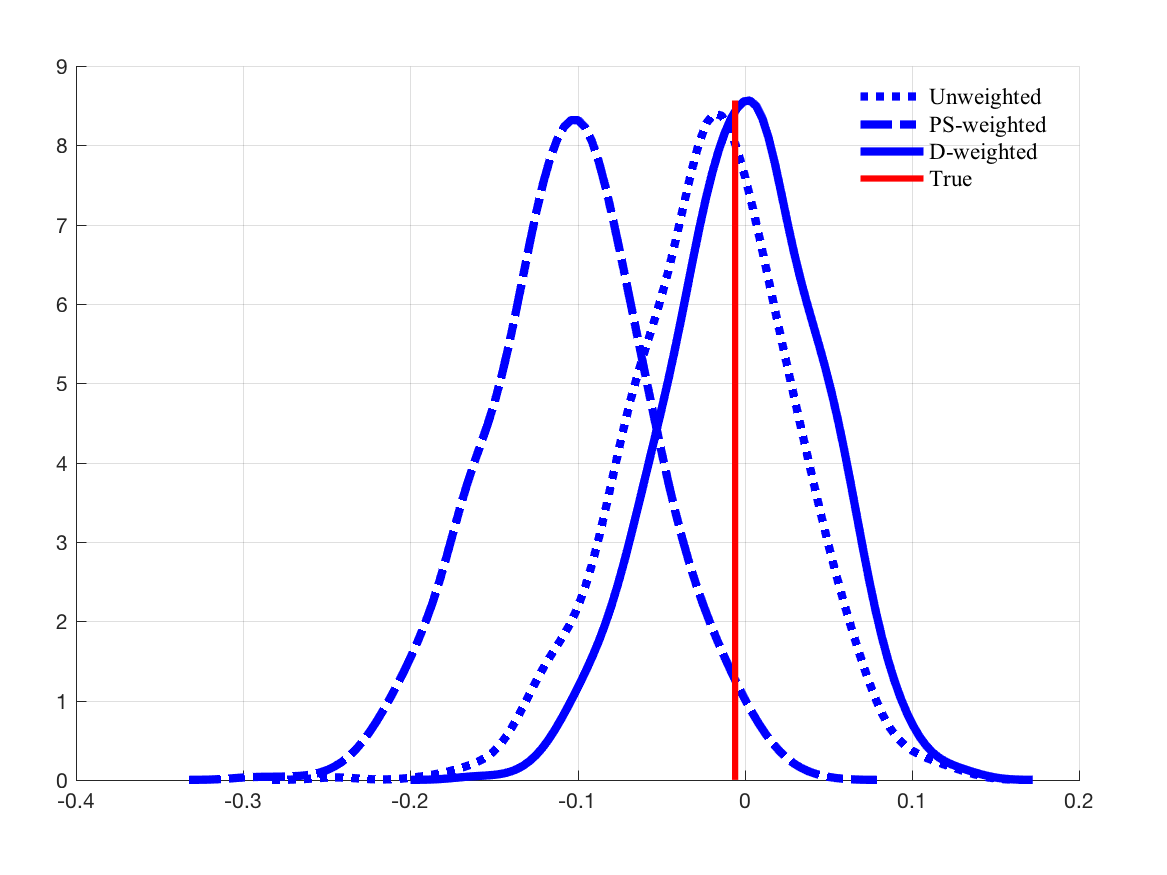}
		\end{minipage}\begin{minipage}{0.50\columnwidth}
			\caption*{{\small c) $\tau=0.75$}} \vspace{-1em}
			\includegraphics[scale=0.45]{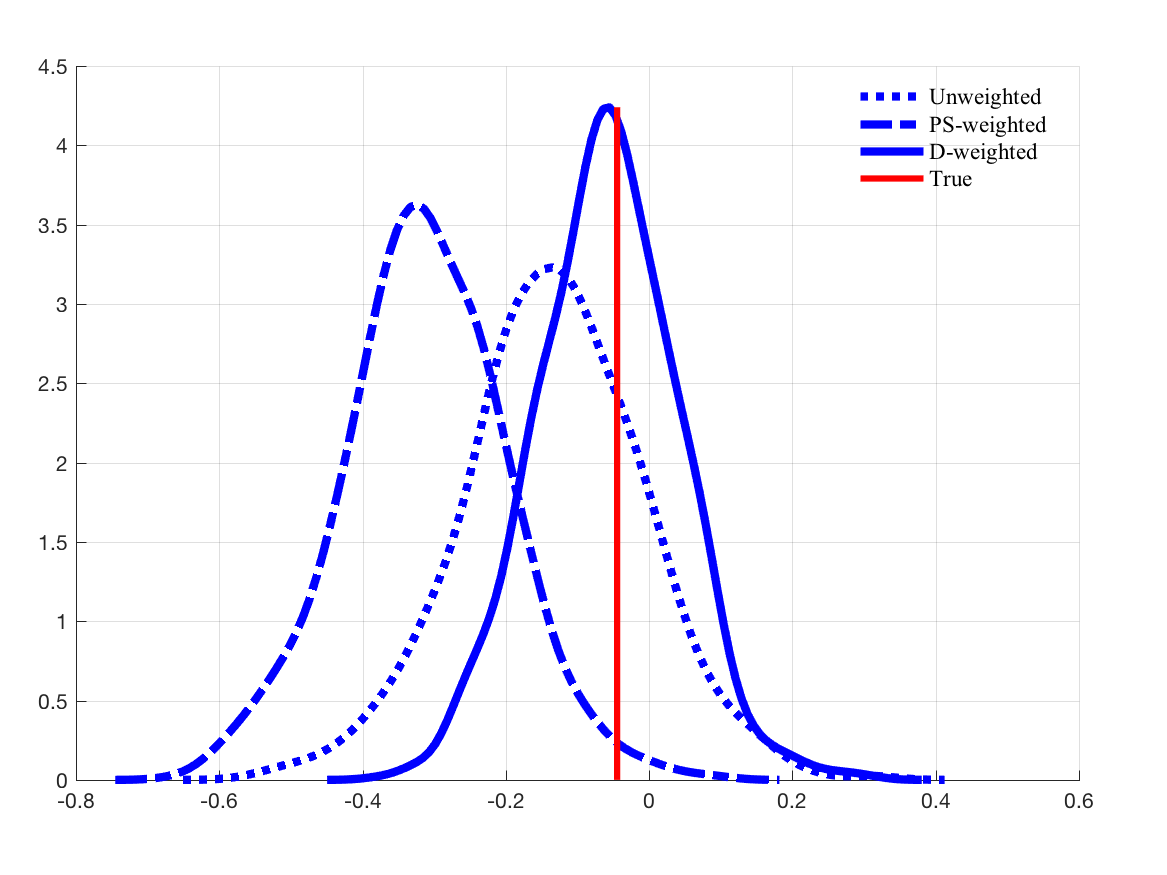}
		\end{minipage}
		
		\begin{minipage}{\columnwidth}
			\begin{spacing}{0.9}
				{\footnotesize \textit{Notes:} This figure plots the empirical distributions of the unweighted, ps-weighted, and d-weighted UQTE estimates using 1,000 Monte Carlo simulation draws of sample size 5,000.  The average treated sample is $N_1 = 5,000\times 0.41\times 0.38 = 779$ and average control sample is $N_0 = 5,000\times (1-0.41)\times 0.38 = 1,121$. The unweighted estimator does not weight the observed data. The ps-weighted estimator weights to correct only for nonrandom assignment and the d-weighted estimator weights by both the treatment and missing outcomes propensity score models to deal with nonrandom assignment and missing outcome problems. \par}
			\end{spacing}
		\end{minipage}
	\end{figure}	
	
	\clearpage
	\begin{figure}[H]
		\centering
		\caption{Empirical distribution of estimated UQTE for N=5,000 when weights are correct}
		\begin{minipage}{0.50\columnwidth}
			\caption*{{\small b) $\tau=0.50$}} \vspace{-1em}
			\includegraphics[scale=0.45]{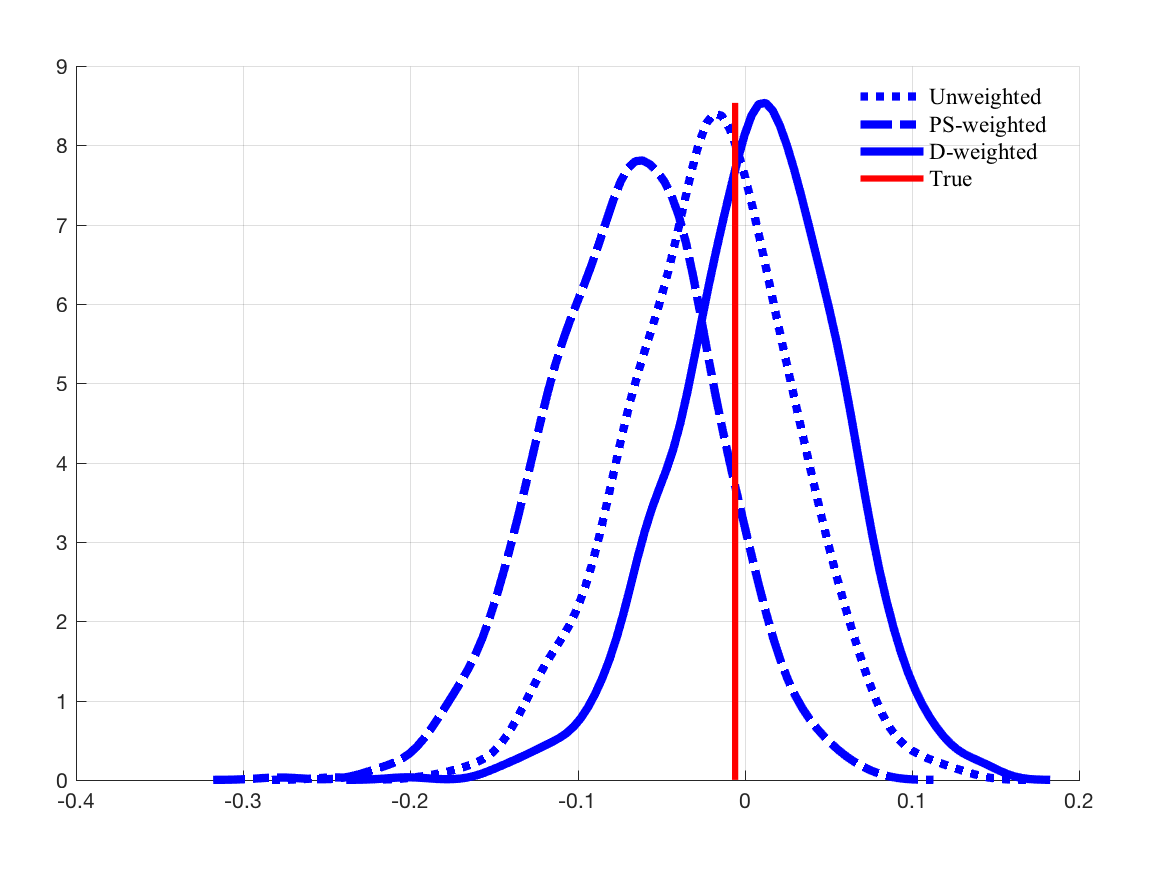}
		\end{minipage}\begin{minipage}{0.50\columnwidth}
			\caption*{{\small c) $\tau=0.75$}} \vspace{-1em}
			\includegraphics[scale=0.45]{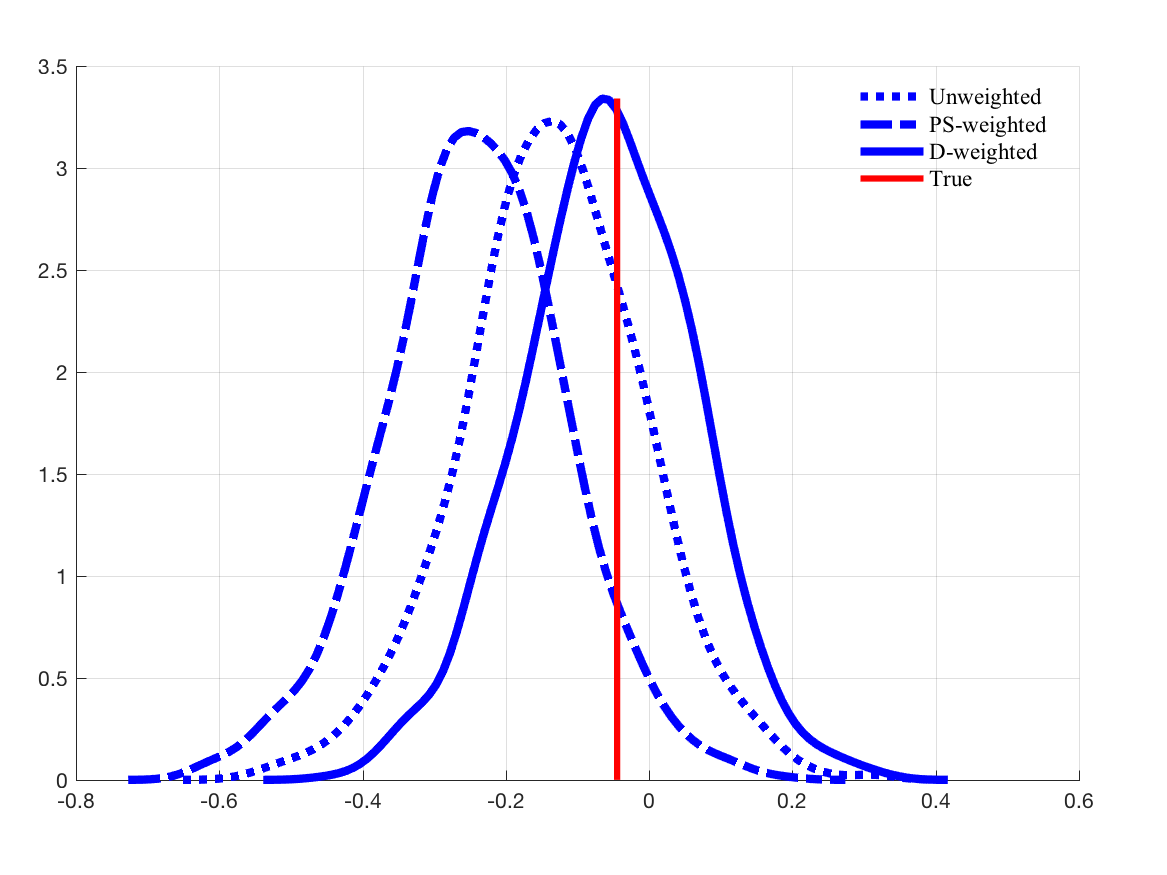}
		\end{minipage}
		\begin{minipage}{\columnwidth}
			\begin{spacing}{0.9}
				{\footnotesize \textit{Notes:} This figure plots the empirical distributions of the unweighted, ps-weighted, and d-weighted UQTE estimates using 1,000 Monte Carlo simulation draws of sample size 5,000.  The average treated sample is $N_1 = 5,000\times 0.41\times 0.38 = 779$ and average control sample is $N_0 = 5,000\times (1-0.41)\times 0.38 = 1,121$. The unweighted estimator does not weight the observed data. The ps-weighted estimator weights to correct only for nonrandom assignment and the d-weighted estimator weights by both the treatment and missing outcomes propensity score models to deal with nonrandom assignment and missing outcome problems. \par}
			\end{spacing}
		\end{minipage}
	\end{figure}	
\end{document}